\documentclass[a4paper,12pt]{article}

\usepackage{fancyhdr}
\usepackage[a4paper, left=2.5cm, right=2.5cm, top=2.5cm, bottom=2.8cm]{geometry}
\usepackage{setspace}
\fancypagestyle{plain}{ 
  \fancyhf{}
  \cfoot{\thepage} 
  }

\usepackage[utf8]{inputenc}
\usepackage[english]{babel}
\usepackage{csquotes} 

\usepackage{titlesec}
\titleformat{\paragraph}[hang]{\normalfont\normalsize\bfseries}{\theparagraph}{1em}{}

\titlespacing*{\paragraph}{0pt}{3.25ex plus 1ex minus .2ex}{0.5ex plus .2ex}

\newcommand{\subsubsubsection}{\paragraph}

\usepackage{multicol}

\usepackage[pdftex]{graphicx}
\usepackage{amsmath}
\usepackage{multirow} 
\usepackage{tablefootnote} 
\usepackage{threeparttable} 
\usepackage[format=plain,labelsep=period,bf]{caption} 

\usepackage[
    backend=biber,       
    style=numeric-comp,  
    sorting=none,        
    sortcites=true,      
    giveninits=true,     
    maxbibnames=99,      
    uniquename=init,
]{biblatex}

\let\cite\parencite 

\DeclareNameAlias{default}{given-family} 
\newbibmacro*{labelnamepunct}{\addcolon\space} 

\DeclareFieldFormat{date}{\mkbibbold{#1}} 
\DeclareFieldFormat{pages}{#1} 
\DeclareFieldFormat{journaltitle}{\mkbibemph{#1}} 
\DeclareFieldFormat{title}{\mkbibemph{#1}} 

\DeclareFieldFormat[article]{volume}{\mkbibemph{#1}}

\DeclareBibliographyDriver{article}{%
  \usebibmacro{bibindex}%
  \usebibmacro{begentry}%
  \printnames{author}%
  \usebibmacro{labelnamepunct}%
  \printfield{journaltitle}%
  \setunit{\addspace}%
  \printtext[parens]{\usebibmacro{date}}%
  \setunit{\addcomma\space}%
  \printfield{volume}%
  \setunit{\addcomma\space}%
  \printfield{pages}%
  \usebibmacro{finentry}
}

\renewbibmacro*{publisher+location+date}{%
    \printlist{publisher}%
    \setunit*{\addcomma\space}%
    \printlist{location}%
    \setunit*{\addcomma\space}%
    \usebibmacro{date}
}
\DeclareBibliographyDriver{book}{%
  \usebibmacro{bibindex}%
  \usebibmacro{begentry}%
  \printnames{author}%
  \usebibmacro{labelnamepunct}%
  \usebibmacro{title}%
  \setunit{\addspace}%
  \printtext[parens]{%
    \usebibmacro{publisher+location+date}%
  }%
  \usebibmacro{finentry}%
}

\DeclareBibliographyDriver{inbook}{%
  \usebibmacro{bibindex}%
  \usebibmacro{begentry}%
  \usebibmacro{author/translator+others}%
  \setunit{\printdelim{nametitledelim}}\newblock
  \usebibmacro{title}%
  \newunit
  \printlist{language}%
  \newunit\newblock
  \usebibmacro{byauthor}%
  \newunit\newblock
  \usebibmacro{in:}%
  \usebibmacro{bybookauthor}%
  \newunit\newblock
  \usebibmacro{maintitle+booktitle}%
  \newunit\newblock
  \usebibmacro{byeditor+others}%
  \newunit\newblock
  \printfield{edition}%
  \newunit
  \iffieldundef{maintitle}
    {\printfield{volume}%
     \printfield{part}}
    {}%
  \newunit
  \printfield{volumes}%
  \newunit\newblock
  \printfield{series}%
  \newunit\newblock
  \printfield{note}%
  \newunit\newblock
  \printtext[parens]{%
    \usebibmacro{publisher+location+date}%
    \setunit{\addcomma\space}%
    \usebibmacro{chapter+pages}%
    \setunit{\addcomma\space}%
  }%
  \newunit\newblock
  \usebibmacro{addendum+pubstate}%
  \setunit{\bibpagerefpunct}\newblock
  \usebibmacro{pageref}%
  \newunit\newblock
  \iftoggle{bbx:related}
    {\usebibmacro{related:init}%
     \usebibmacro{related}}
    {}%
  \usebibmacro{finentry}}

\usepackage{hyperref} 
\usepackage{soul} 
\usepackage{chemformula,mhchem,array}
\usepackage{braket}
\newcommand{\Romannum}[1]{\uppercase\expandafter{\romannumeral #1\relax}}
\usepackage[utf8]{inputenc} 
\usepackage{textcomp}         
\usepackage{xurl}
\usepackage{hyperref}
\usepackage{parskip}

\usepackage[
    font=small,       
    labelfont={bf,it},       
    textfont=normalfont,     
    singlelinecheck=false    
]{caption}

\begin{document}

\begingroup
\begin{titlepage}
    \begin{figure}[t!]
        \centering
        \includegraphics[scale=0.25]{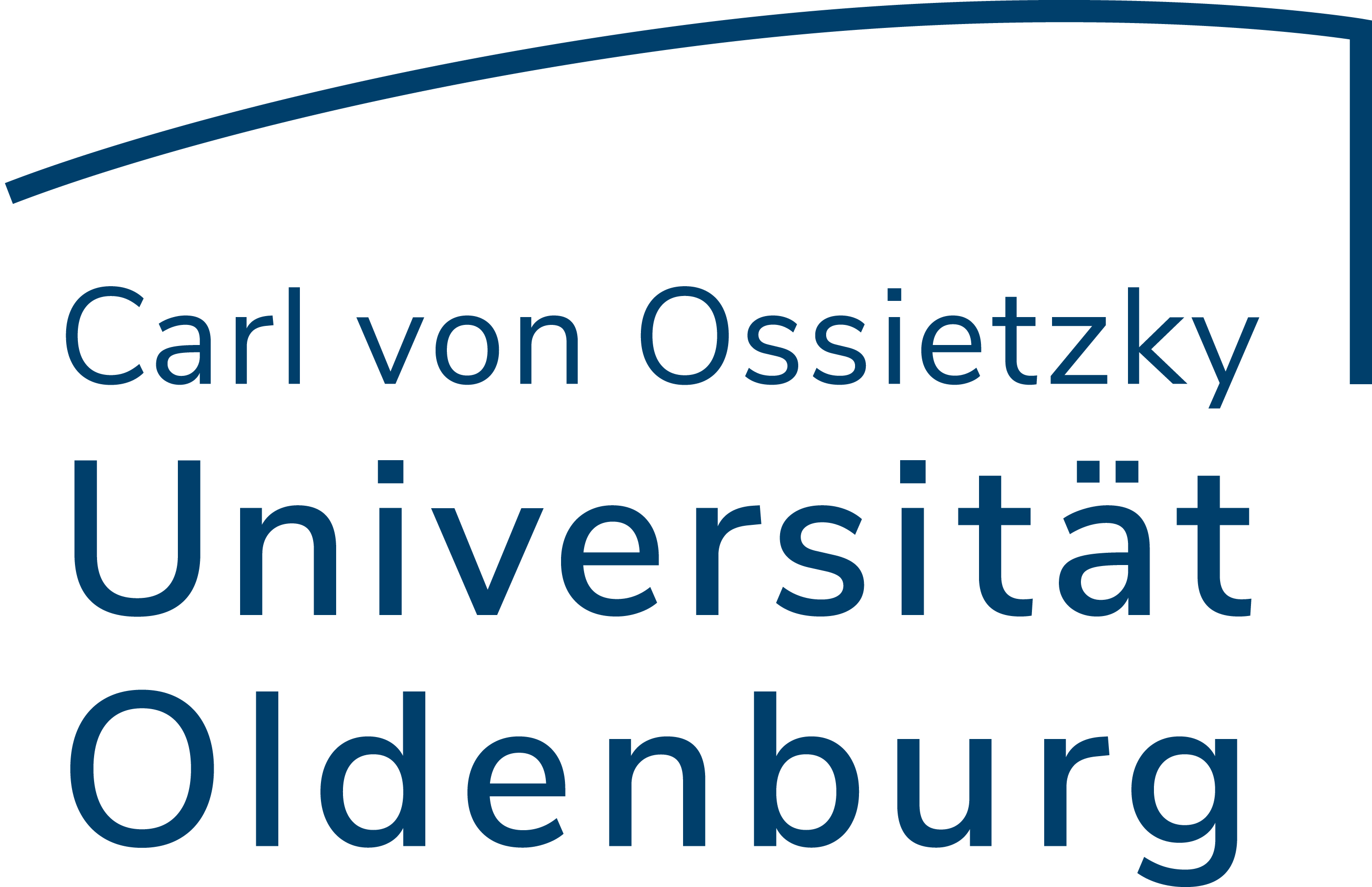}
        \label{fig:my_label}
    \end{figure}
    \begin{center}
        Institute of Chemistry \\
        \vspace{1cm}
        Master's Program Chemistry \\
        \vspace{1cm}
        \large\textbf{Master's Thesis}\\
        \vspace{1cm}
        \textbf{A Theoretical Investigation of the Thermal and Photochemical Mechanisms of Ethylbenzene Dehydrogenation on Rutile \ch{TiO2}(110)}\\
        \vspace{1cm}
        \normalsize
        Submitted by: Nico Yannik Merkt\\
        \vspace{1cm}
        Matriculation Number: 3994107\\
        \vspace{5cm}
        Supervising reviewer: Prof. Dr. Thorsten Klüner\\
        \vspace{0.5cm}
        Second reviewer: Dr. Lars Mohrhusen\\
        \vspace{1cm}
        Oldenburg, 11.03.2026
    \end{center}
\end{titlepage}
\endgroup

\thispagestyle{plain}
\setcounter{page}{2}
\tableofcontents

\newpage
\thispagestyle{plain}
\listoffigures

\newpage
\thispagestyle{plain}
\listoftables

\newpage
\thispagestyle{plain}
\clearpage
\phantomsection
\addcontentsline{toc}{section}{Acknowledgements}
\paragraph*{\Large{Acknowledgements}}\mbox{}\vspace{0.5cm}\\
After finishing this work my special thanks goes to Prof. Dr. Thorsten Klüner for giving me the opportunity to write this thesis in his working group. Furthermore, I would like to express my gratitude to Dr. Lars Mohrhusen for his supervision as second examiner in this thesis. Further, I appreciate the introduction and practical input to this topic by Dr. Fangliang Li and Prof. Dr. Qing Guo, as well as their constructive response to my related questions. I would also like to thank my family who supported and helped me during my studies. Additionally, I would like to thank the Theoretical Chemistry working group in Oldenburg in general for their assistance with questions and their good reception in the working group. Finally, I would like to appreciate the team of the HPC cluster ROSA of the University of Oldenburg and Robert Röhse for their technical support.

\newpage
\thispagestyle{plain}
\section{Introduction} \label{1}
The catalytic dehydrogenation of ethylbenzene (EB) to styrene is one of the most significant processes in the current chemical industry. Styrene serves as the essential precursor for a wide range of polymers, such as polystyrene (PS) and acrylonitrile-butadiene-styrene (ABS), which are widespread in products ranging from packaging to automotive parts. Currently, the established industrial process relies on thermal dehydrogenation at very high temperatures (550–650°C) over iron oxide catalysts. This process accounts for 80–90\% of the global styrene production.\supercite{Fan, Soares2025} However, this method has significant disadvantages, including immense energy consumption and undesirable side reactions which are reducing the efficiency of the reaction and the catalyst longevity.\supercite{Fan, Soares2025, Zhou2022, Lai2023, Sanz2015}

Photocatalytic dehydrogenation on semiconductor materials, such as titanium dioxide (\ch{TiO2}), is a promising and more sustainable alternative to the thermal reaction. This process allows the reaction to occur under significantly milder conditions.\supercite{Lin2020, Lai2022, Lai2023, Li2022_JACS} While \ch{TiO2}-based catalysts show promising selectivity, essential mechanistic details are unknown. In particular, details related to overcoming high activation barriers at low temperatures are missing.\supercite{Lin2020, Lai2022} Experimental findings suggest that the dehydrogenation is kinetically limited by a high activation energy of the second C-H bond cleavage at the $\beta$-carbon atom.\supercite{Lai2022, Lai2023, Li2022_Lett} Photocatalytic approaches offer the possibility of bypassing this barrier. However, the efficiency depends heavily on the incident wavelength and the oxidation state of the catalyst surface.\supercite{Lai2023, Li2022_Lett} This finding contradicts the assertion of the simple photocatalysis model that hot electrons/holes rapidly relax to the band edge and excess energy is not utilized effectively. 

The central motivation of this work is the necessity to understand these complex thermal and photochemical reaction pathways. First, the applied "charged cluster" surface model is validated by comparing calculated electronic properties and adsorption energies with temperature programmed desorption (TPD) data and literature benchmarks.\supercite{Gerhards, Gerhards2021, Gerhards2022} For this comparison, a dual methodological approach is employed. Density functional theory (DFT) is used for extensive geometry optimizations, while a multi reference method (CASSCF) is utilized to accurately describe the static correlation and the topology of excited state energy profile. All quantum chemical simulations presented in this work were performed using the program ORCA\supercite{Neese2012, Neese2025}.

A key challenge addressed in this study is the high dimensionality of the chemical space. With 3$N$-5 ($N$ - number of atoms) degrees of freedom, a complete mapping of the potential hyperplane is computationally impossible for the large cluster models required to simulate the \ch{TiO2}(110) surface. Therefore, the investigation focuses on a systematic evaluation of a one dimensional energy profile along the predicted minimum energy path (MEP), identified via DFT and validated via complete active space self consistent field (CASSCF) calculations. This allows for a detailed characterization of the reaction mechanism, distinguishing between proton-coupled electron transfer (PCET) and direct hydrogen atom transfer (HAT). Additionally, this thesis examines the "hot hole" theory by analyzing how higher photon energies enable the system to navigate through excited electronic manifolds and evade the energy barriers of the ground state. The work concludes with an examination of the oxidized surface.

\section{Fundamentals and State of Research} \label{2}
Converting alkanes into olefins is an important process in the chemical industry. The main purpose of olefins is their function as monomers in subsequent polymerization processes. The dehydrogenation of ethylbenzene (EB) to styrene is a key reaction in this context. Styrene is an essential monomer for the production of a variety of polymers, including polystyrene (PS), styrene-butadiene rubber (SBR), and acrylonitrile butadiene styrene (ABS). These polymers are used in countless everyday products, ranging from packaging in form of styrofoam to car parts and electronics housings.\supercite{Fan, Soares2025, Lai2022}

\subsection{Significance in Industry and Science} \label{2.1}
\begin{figure}[h]
    \centering
    \includegraphics[scale=0.75]{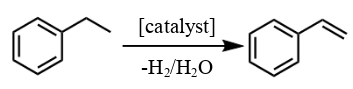}
    \caption{The chemical equation of the industrial synthesis of styrene from ethylbenzene.}
    \label{fig:1}
\end{figure}

Currently, about 80-90\% of the global styrene production is covered by the direct dehydrogenation of ethylbenzene in the chemical industry.\supercite{Fan, Soares2025} Direct dehydrogenation is typically conducted at very high temperatures between 550-650°C over a potassium-promoted iron oxide catalyst and in the presence of superheated steam (reaction equation is shown in figure \ref{fig:1}). In this process, excess steam is used as a heat carrier for the highly endothermic reaction ($\Delta H~\approx$~129.7~kJ/mol~$\approx$~1.34~eV). It also serves as a diluent to shift the reaction equilibrium in favor of the products and to reduce coke formation. Despite its established application, the process has serious disadvantages, mainly the immense energy consumption due to the necessary high temperatures. Also, these high temperatures promote undesirable side reactions, such as thermal cracking and coke formation, leading to catalyst deactivation and a shift in selectivity towards side products. The reaction is also constrained by its specific thermodynamic equilibrium, which limits the attainable conversions. These limitations motivate the search for alternative, ideally less complex catalytic pathways that enable the conversion under milder conditions. The future development of sustainable technologies is dependent on a fundamental understanding of the underlying thermal and photochemical mechanisms.\supercite{Fan, Soares2025, Zhou2022, Lai2023, Sanz2015}

These industrial challenges are the driving force behind this research. From a literature perspective, the central challenge lies in the selective activation of the kinetically inert C($sp^3$)-H bonds of the ethyl group. The industry relies on the thermal reaction with complex catalysts, but research focuses on well-defined models, such as the rutile (110) surface, to understand elementary reaction steps.\supercite{Wang2022, Lai2023} A particularly promising alternative to the thermal approach is photocatalysis. It offers the potential to provide the necessary activation energy through light instead of thermal energy, thus enabling the reaction under significantly milder conditions.\supercite{Lin2020, Li2022_Lett, Lai2022, Lai2023} On semiconductor catalysts like \ch{TiO2}, light irradiation can generate electron-hole pairs through the excitation of an electron. Electron-hole pairs are likely to be generated in the bulk phase and subsequently are transferred and trapped on the surface.\supercite{Henderson2011} On the surface, these highly reactive species can directly abstract hydrogen atoms from hydrocarbons, a process known as hydrogen atom transfer (HAT). The examination of such photochemical pathways on model surfaces is essential to clarify the exact route of C-H activation and to formulate new approaches for this reaction.\supercite{Zhou2022, Li2022_JACS, Lin2020}	

\subsection{Titanium Dioxide as Catalyst for Dehydrogenations} \label{2.2}
In the search for catalysts for photocatalytic ethylbenzene dehydrogenation, titanium dioxide (\ch{TiO2}) has proven to be a promising material. \ch{TiO2} is a cost effective, chemically stable semiconductor with well known surface properties. This makes its surface an attractive candidate for heterogeneous catalytic applications.\supercite{Zhou2022, Li2022_JACS} A particularly well known form of \ch{TiO2} is the commercial catalyst P25, a mixture of its main crystalline phases, anatase and rutile. Experimental studies have demonstrated the exceptional performance of P25 in the thermocatalytic dehydrogenation of ethylbenzene, achieving high selectivities to styrene of over 95\%.\supercite{Lin2020, Lai2022}

The high activity of P25 is due to its mixed-phase nature, but this also complicates the investigation of reaction mechanisms. This is especially the case in investigations through quantum chemical methods. However, the literature on ethylbenzene dehydrogenation indicates that the specific crystal phase plays a subordinate role. Due to the observed high selectivity on P25 literature \cite{Lin2020} suggests that the reaction pathway on the different \ch{TiO2} surfaces is very similar. This assumption is of considerable importance as it legitimizes the use of a well defined single crystal surface as a representative model system to clarify the elementary steps of the catalytic reaction.

\begin{figure}[t!]
    \centering
    \includegraphics[scale=0.65]{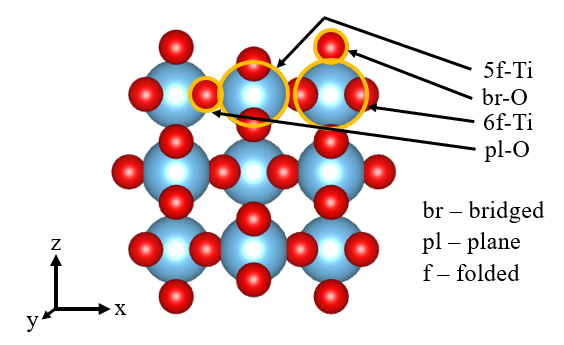}
    \caption{Representation of rutile(110) bulk. The blue atoms represent titanium, and the red atoms symbolize oxygen. The orange-circled atoms are the relevant oxygen and titanium species.\supercite{Henderson2011, Schneider2014}}
    \label{fig:2}
\end{figure}

In this context, the rutile phase of \ch{TiO2}, specifically its (110) surface, was chosen as the model system for this work. This choice is justified and supported by the following fundamental facts. Rutile is the thermodynamically most stable polymorph of titanium dioxide and its (110) orientation is one of its most studied crystal facets.\supercite{Wang2022, Zhang2006} The bulk of rutile consists of a tetragonal crystal system in which the titanium atoms are octahedrally (6f-Ti) and the oxygen atoms are trigonally coordinated. The \ch{TiO2} octahedra are distorted and interconnected in various ways, leading to the many individual structures and energetic properties of \ch{TiO2}. Furthermore, the rutile (110) surface structure consists of rows of twofold coordinated bridging oxygen atoms (O$_{br}$) and coordinatively unsaturated, fivefold coordinated titanium atoms (5f-Ti) (see figure \ref{fig:2}).\supercite{Gerhards, Zhou2022, Chen2019} These undercoordinated atoms function as primary adsorption and reaction centers.

\begin{figure}[b!]
    \centering
    \includegraphics[scale=0.8]{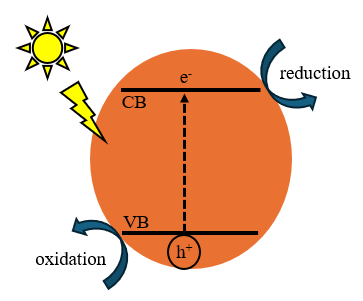}
    \caption{Schematic representation of the underlying step of a photochemical reaction. In this step, a photon ($h\cdot\nu$) excites the semiconductor material, transferring an electron (e$^{-}$) to the conduction band (CB) and leaving a hole (h$^{+}$) in the valence band (VB).\supercite{Schneider2014, Henderson2011}}
    \label{fig:3}
\end{figure} 	

In addition to these structural features, the electronic properties of rutile are crucial for its photocatalytic potential. With a band gap of about 3.03~eV \supercite{Amtout1995, Tang1995}, the material is capable of absorbing ultraviolet light.\supercite{Schneider2014} These characteristic enable the photocatalytic process, which can be divided into several fundamental steps. The absorption of a photon with an energy greater than the band gap leads to the excitation of an electron (e$^-$) from the valence band (VB) to the conduction band (CB). This leaves a positive charged hole (h$^+$) in the valence band (see figure \ref{fig:3}). In order to prevent immediate recombination, the newly formed charge carriers must migrate to the catalyst surface and be separated. One effective method for separating these charge carriers is the use of a second catalyst component, such as a second semiconductor or a transition metal, which leads to a so-called heterojunction.\supercite{Low2017} For example, by introducing a second semiconductor with a higher CB or lower VB, the charge carriers will accumulate on opposite semiconductors. Thus, heterojunction would mainly lead to an increase in the catalysts efficiency. As this study focuses on the fundamental mechanism, a catalyst mixture will not be considered for the sake of simplicity. Once at the surface, the charge carriers are transferred to adsorbed molecules. The electron in the CB initiates a reduction reaction, while the hole in the VB triggers an oxidation reaction.\supercite{Schneider2014, Low2017} For the dehydrogenation reaction studied here, the hole is particularly important as it has the ability to abstract hydrogen atoms from adsorbed ethylbenzene. This combination of a well defined structure and fundamental photocatalytic activity justifies the rutile-\ch{TiO2}(110) surface as an excellent model system.

\subsection[Adsorption/Thermal Activation of Ethylbenzene on \ch{TiO2}(110)]{Adsorption/Thermal Activation of Ethylbenzene on\\ \ch{TiO2}(110)} \label{2.3}
After describing the fundamental properties of titanium dioxide as a catalyst, the first and crucial step for any catalytic reaction is the adsorption of the reactant molecule onto the surface of the catalyst. In the case of ethylbenzene on rutile-\ch{TiO2}(110), studies using scanning tunneling microscopy (STM) and density functional theory (DFT) show that ethylbenzene is weakly physisorbed at low temperatures. The adsorption energy is approximately -1~eV, and desorption occurs at about 245 K according to temperature programmed desorption spectroscopy (TPD).\supercite{Chen2019, Li2022_Lett, Lai2022} During this adsorption, the aryl group of ethylbenzene (EB) preferentially aligns over a row of 5f-Ti atoms, and the ethyl group is slightly bent upwards over the neighboring O$_{br}$ row. STM images confirm this "cashew-like" adsorption geometry.\supercite{Lin2020}

The primary driving force for the binding is a dative Lewis acid-base interaction between the $\pi$-electron system of the aromatic ring and the undercoordinated 5f-Ti centers, which act as Lewis acids.\supercite{Chen2019, Henderson2011} The physisorbed molecules exhibit high mobility on the surface and can be manipulated by the STM tip, confirming a weak bond and thus a low adsorption energy.\supercite{Lin2020}

The thermal activation of adsorbed ethylbenzene, i.e., its dehydrogenation through the application of heat alone, is an inefficient process requiring high temperatures. Mechanistically, this ground state reaction is driven by the Lewis acid-base pair character of the \ch{TiO2} surface. Unlike radical reactions, the thermal C-H bond cleavage typically follows a heterolytic pathway, often described as proton-coupled electron transfer (PCET). Here, the under-coordinated sites are acting as a Lewis acid stabilizing the electron-rich ethylbenzene fragment, while the basic bridging oxygen (O$_{br}$) abstracts the proton (H$^+$).\supercite{Zhou2022, Chen2019} Due to the lack of open-shell radical species on the stoichiometric surface, overcoming the activation barrier for this concerted bond breaking requires significant thermal energy.\supercite{Lai2022, Lai2023, Li2022_JACS}

Regarding the regioselectivity of this first step, literature discusses cleavage at either the $\alpha$- \supercite{Lai2022, Lai2023, Li2022_Lett} or $\beta$- \supercite{Lin2020} carbon atom of the ethyl group. While earlier studies suggested $\beta$-cleavage, more recent findings point towards a preference for $\alpha$-C-H dehydrogenation, based on the identification of specific coupling products (dimers) formed from phenylethyl intermediates.\supercite{Lai2022, Lai2023, Li2022_Lett} Consequently, the initial thermal process results in a surface-bound phenylethyl intermediate which is stabilized by the titanium center and a hydroxyl group at the bridging oxygen row. Notably, this activation step is significantly facilitated by excessive surface oxygen species, such as peroxo-\ch{TiO2^{2-}} on oxidized \ch{TiO2}(110), allowing the reaction to proceed at temperatures below 285 K.\supercite{Lai2022}

The rate determining step of the overall reaction appears to be the subsequent cleavage of the second C-H bond, which occurs at the $\beta$-carbon atom. This second dehydrogenation step has a high activation barrier. In oxidative dehydrogenation with peroxo-\ch{TiO2^{2-}} this high barrier is known to be about 1.2~eV.\supercite{Lin2020} Therefore, temperatures of around 400 K (approximately 130 °C) or higher are needed for significant reaction rates.\supercite{Lin2020, Lai2022} The high thermal requirement for the second dehydrogenation step is the main reason for the low efficiency of purely thermal catalysis. Excess oxygen makes milder conditions possible. This however significantly affects the selectivity towards styrene, reducing the reachable percentage from $>$95\% to about 37\%. Furthermore, higher temperatures improve the desorption of the product styrene, which has its desorption peak in TPD at 435~K, binding more strongly to \ch{TiO2}(110) than ethylbenzene.\supercite{Li2022_Lett} The improved desorption of styrene is also important for the practical application of the process to prevent potential product poisoning of the surface. Thus, the necessity of alternative activation pathways, such as photocatalysis, becomes evident in order to overcome this kinetic hurdle and enable efficient styrene synthesis at lower temperatures.

\newpage

\subsection{Photochemical Activation of Hydrocarbons on \ch{TiO2}(110)} \label{2.4}
In contrast to the purely thermal activation, which requires high temperatures to overcome activation barriers, photochemical activation offers an alternative reaction pathway that can proceed under milder conditions. The fundamental idea of photocatalysis on semiconductors like titanium dioxide is the utilization of light energy to generate reactive charge carriers, electron-hole pairs. The hole (h$^+$), a defect electron in the valence band of the catalyst, is crucial for the hydrogenation reactions of hydrocarbons. This hole can act as a strong oxidizing agent and initiate the cleavage of inert C-H bonds via the so-called HAT mechanism.\supercite{Schneider2014, Zhou2022}

\subsubsection{Reaction Mechanism} \label{2.4.1}
The photocatalytic mechanism of C-H bond activation on \ch{TiO2}(110) differs fundamentally from common thermal processes. While thermal activation often proceeds via heterolytic cleavage pathways such as the proton-coupled electron transfer (PCET), the photocatalytic reaction is dominated by a homolytic mechanism, the hydrogen atom transfer (HAT).\supercite{Zhou2022, Li2022_Lett}

\begin{figure}[h!]
    \centering
    \includegraphics[scale=0.86]{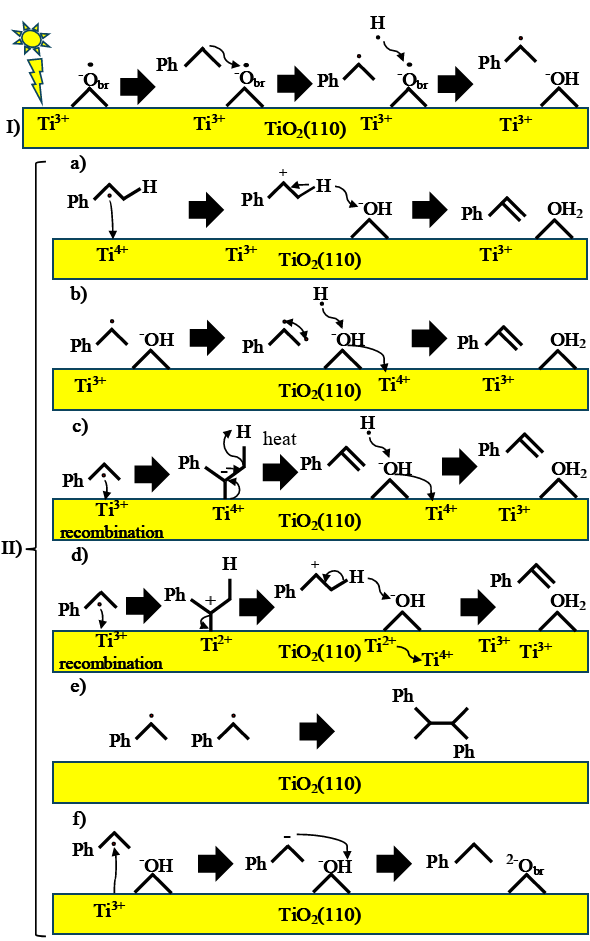}
    \caption{A proposed photocatalytic reaction mechanism for the dehydrogenation of ethylbenzene to styrene. This mechanism was created using information from source \supercite{Chen2019, Lai2022, Lai2023, Li2022_Lett, Lin2020}.}
    \label{fig:4}
\end{figure}

The driving force of the photochemical mechanism is the absorption of a photon by the \ch{TiO2} crystal lattice. This energy excites an electron from the VB, which is primarily composed of the occupied O(2p) orbitals, into the empty CB, which is mainly formed by the unoccupied Ti(3d) orbitals. The result is a spatially separated electron-hole pair (e$^-$/h$^+$). The positively charged hole (h$^+$) preferentially localizes on a bridging oxygen atom (O$_{br}$) of the surface, transforming a lattice O$^{2-}$ ion into a highly reactive, paramagnetic oxygen radical anion (\ch{O_{br}^{.-}}). Simultaneously, the electron (e$^-$) typically localizes on a nearby 5f-Ti atom, which is consequently reduced from Ti$^{4+}$ to Ti$^{3+}$. This highly reactive \ch{O_{br}^{.-}} radical acts as a strong oxidizing agent and can abstract a neutral hydrogen atom (a proton and its electron) directly without a high energy barrier from heterolytic splitting of a C-H bond of the adsorbed ethylbenzene. In contrast, the thermal pathway is kinetically hindered because the closed-shell O$_{br}^{2-}$ ion in the electronic ground state lacks this radical character. Consequently, it cannot easily abstract a neutral hydrogen atom but must rely on the concerted, yet energetically demanding, PCET mechanism involving adjacent titanium sites. This first HAT step leads to the formation of two radicals, a secondary phenylethyl radical and a hydrogen radical (\ch{H^{.}}), which remain on the surface. The \ch{H^{.}} and the \ch{O_{br}^{.-}} can then combine to form an OH group while retaining the negative charge.\supercite{Zhou2022, Wang2022, Lin2020, Li2022_Lett} This part of the reaction is described as Step \Romannum{1} in figure \ref{fig:4}. This HAT process (Step \Romannum{1}) has been demonstrated with methane and ethane as examples, where the activation barrier for the photochemical C-H cleavage was lowered by about 70\% compared to thermal catalysis.\supercite{Zhou2022}

\clearpage

The selectivity of the reaction critically depends on the position of the initial HAT step and the subsequent resulting phenylethyl radical. The literature suggests that the kinetically favored initial attack occurs at the $\alpha$-carbon atom yielding the more stable, secondary benzylic radical, rather than the sterically more accessible $\beta$-carbon atom. Regardless of the initial point of attack, the formation of styrene requires a second dehydrogenation step. According to the literature, a single photogenerated hole is sufficient to cleave two C-H bonds consecutively, highlighting the efficiency of the photochemical pathway.\supercite{Lai2023}

After the initial photo-oxidation step the surface holds a hydroxyl group, a phenylethyl radical, and a reduced Ti$^{3+}$ center. The subsequent reaction pathway determines the final product distribution. Several competing pathways emerge from this key intermediate state, including four distinct pathways (Step \Romannum{2}a-d) leading to the formation of styrene. For the illustration of these subsequent routes, please refer to figure \ref{fig:4}.

In the first pathway, \Romannum{2}a, a sequential electron-proton transfer is considered. Here, the phenylethyl radical can undergo further oxidation by transferring its unpaired electron to an adjacent Ti$^{4+}$ site on the surface, forming a second Ti$^{3+}$ center. This step results in the formation of a transient phenylethyl cation (\ch{C8H9+}). Due to its high acidity, this cation immediately undergoes deprotonation at the $\beta$-carbon. The strongest base available on the surface is the adjacent hydroxyl group, which abstracts the proton to form a water molecule, while the C-C double bond of the product styrene is formed.

Pathway \Romannum{2}b follows a concerted oxidative HAT mechanism. Here, the remaining oxidative potential of the system after step \Romannum{1} drives a second HAT from the $\beta$-position of the phenylethyl radical directly to the neighboring hydroxyl group. As the new O-H bond in the forming water molecule is established, the C-C double bond of styrene is simultaneously created, and the leftover radicals electron is transferred to the surface, generating a second Ti$^{3+}$ center.

Pathway \Romannum{2}c is another possible route, i.e. a thermally activated anion elimination, which explains the experimentally observed temperature dependence of the second dehydrogenation step. Initially, the phenylethyl radical recombines with the electron from the nearby Ti$^{3+}$ center, forming a stable, chemisorbed phenylethyl anion (\ch{C8H9-}) and re-oxidizing the titanium site to Ti$^{4+}$. This carbanionic intermediate is stable at low temperatures but can undergo an elimination reaction upon heating (e.g., during a TPD experiment $>$400 K). This step cleaves the C-H bond at the $\beta$-position, releasing the final styrene product and a \ch{H^{.}}. The \ch{H^{.}} then reacts with a surface Ti species, donating an electron and forming water with the hydroxyl group.

A more detailed perspective is offered by the organometallic intermediate pathway \Romannum{2}d. Instead of remaining as a weak physisorbed radical, the phenylethyl radical (\ch{C8H9^{.}}) can directly couple with the adjacent, paramagnetic Ti$^{3+}$ center. The two unpaired electrons form a new, covalent carbon-titanium bond, resulting in a stable, chemisorbed organometallic intermediate (\ch{C8H9-Ti(IV)}). This species represents a significant thermodynamic sink, potentially stabilizing the product of the first HAT step. The final release of styrene from this stable intermediate then requires the heterolytic cleavage of the C~-~Ti bond, which can proceed in two distinct ways. In the first variant, the bond breaks such that the electron pair remains at the titanium center, releasing a phenylethyl cation (\ch{C8H9+}) and  forming a Ti$^{2+}$ ion on the surface. The second electron then is distributed on the surface leading to a second Ti$^{3+}$ center. Alternatively, if the bond cleaves with the electron pair moving to the carbon, a phenylethyl anion (\ch{C8H9-}) and a Ti$^{4+}$ site are formed, leading directly to the carbanionic intermediate central to the mechanism described in \Romannum{2}c. Therefore, this organometallic pathway serves as a crucial branching point that can lead the reaction towards either a spontaneous or a thermally activated route to styrene, depending on the relative energetics of the C-Ti bond cleavage. These pathways to styrene are in direct competition with non-selective side reactions. 

In pathway \Romannum{2}e, the experimentally observed coupling product 2,3-diphenylbutane can be formed.\supercite{Lai2022} This indicates the formation of $\alpha$-phenylethyl radicals on the surface. If the concentration of these radicals is sufficiently high, two of the radicals can combine to form a dimer before the second dehydrogenation occurs. 

Lastly, in pathway \Romannum{2}f, the phenylethyl radical can also recombine with the electron at the Ti$^{3+}$ center to form a carbanion, which could subsequently be protonated by the surface hydroxyl group, thereby regenerating the original ethylbenzene molecule. This pathway represents a deactivation or back reaction route that reduces the overall efficiency of styrene formation. The entire proposed mechanism, which considers these competing pathways, is illustrated in figure \ref{fig:4}.

\subsubsection{Role of the Photon and Wavelength Dependence} \label{2.4.2}
The traditional model of \ch{TiO2} photocatalysis assumes that photogenerated charge carriers release their excess energy extremely quickly to the crystal lattice through thermal relaxation and accumulate at the band edges. In this scenario, reactivity depends solely on the number of charge carriers generated and not on the energy of individual photons, given that the energy exceeds the band gap.\supercite{Xu2021, Lai2023}

However, recent studies on the dehydrogenation of hydrocarbons contradict this simplified model. Research has shown that the reaction rate depends significantly on the wavelength of the incident light, especially in the dehydrogenation of ethylbenzene. A direct comparison of irradiation at 343~nm (3.6~eV) and 257~nm (4.8~eV) showed that the rate of the initial, kinetically simple $\alpha$-C-H cleavage step is nearly identical for both wavelengths. Moreover, the second rate determining $\beta$-C-H cleavage steps rate is higher when 257~nm wavelength irradiation is used, which leads to a significantly increased styrene yield.\supercite{Lai2023}

This observation is explained in literature by the concept of "hot" or "deep" holes. Photons with higher energy generate charge carriers that are far from the band edges and possess excess kinetic energy. These non-thermalized charge carriers can react directly with adsorbates before they relax. Such "deep" holes are stronger oxidizing agents than the holes thermalized at the valence band edge. Therefore, they overcome higher activation barriers more efficiently, such as the second C-H bond cleavage in ethylbenzene.\supercite{Lai2023, Xu2021} Another possibility is that the entire reaction occurs in an higher excited state, which is achieved through the excess energy. Thus, the energy of the photon plays a crucial role that goes beyond the mere generation of charge carriers.

\subsubsection{Influence of Surface Properties: Oxidized Surface and HAT Mechanism} \label{2.4.3}
The photocatalytic activity is not only dependent on the energy of the charge carriers but also significantly on the specific adsorption sites and the chemical state of the \ch{TiO2}(110) surface. The two primary adsorption centers are the titanium (5f-Ti) and the bridging oxygen rows (O$_{br}$). Adsorption of alkanes on 5f-Ti sites in the ground state is energetically favorable. The bridging oxygen atoms are the primary sites where photogenerated holes are localized to form the highly reactive oxygen radicals necessary for the HAT mechanism.\supercite{Wang2022, Zhou2022} As explained in literature \cite{Li2022_Lett}, the HAT mechanism produces water as a byproduct. This creates defect sites in the O$_{br}$ rows, resulting in catalyst degeneration.

Experimental studies show that the photocatalytic dehydrogenation of propane or ethylbenzene on a stoichiometric, reduced R-\ch{TiO2}(110) surface is extremely inefficient or even completely absent.\supercite{Lai2022, Li2022_JACS} Only a pre-oxidation of the surface, for example through oxygen dosing, leads to higher catalytic activity. The pre-oxidation leads to new, highly reactive oxygen species, which are atomic oxygen bound to 5f-Ti centers (O$_{Ti}$) or peroxo species (\ch{O2^{2-}}). These new species increase activity due to a regenerative effect on previously mentioned catalyst loss. Another possibility is that these adsorbed species act as the actual "electron scavengers" or "proton acceptors". They attract the photogenerated electrons and are thereby converted into radical species (e.g., \ch{OTi^{-.}}), which then function as the primary hydrogen abstractors in the HAT mechanism. The surface condition, particularly the oxidation state, is crucial for the generation of the specific active sites required for efficient photocatalytic C-H activation.\supercite{Lai2022, Li2022_JACS, Wang2022}

\newpage
\thispagestyle{plain}
\section{Quantum Chemical Investigation Methods} \label{3}
In order to develop a mechanistic understanding of the thermal and photocatalytic dehydrogenation of ethylbenzene to styrene, extensive quantum chemical simulations are necessary. This chapter provides the theoretical foundations required for the interpretation of surface processes and the analysis of reaction mechanisms. First, the fundamental approaches for calculating the electronic structure of systems in the ground state and in electronically excited states, will be described. Subsequently, the surface model used is explained in detail. All calculations presented in this work were performed using the program ORCA\supercite{Neese2012, Neese2025} version 6.0.1. 

\subsection{Calculation of the Electronic Structure} \label{3.1}
The accurate description of chemical reactions requires a precise calculation of the electronic structure of all involved species. Since the thermal reaction occurs in the electronic ground state, while the photocatalytic pathway is initiated by electronic excitation, different theoretical approaches are necessary for the respective processes. The following chapters describe the challenges of selecting appropriate methods for describing the ground and excited states.

\subsubsection{The Quantum Chemical Many-Body Problem} \label{3.1.1}
The basis of any quantum chemical calculation is the solution of the time-independent Schrödinger equation, which describes the relationship between the energy and the wave function of a system via:
\begin{equation}
    \hat{H}\Psi_i(x_1, x_2, ..., x_N, R_1, R_2, ..., R_M) = E_i\Psi_i(x_1, x_2, ..., x_N, R_1, R_2, ..., R_M).
	\label{eq.1}
\end{equation}
$\hat{H}$ is the Hamiltonian operator, $\Psi$ is the wave function, and $E$ is the total energy of the system. $R$ and $x$ represent the coordinates of the $M$ nuclei and $N$ electrons, respectively. The Hamiltonian operator includes all kinetic and potential energy terms of the atomic nuclei and electrons. It can be formulated as follows:
\begin{equation}
    \hat{H} = -\frac{1}{2}\sum_{i=1}^{N}\nabla^{2}_{i} - \frac{1}{2}\sum_{A=1}^{M}\frac{1}{M_A}\nabla^{2}_{i} - \sum_{i=1}^{N}\sum_{A=1}^{M}\frac{Z_A}{r_{iA}} + \sum_{i=1}^{N}\sum_{j>i}^{N}\frac{1}{r_{ij}} + \sum_{A=1}^{M}\sum_{B>A}^{M}\frac{Z_AZ_B}{R_{AB}},
	\label{eq.2}
\end{equation}
where $i$ and $j$ are indices for the $N$ electrons, $A$ and $B$ are the indices for the $M$ nuclei and $M_A$ and $Z_A$ stand for the mass and charge of the nucleus, respectively. Due to the complex interactions between all particles, an exact solution of the wavefunction via this eigenequation (eq. \ref{eq.1}) is only possible for very small systems with few electrons.\supercite{Koch}

To make the problem manageable for molecular systems, the fundamental Born-Oppen\-hei\-mer approximation is applied. It is based on the large mass difference between electrons and nuclei, which allows for the decoupling of their motions. The nuclei are considered quasi-stationary, so the electronic Schrödinger equation can be solved for a fixed nuclear arrangement:
\begin{equation}
    \hat{H}_{elec} \Psi_{elec} = E_{elec} \Psi_{elec}.
	\label{eq.3}
\end{equation}
The electronic energy $E_{elec}$ depends parametrically on the nuclear coordinates and defines the potential energy surface (PES) on which chemical reactions take place. The total energy of the system $E_{tot}$ is obtained by adding the constant nuclear-nuclear repulsion term to the electronic energy. The ground state energy can then be determined by applying the variational principle by minimizing the energy.\supercite{Koch, Born1927}

Despite this simplification, the exact solution of the electronic Schrödinger equation remains a challenge due to the repulsive interactions between the electrons, an effect known as electron correlation. This is because it is still a 3$N$-dimensional problem.

\subsubsection{LCAO Approach and Basis Sets} \label{3.1.2}
The numerical solution of the quantum chemical equation requires the representation of the molecular wave function by a set of mathematical functions. Programs for periodic solid state calculations like VASP\supercite{Kresse1_1996, Kresse2_1996} typically use a plane wave basis set. This is ideal for describing the periodic electron density in a crystals. However, quantum chemical programs for molecular systems like ORCA\supercite{Neese2012, Neese2025} follow a fundamentally different approach. The standard method implemented in ORCA and similar programs is the LCAO (linear combination of atomic orbitals) method. This approach is chemically intuitive, as it conceives the molecular orbitals (MOs) $\phi_i$ as a linear combination of $M_{basis}$ atom-centered basis functions ($\chi_\alpha$), which mimic the well known atomic orbitals (e.g., s, p, d orbitals). Each molecular orbital is represented by the equation:
\begin{equation}
    \phi_i = \sum_{\alpha = 1}^{M_{basis}} c_{\alpha i} \chi_{\alpha},
	\label{eq.4}
\end{equation}
where the expansion coefficients $c_{\alpha i}$ are optimized in the self consistent field (SCF) procedure to minimize the total energy of the system. The entirety of the basis functions $\chi_{\alpha}$ used is referred to as the basis set.\supercite{Jensen, Cramer}

In this work, the solid-state optimized POB-TZVP\supercite{Peintinger2013} basis sets were used exclusively for the description of the AOs (atom orbitals). This basis set of triple-zeta quality offers a high degree of flexibility for describing the wave function, as it provides three basis functions per valence electron. Additionally, it includes polarization functions, which allow for an anisotropic deformation of the electron cloud in response to the molecular environment. This property is particularly crucial for the correct description of interactions at surfaces, polarized C-H bonds during activation and the delocalized electron density in radical species. The choice of this basis set represents a compromise between high accuracy and reasonable computational cost.

\subsubsection{Ground State Properties} \label{3.1.3}
To investigate systems in the electronic ground state, such as those present in the thermal reaction pathway, density functional theory (DFT) is the method of choice. The theoretical foundation of DFT is formed by the Hohenberg-Kohn theorems, which prove that the total energy of a system in its ground state is a unique functional of its electron density $\rho(r)$.\supercite{Hohenberg1964, Jensen, Parr, Koch} The total energy of the system can be decomposed into various terms as follows:
\begin{equation}
    E_{DFT} [\rho(r)] = T_s [\rho(r)] + E_{ne} [\rho(r)] + J [\rho(r)] + E_{xc} [\rho(r)],
	\label{eq.5}
\end{equation}
which include the kinetic energy ($T_s$), the nuclear-electron attraction ($E_{ne}$), the classical coulomb repulsion ($J$), and the exchange correlation ($E_{xc}$) terms. This approach reduces the complexity of the quantum mechanical problem from 3$N$ dimensions to just three dimensions (three spatial dimensions), which leads to a significant reduction in computational cost. The terms $E_{ne}$ and $J$ can be solved with: \supercite{Hohenberg1964, Jensen}
\begin{equation}
    E_{ne} [\rho(r)] = -\sum_{\alpha = 1}^{K} \int\frac{Z_{\alpha}\rho(r)}{|R_{\alpha}-r|}dr,
	\label{eq.6}
\end{equation}
\begin{equation}
    J [\rho(r)] = \frac{1}{2} \iint\frac{\rho(r)\rho(r')}{|r-r'|}.
	\label{eq.7}
\end{equation}
	The practical implementation of DFT is carried out through the Kohn-Sham approach, which maps the complex interacting many-body problem onto an effective, non-interacting system that possesses the same electron density.\supercite{Jensen, Kohn1965} The kinetic energy $T_s$ of the non-interacting electrons can then be calculated by the sum over the Kohn-Sham orbitals $\varphi_i$:
\begin{equation}
    T_s [\rho(r)] = \sum_{i = 1}^{N} \braket{\varphi_i|\frac{1}{2}\nabla_i^2|\varphi_i},
	\label{eq.8}
\end{equation}
where $\nabla$ corresponds to the Laplace operator. The greatest challenge and the source of the approximation now lies in the exact formulation of the exchange correlation functional $E_{xc} [\rho(r)]$, whose mathematical form is:
\begin{equation}
    E_{xc} [\rho(r)] = (T [\rho(r)] - T_S [\rho(r)]) + (E_{ee} [\rho(r)] - J [\rho(r)]).
	\label{eq.9}
\end{equation}
$E_{xc}$ includes the difference between the exact kinetic energy ($T$) and the energy of the non-interacting system ($T_s$), as well as the difference between the electron-electron interaction ($E_{ee}$) and the classical coulomb repulsion ($J$).\supercite{Jensen}

\begin{figure}[h]
    \centering
    \includegraphics[scale=0.65]{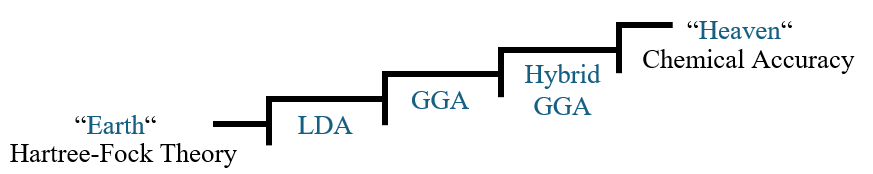}
    \caption{Jacob's ladder illustrates the various functionals (LDA, GGA and Hybrid GGA) that leading to enhanced accuracy in calculations. The image is adapted from \supercite{Awoonor2018, Perdew2001}.}
    \label{fig:5}
\end{figure}

Modern $E_{xc}$ functionals are often classified hierarchically on the so-called "Jacob's ladder".\supercite{Awoonor2018, Perdew2001} In this work, the PBE (Perdew-Burke-Ernzerhof) functional, which belongs to the family of the generalized gradient approximation (GGA), is used as a basis.\supercite{PBE1996} Its functional can be defined as follows: \supercite{Parr, Koch}
\begin{equation}
    E_{xc}^{GGA}[\rho(r)] = \int\rho(r)\epsilon_{xc}[\rho(r), \nabla\rho] dr,
	\label{eq.10}
\end{equation}
GGA considers not only electron density, but also its gradient. Including the gradient significantly improves the description of inhomogeneous electron densities, such as those found in molecules and on surfaces. Consequently, GGA is ranked higher than LDA (local density approximation) on the Jacobs ladder, as LDA only considers the density locally. For more accurate calculations, the hybrid functional PBE0 \supercite{Perdew1996, Silverstein2010} is considered. This functional mixes a fixed portion of 25\% exact Hartree-Fock exchange into the GGA functional. This approach reduces the self interaction problem typical for many DFT functionals and leads to a more reliable description, as it corrects some of the systematic errors of pure GGA functionals. For example, it corrects the underestimation of band gaps and activation barriers, but leads to significantly higher computational cost. In the end, it is necessary to determine which function is best suited to the system under consideration since the Jacob's Ladder does not always apply. Therefore, when available, a comparison to experimental data is necessary. 

Another crucial aspect in modeling adsorption processes is the correct description of long-range van der Waals (vdW) interactions. Standard DFT functionals like PBE and PBE0 neglect these dispersive forces, which are crucial for the physisorption of molecules like ethylbenzene on a surface. To address this deficiency, empirical corrections such as Grimme's D3 method \supercite{PBE1996, Grimme2010} are added. This correction adds an energy dependent term that approximately describes dispersion interactions without significantly increasing the computational cost.

\subsubsection{Excited States Properties} \label{3.1.4}
The description of photocatalytic processes requires the explicit treatment of electronically excited states. Such states, as well as situations with degenerate or quasi-degenerate orbitals during bond breaking, often exhibit a strong multi reference character. This condition, known as static correlation, cannot be accurately described by single determinant methods like Kohn-Sham DFT.

\begin{figure}[h]
    \centering
    \includegraphics[scale=0.8]{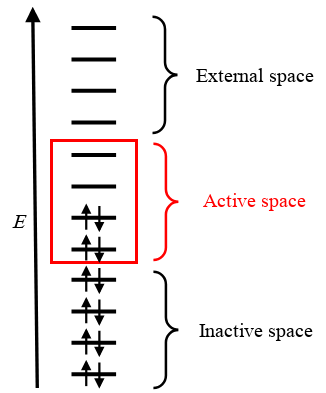}
    \caption{Visualization of the inactive, external and active orbital spaces in a CASSCF calculation.\supercite{Cramer}}
    \label{fig:40}
\end{figure}

A qualitatively correct description of such systems is enabled by the complete active space self consistent field (CASSCF)\supercite{Roos1980, Cramer} method. In CASSCF, the electronic wave function of a quantum chemical system is represented as a linear combination of configuration state functions (CSFs):
\begin{equation}
    \ket{\Phi_I^S} = \sum_{k}C_{kI}\ket{\phi_i^S}.
	\label{eq.11}
\end{equation}
Here, $S$ denotes the total spin, $I$ the state (ground or excited state), and $C_{kI}$ the respective expansion coefficients. The $\ket{\phi_i^S}$ are individual CSFs. Depending on the spin and point group symmetry, these can contain multiple slater determinants. The underlying molecular orbitals $\psi_i(r)$ are written as a linear combination of basis functions $\varphi_{\mu}(r)$:
\begin{equation}
    \psi_i(r) = \sum_{\mu}c_{\mu i}\varphi_{\mu}(r).
	\label{eq.12}
\end{equation}
The CASSCF procedure iteratively optimizes both the structure of the molecular orbitals (MO coefficients) and the weighting of all CSFs (configuration interaction (CI) coefficients). The energy of the state $\Phi_I^S$ is given by the expectation value of the Hamiltonian:
\begin{equation}
    E(c,C) = \frac{\braket{\Phi_I^S|\hat{H}|\Phi_I^S}}{\braket{\Phi_I^S|\Phi_I^S}}.
	\label{eq.13}
\end{equation}
The optimization is carried out according to the variational principle, i.e., by solving under the following conditions:
\begin{equation}
    \frac{\partial E(c, C)}{\partial C_{kI}} = 0,
	\label{eq.14}
\end{equation}
\begin{equation}
    \frac{\partial E(c, C)}{\partial c_{\mu i}} = 0.
	\label{eq.15}
\end{equation}
In this process, the orbital space is divided into three groups. Figure \ref{fig:40} shows a visualization of the inactive, external, and active orbital spaces. The inactive space contains the core and other chemically inert orbitals, which are doubly occupied in all configurations. The external space deals with the high energy, virtual orbitals that remain unoccupied in all configurations. The active space contains the orbitals crucial for chemical reactivity and electronic excitation. Here, a defined number of active electrons (n) is distributed among the active orbitals (m) in all possible ways. A full CI calculation is performed within the active space, meaning all possible configurations are considered. This is also referred to as a CAS(m,n) calculation. The selection of the active space is crucial for the quality of the calculation and is based on chemical intuition. Particularly $\pi$-systems, biradicals, transition states, or photochemical reactions benefit from a carefully chosen active space. Since the number of CSFs increases exponentially with the number of orbitals and electrons, typically only small CASSCF calculations with 14 or fewer orbitals are possible. Nevertheless, this allows for a flexible and physically correct description of static electron correlation.\supercite{Roos1980, Cramer}

However, the CASSCF method neglects dynamic correlation, which describes the short range repulsion effects between electrons. To account for this and in order to obtain quantitatively accurate energies, a second-order perturbation theory, by means of the $N$-electron valence state perturbation theory (NEVPT2) \supercite{Angeli2001_1, Angeli2001_2, Angeli2002}, is applied to the CASSCF wave function. The combined CASSCF/NEVPT2 approach is considered a highly accurate method for calculating the absolute and relative energies of ground and excited states. It is capable of reliably describing the topology of potential energy surfaces, the location of excited states, transition states and the geometries of conical intersections. In this work, this approach is therefore used as the reference method to validate the results obtained from DFT and to elucidate the complex photochemical reaction pathways.\supercite{Guo2021}

\newpage

\subsection{Surface Model} \label{3.2}
While the methods described in Chapter \ref{3.1} provide the theoretical basis for the calculation of the electronic structure, their application to dehydrogenation on a solid state surface requires an appropriate model of the catalyst itself. The creation of such a model for quantum chemical investigation of \ch{TiO2}(110) presents a challenge, which is addressed in the following. While periodic calculations that use periodic boundary conditions (PBC) \supercite{Bloch1929} can accurately represent the quasi-infinite character of a crystal, they are unsuitable for computationally intensive multi reference methods. An established alternative is the use of finite cluster models, wherein a small section of the surface represents the reactive region.

However, a disadvantage of standard cluster models is the emergence of unsaturated valences at the cluster edges, the so-called "dangling bonds". These lead to the formation of unphysical, artificial states within the materials band gap, which can cause inaccurate results and convergence problems, especially when studying excited states or radical species. To solve the problem of dangling bonds, two solutions have been presented in the literature. The most common solution is the saturation of the "dangling bonds" with hydrogen atoms.\supercite{Casarin2005, Casarin1998, Staemmler, He2010} However, this transforms the surface section into an isolated, neutral molecule, thereby sacrificing crucial physical properties of the solid. The biggest problem is the complete loss of the long-range electrostatic Madelung potential, which leads to incorrect orbital energies and an inaccurate description of the adsorbate-surface interaction. Moreover, the artificial chemical bonds at the cluster edge create an unphysical charge distribution and geometric distortions, which can falsify the electronic structure of the reactive center. For this reason, an alternative solution was chosen for this thesis. In the literature, this method is referred to as the "charged cluster" model. \supercite{Gerhards, Teusch2020, Petersen2020_1, Petersen2020_2, Mitschker2015, Casarin1998, Kubas2016, Berger2014, Bredow1998, Giordano2001, Pacchioni1994, Staemmler, Kick2019} This approach combines the advantages of a local cluster description with the simulation of the global crystal potential.

The construction of the model is based on the rules of Casarin $et.$ $al.$ \cite{Casarin1998} for the local bonding situation in \ch{TiO2}, where each titanium atom formally contributes $\frac{2}{3}$, and each oxygen atom $\frac{4}{3}$ of the two involved electrons to each Ti-O bond. The central idea of the model is to abandon stoichiometry to achieve complete electronic saturation. The cluster is cut from the crystal lattice in such a way that it terminates exclusively with undercoordinated oxygen atoms at its edges. To compensate for the missing valence electrons of the missing titanium bonding partners, a negative total charge is assigned to the cluster. This charge corresponds exactly to the number of valence electrons that would be provided by the titanium atoms in the same region in a stoichiometric equivalent. This makes the cluster electronically isovalent to its stoichiometric counterpart, thereby ensuring that the energy levels relevant to the chemical processes are represented in a physically correct manner. For the present investigations, the largest of the models validated by L. Gerhards \cite{Gerhards, Gerhards2021, Gerhards2022}, the \ch{Ti27O54[O34]^{-68}} cluster, is used to ensure an adequate description of the interactions even with larger adsorbates.

\begin{figure}[h]
    \centering
    \includegraphics[scale=0.8]{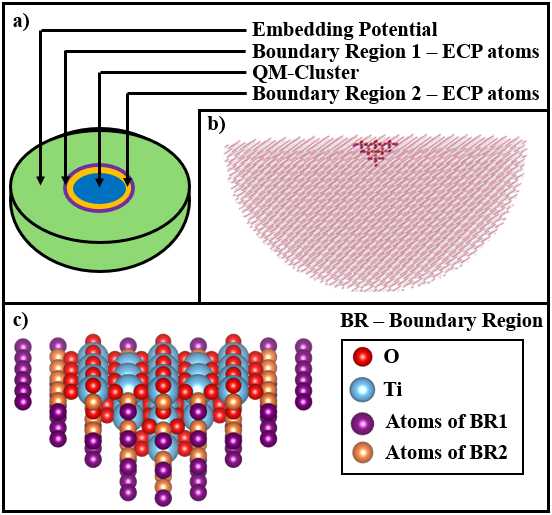}
    \caption{a) All four Regions of the Embedded cluster model are shown. b) Cluster embedded in a half-spherical field of point charges.\supercite{Gerhards} c) All atoms with and without charge that are treated in  all calculations. (Color scheme: red – oxygen, blue – titanium, orange – effective core potential (boundary region 2), violet - effective core potential (boundary region 1), green/see-through - point charge field) \supercite{Gerhards, Gerhards2021, Gerhards2022}}
    \label{fig:6}
\end{figure}

A crucial further step for practical calculations is the embedding of this charged quantum cluster in an environment which simulates the electrostatic potential of the rest of the crystal. This is realized by three components. One part, is a half spherical point charge field (PCF) of 71247 point charges, which are placed at the lattice positions of the ideal crystal and reproduce the long-range Madelung potential. The PCF value assigned to oxygen atoms is +1.00, while titanium atoms are assigned a value of -2.00. Then, there is a first boundary region consisting of 66 effective core potentials (ECPs), located between the PCF and the charged active cluster. This region consists of fixed charges at Ti atom positions in the crystal grid of \ch{TiO2} with a charge of +2.00 and is described by the SDD potential. A second boundary region is formed by another 47 ECPs, positioned between the quantum cluster and the first boundary region. This region also consists of fixed charges, described by the SDD potential, located at Ti atom positions. These ECPs prevent an unphysical over-polarization of the clusters electron cloud by the PCF. The entire system is visualized in figure \ref{fig:6}. To ensure the electrical neutrality of the total system, the charge of the ECPs in the second boundary region is adjusted according to:
\begin{equation}
    Z_{ECP} = \frac{|Z_{cluster} + Z_{PCF}|}{N_{ECP}},
	\label{eq.16}
\end{equation}
so that it exactly compensates for the negative charge of the quantum cluster and the charge of the PCF. $Z_{ECP}$ represents the charge of each neighboring ECP, while $Z_{cluster}$ is the charge of the negatively charged cluster, and $Z_{PCF}$ is the charge of the PCF without the modified ECPs. $N_{ECP}$ represents the number of modified ECPs required to distribute the charge homogeneously, ensuring complete neutrality of the entire system and accounting for coulomb effects over large distances. The charge of the ECPs is approximately 2.47.\supercite{Gerhards, Gerhards2021, Gerhards2022, Teusch2020, Petersen2020_2}

The reliability of this "charged cluster" approach was extensively validated in the works of L. Gerhards \cite{Gerhards, Gerhards2021, Gerhards2022} through a systematic comparison with periodic reference calculations under identical theoretical conditions (PBE functional, POB-TZVP basis set). The results of this validation demonstrate excellent agreement in all relevant physical and electronic properties.

A critical parameter for the description of photocatalytic processes is the band gap. The calculations of the cluster yielded a band gap of 2.17~eV with the PBE functional and a value of 4.48 eV with PBE0. These values deviate by only 0.44~eV (PBE) and 0.56~eV (PBE0) from the corresponding values of the periodic slab model calculations, which represents a remarkably good quantitative agreement.\supercite{Gerhards} The chosen cluster with the PBE functional is closer to the real band gap than the periodic slab model calculation. The opposite is true for the PBE0 functional.

Equally crucial is the correct reproduction of the local geometry at the active center. After a partial optimization of the inner cluster atoms, the geometric deviation from the periodic model was quantified using the root mean square error (RMSE). A low RMSE value of only 0.021~Å was determined for the cluster, which confirms that the model reproduces the surface structure without significant distortions.\supercite{Gerhards}

The charge distribution was also compared using a Mulliken population analysis.\supercite{Mulliken1955} The average charges of the titanium (+2.06~e) and oxygen atoms (-1.10~e) in the cluster deviate only minimally from the values of the periodic model (+2.02~e / -0.97~e). This shows that the electrostatic environment of the cluster correctly simulates the polarization of the atoms in the solid state.\supercite{Gerhards}

Moreover, it is essential to assess the quality of the electronic structure in proximity to the Fermi energy. The analysis of the frontier orbitals (HOMO and LUMO) for the cluster confirms that they have the correct physical character. The HOMO is, as expected, primarily composed of O(2p) orbitals, while the LUMO is dominated by Ti(3d) orbitals.\supercite{Gerhards} Edge-localized "edge states" that are artificial and within the band gap do not appear in this case, unlike in stoichiometric clusters.

In summary, this quantitative agreement in band gap, geometry, charge distribution, and the nature of the frontier orbitals proves that the "charged cluster" model represents a robust and physically correct basis. With its 228 defined atoms (including atoms with ECP), it has a manageable computational cost and shows its only weakness in the under- (by 0.86~eV with PBE) or overestimation (by 1.45~eV with PBE0) of the band gap.

\subsection{Analysis of Reaction Mechanisms and Properties} \label{3.3}
Based on the quantum chemical methods and the validated surface model presented in the preceding sections, the reaction mechanisms of ethylbenzene dehydrogenation can be investigated in detail. The following chapters describe the procedures and definitions used to analyze the reaction pathways, calculate relevant energetic values, and to characterize the crucial chemical processes.

\subsubsection{Reaction Pathways and Activation Energies} \label{3.3.1}
A chemical reaction can be understood as the movement of atoms on a multidimensional potential energy surface (PES), which represents the energy of the system as a function of its nuclear coordinates. Local minima on this surface correspond to stable and metastable species, such as reactants, products, and intermediates. The minimum energy path (MEP) connecting two of such minima passing through a transition state (TS). A TS is defined as a first-order saddle point, i.e., a maximum along the reaction coordinate and a minimum in all other directions. The energy difference between the transition state and the reactants is called the activation energy ($E_A$) and is the key factor for describing the reaction kinetics.\supercite{Jónsson}

The nudged elastic band (NEB) method is applied to locate the MEP and the corresponding transition state. This method optimizes a chain of states (so-called "images") that is spanned between the known reactant and product geometries. By minimizing the forces on the images, the reaction path on the PES is traced without having to make an $a$ $priori$ assumption about the reaction coordinate. The highest point of the resulting energy profile provides a good starting structure for a subsequent, precise transition state optimization with algorithms such as the Quasi-Newton method.\supercite{Jónsson, Ásgeirsson2021}

The crucial step in validating a found transition state is the calculation of vibrational frequencies. As indicated above, a true transition state is mathematically defined as a first-order saddle point. This manifests itself in the vibrational frequencies. A correct transition state must have exactly one imaginary frequency. The normal mode of this vibration describes the systems movement over the energy barrier, i.e., the breaking and forming of the involved chemical bonds. All other frequencies must be real.\supercite{Jensen}

\subsubsection{Definition of Relevant Energies} \label{3.3.2}
In order to quantitatively describe the thermodynamics of surface reactions and to evaluate the stability of adsorbed states, intermediates and products the relative energy differences are calculated. The adsorption energy ($E_{ads}$) and desorption energy ($E_{des}$) are a measure of the strength of the interaction between a molecule and the catalyst surface. It is calculated according to the following equation:
\begin{equation}
    E_{ads} = E_{sys} - E_{surf} - E_{mol}.
	\label{eq.17}
\end{equation}
Here, $E_{sys}$, $E_{surf}$, and $E_{mol}$, are the total energies of the adsorbate-surface complex, the clean surface, and the isolated molecule in the gas phase, respectively. A negative adsorption energy indicates a thermodynamically favorable, exothermic process. $E_{des}$ corresponds to the adsorption energy but with the opposite sign.

When calculating the total adsorption energies, corrections must be taken into account to obtain physically meaningful values. The zero-point energy (ZPE), obtained from a frequency calculation, corrects the purely electronic energy for the vibrational energy of the nuclei in the ground state. It is required to use the actual reaction energy ($E_{reac}$), which corresponds to the sum of $E_{ZPE}$ and the total energy. Furthermore, the basis set superposition error (BSSE) occurs when calculating adsorption energies. This numerical error arises from the artificial stabilization of the adsorbate-surface complex, as the adsorbate can use the basis functions of the surface and vice versa. The BSSE is corrected using the counterpoise method developed by Boys and Bernardi, in which the energies of the fragments are calculated in the basis set of the entire complex.\supercite{Boys1970}
\begin{equation}
    E^{corr}_{ads} = E_{ads,ZPE} + [E^{geom}_{mol} - E^{ghost}_{mol} + E^{geom}_{surf} - E^{ghost}_{surf}].
	\label{eq.18}
\end{equation}
Here, $E^{corr}_{ads}$ corresponds to the corrected adsorption energy, and $E_{ads,ZPE}$ reflects to the adsorption energy calculated according to equation \ref{eq.17} using the reaction energies. $E^{geom}_{mol}$ and $E^{geom}_{surf}$ correspond to the previous total energies of the adsorbate and the surface, respectively. $E^{ghost}_{mol}$ and $E^{ghost}_{surf}$ correspond to the total energies of the adsorbate and the surface, respectively, in the presence of the other system as ghost functions in the adsorption geometry. The combination of ZPE and BSSE correction is not exact, and other errors, such as the basis set incompleteness error (BSIE) \supercite{Dononelli2021} could still influence the adsorption value.

\subsubsection{Spin-Unrestricted Formalism} \label{3.3.3}
Since both the thermal and the photocatalytic mechanisms are presumably of a radical nature, the spin-unrestricted formalism (UKS for DFT) is necessary. This approach allows $\alpha$- and $\beta$-spin orbitals to assume different spatial distributions, and thus radicals can be represented. However, this often leads to spin contamination, where the resulting wave function is no longer a pure eigenfunction of the total spin operator $S^2$. The wave function then becomes contaminated with contributions from states of higher multiplicity (e.g., quintet contributions in a triplet calculation). A heavily contaminated wave function is physically meaningless and can lead to significant errors. Therefore, the expectation value $\braket{S^2}$ must always be checked and should be close to the theoretical value of $S(S+1)$, i.e., 0.0 for a singlet, 0.75 for a doublet, or 2.0 for a triplet, to ensure the reliability of the results.\supercite{Jensen}

The thermal reaction pathway is investigated on the PES of the electronic ground state. The cleavage of a C-H bond, for instance through the homolytic abstraction of a hydrogen atom, leads to a system with zero or two unpaired electrons, resulting in a singlet or triplet spin state. For example, one unpaired electron remains on the phenylethyl radical, and the surface possesses the additional remaining electron. Furthermore, the photocatalytic reaction pathway begins with an electronic excitation. Absorption of light excites the surface into a triplet state, demonstrating the necessity of the unrestricted formalism.
	
\subsubsection{Characterization of Electronic Excitation} \label{3.3.4}
Photochemical activation begins with the absorption of light, which leads to an electronic excitation from the electronic ground state to an electronically excited state. In this work, the excitation process via different ways is checked, such as pure local excitation of the surface or as a charge transfer from the adsorbate to the surface. To quantify the energy of this process and evaluate its plausibility, both vertical and relaxed excitation energies are being calculated. The vertical excitation energy, $\Delta E_{vertical}$, corresponds to the energy difference between the electronic ground state and the excited state, both calculated at the geometric ground state (the optimized equilibrium geometry of the electronic ground state). This simulates the immediate moment of light absorption according to the Franck-Condon principle, where the nuclei remain stationary. The relaxed excitation energy, $\Delta E_{relaxed}$, is the energy difference between the geometric ground state and the most energetically favorable point on the potential energy surface of the excited state after its geometric relaxation. This value represents the energy of the system once the nuclei have adjusted to the new electronic configuration.\supercite{Schneider2014, Henderson2011}

The analysis of the molecular orbitals involved provides insight into the character of the excitation. If both the source and the target orbital are primarily localized on the adsorbate or the surface, respectively, it is a local excitation of the molecule. In this case, the exchange of electron density between the adsorbate and the surface is minimal. If the source orbital is primarily localized on the adsorbate and the target orbital is on the \ch{TiO2} surface (e.g., in the d-orbitals of the Ti ions), it is an adsorbate to surface charge transfer. In the reverse case, it is a surface to adsorbate charge transfer. Such charge transfer states are often crucial for photocatalytic processes, as they lead to a radical species on the molecule (e.g., a radical cation) and an altered charge distribution on the surface, which can initiate subsequent chemical reactions.\supercite{Henderson2011}

\subsubsection{CASSCF Procedure and Choice of Active Space} \label{3.3.5}
Selecting the active space is the most crucial step in a CASSCF calculation. For this reaction, the CAS(m,n) space consists of the minimum number of orbitals that actively participate in the entire reaction process, not an arbitrary set of orbitals. The space must be chosen in a way that allows it to consistently describe the entire reaction, from adsorption to the product. The construction of this space is best understood by examining the four fundamental chemical processes that must be modeled.
	
The first step in the reaction is the photoexcitation of the surface. A photon excites an electron from a fully occupied O(2p) orbital (representative of the valence band) into an empty Ti(3d) orbital (representative of the conduction band). Thus, the O(2p) orbital acts as an electron donor, creating a reactive hole, while the Ti(3d) orbital acts as an electron acceptor, forming a Ti$^{3+}$ center. Although no new hole is created by photoexcitation in the second reaction step, this orbital pair is still crucial because it enables a correct description of the electronic structure of the surface and its defects, such as negatively charged OH groups. Additionally, a Ti(3d) orbital corresponds to the acceptor orbital for the second electron transferred from the organic molecule. This process requires two electrons in two orbitals.

The second step of the reaction involves cleaving two C-H bonds. To correctly describe this bond break, the active space for each bond must contain the fully occupied bonding $\sigma$ orbital and the corresponding empty antibonding $\sigma$* orbital. Since two C-H bonds are cleaved during the reaction, the active space must include two $\sigma$(C-H) and two $\sigma$*(C-H) orbitals. This adds four electrons in four orbitals to the active space.

The third part corresponds to the role of the $\pi$ system. First, ethylbenzene is adsorbed through the interaction of its $\pi$ system with the surface. Next, in the intermediate stage, the resulting phenylethyl radical is stabilized by delocalizing its unpaired electron into the $\pi$ system. The product, styrene, is characterized by its extended conjugated $\pi$ system at the end of the reaction. To correctly capture these central electronic effects, the space must contain the decisive frontier orbitals of the $\pi$ system. In this case, the two highest occupied $\pi$ orbitals and the two lowest unoccupied $\pi$* orbitals are used. This corresponds to an additional contribution of four electrons in four orbitals.

The fourth part is equal to the two unpaired electrons found along the reaction path that leads to two single occupied molecular orbitals (SOMOs). In the intermediate, these electrons correspond to the radical center on the phenylethyl molecule and the first Ti$^{3+}$ center on the surface. In the product, the system changes and the two SOMOs correspond to the two newly formed Ti$^{3+}$ centers. To correctly model these open-shell states across the entire pathway, two electrons must be contained in two specific SOMOs in the active space.

Taken together, these four building blocks create a 12 electron, 12 orbital space. Thus, a CAS(12,12) is the smallest complete model capable of releasing and absorbing electrons as required by the complex physics of the entire photocatalytic reaction for all pathways form figure \ref{fig:4}. 

Since CAS(12,12) calculations are close to the limit of what computationally is feasible, this work uses the Resolution of Identity (RI) approximation to speed up the calculations. The RI method is an efficient approximation that reduces the cost of the two-electron repulsion integral calculations. Rather than calculating complex four-center integrals directly, the RI approximation breaks them down into simpler, faster three- and two- center integrals. This is achieved by introducing an auxiliary basis set optimized to accurately represent the products of orbital pairs. Thus, the electron density is "fitted" with this auxiliary basis set. The resulting speed increase is significant, while the error introduced in the total energy is usually minimal. \supercite{Eichkorn1995, Neese2003, Weigend2002}

The general RI approximation is commonly used to calculate Coulomb terms in the DFT method. However, it plays a more specific role in CASSCF calculations when the keyword Trafostep RI is used. Solving the CASSCF equations requires transforming the large number of two electron integrals from the basis of atomic orbitals (AOs) to the smaller, more relevant basis of molecular orbitals (MOs). This step, known as integral transformation, scales unfavorably with system size (to the fifth power) and is often the most computationally intensive part of the entire CASSCF procedure. Trafostep RI applies the RI approximation to this most expensive step, speeding up the transformation. Using Trafostep RI replaces a specific part of the CASSCF algorithm with an efficient approximation. This makes it possible to investigate larger systems.\supercite{Neese2003, Weigend2002}

The choice for the auxiliary basis set is def2-TZVP/C because it is closely related to and highly compatible with the primary POB-TZVP basis set. The POB basis sets are a direct development of the widely used def2 family. The POB basis sets represent a re-optimization that improves the description of certain molecular properties without significantly altering the basic structure and mathematical functions of the def2 sets. The /C auxiliary basis set is optimized to fit the products of the primary basis functions. Since the POB-TZVP and def2-TZVP functions are nearly identical, the def2-TZVP/C auxiliary basis set is also a nearly perfect partner for POB-TZVP, despite being tailed for def2-TZVP. Using this explicitly expert optimized basis set circumvents sources of error associated with automatic generation and ensures the RI approximation is performed with maximum accuracy and minimum "numerical noise." This “smooths” the energy hyperplane and is a crucial factor in stabilizing and achieving convergence in such complex and sensitive systems.\supercite{Weigend2005, Hellweg2007, Peintinger2013}

\section{Results and Analysis} \label{4}
After covering the theoretical background, the next step is to verify the described charge cluster model. This is done by comparing experimental to calculated values in a model validation. Next, the reaction will be analyzed starting with the adsorption and electronic excitation and finally, the formation of the product. The main focus is on the photochemical mechanism, but the thermal mechanism will be briefly discussed as well. Finally, the influence of excessive oxygen species on an oxidized rutile \ch{TiO2}(110) surface will be investigated.

\subsection{Model Validation} \label{4.1}
The in chapter \ref{3.2} introduced charge cluster model of \ch{TiO2} and the specific surface model, the \ch{Ti27O54[O34]^{-68}} cluster, validated by L. Gerhards \cite{Gerhards, Gerhards2021, Gerhards2022}, is utilized. However, since L. Gerhards only used this cluster for small adsorbates such as \ch{SO2}, \ch{H2O} or \ch{CH4} verification for EB as an adsorbate is required. To accomplish this,  the adsorbed state of EB on the \ch{TiO2} cluster with the same number of relaxed atoms (15 atoms are relaxed\supercite{Gerhards}) must be examined. This results in no minimum with PBE or PBE0, which is indicated by three or more imaginary frequencies. Thus, a few more atoms must be unrestricted, resulting in a slight loss of periodicity. By relaxing 19 atoms, a minimum is reached with PBE. For PBE0, still three imaginary frequencies $>$~40~cm$^{-1}$ are present. Therefore, only PBE will be applied. 

The reason why PBE0 may not work for EB as an adsorbate is because, unlike PBE, the functional tries to localize the charge rather than distribute it.\supercite{Cohen2008} While this localization is PBE0s strength compared to PBE, the large adsorbate may lead to an instability of the cluster due to the strong coupling between charge localization and local lattice distortion (polaron effect).\supercite{Valentin2006} Another hypothesis is that PBE0, a hybrid functional, places greater weight on the non-additivity of the exchange correlation potential, making it more sensitive to the artificial boundary of the ECPs. PBE artificially stabilizes the system through self-interaction driven delocalization, while PBE0 attempts to solve the physics at the boundary exactly. The rather large EB pushes the charge toward this unphysical boundary. Consequently, the electronic solution undergoes a collapse due to the rigid nature of the fixed ECPs, which lacks the adaptability exhibited by real atoms in terms of accommodating localized charge.\supercite{Sun2016, Cohen2008} The PBE0 failure possibilities mentioned thus far are only initial hypotheses that would require further research for confirmation. This research will not be performed here.

Unlike the exact system introduced by L.~Gerhards, the amount of periodicity lost by relaxing four additional atoms can be determined by examining the band gap. An overview of all band gaps is shown in table \ref{tab.1}. When comparing the values calculated by L.~Gerhards using PBE with the newly calculated value, it is visible that adjusting for four additional relaxed atoms decreases the band gap by 0.21~eV. Therefore, the adjusted models band gap is further away from the literature value than before, but it is still higher by 0.23~eV than the value for a comparable PBC calculation. Similar to the band gap from the original cluster, the adjusted cluster using PBE underestimates the band gap by about 1~eV. This underestimation is common for the PBE functional and even using PBE0 would not close the gap to the experimental value, because PBE0 overestimates the band gap by about 1~eV.\supercite{Cohen2008, Valentin2006, Gerhards} In the end, the calculated band gap of 1.96~eV can be regarded as an acceptable value to evaluate the dehydrogenation of EB.

\begin{table}[h!]
\caption{Band gap values calculated using PBE from L.~Gerhards works \cite{Gerhards, Gerhards2021, Gerhards2022} are compared to the values of the slightly adjusted model used in this study.}
\centering
\label{tab.1}
\begin{tabular}{llll}
\hline
 System & Relaxed atoms & $E_{gap}$ [eV] & source  \\ \hline
  PBC &  & 1.73 & \cite{Gerhards, Gerhards2021, Gerhards2022} \\
  Charge Cluster & 15 & 2.17 & \cite{Gerhards, Gerhards2021, Gerhards2022} \\
  Experimental &  & 3.03 & \cite{Amtout1995, Tang1995} \\
  Charge Cluster & 19 & 1.96 & this work \\ \hline
\end{tabular}
\end{table}

Direct literature \cite{Chen2019, Lai2022, Lai2023, Li2022_Lett, Lin2020} on dehydrogenation of EB to styrene mainly provides TPD temperatures of EB, styrene and side products, as well as DFT calculations for energy barriers. To obtain additional reference values, like adsorption/desorption energies, it is necessary to analyze the TPD temperatures via the complete Redhead\supercite{Redhead1962} equation. The Redhead equation allows one to determine the desorption energy from the temperature of the desorption maximum alone. This is achieved by solving the relationship between the heating rate ($\beta$), the pre-exponential factor ($\nu$), and the peak temperature ($T_{des}$) for first-order reactions via a linear approximation. The equation to obtain desorption energies can be formulated as follows:
\begin{equation}
    E_{des} = R \cdot T_{des} \cdot [ln\frac{\nu \cdot T_{des}}{\beta}-C] .
	\label{eq.19}
\end{equation}
The Redhead equation is primarily used for the desorption of ideal gases as indicated by the ideal gas constant ($R$). However, it can also be used to make a good approximation for ethylbenzene (EB) and styrene. This work uses the general values of $C$ (the Redhead constant) and the vibration frequency of the adsorbed molecule ($\nu$) according to Redhead of 3.64 and $10^{13}$~s$^{-1}$, respectively.\supercite{Redhead1962} While EB and styrene regain many translational and vibrational degrees of freedom during desorption, this may lead to an underestimation compared to the standard for the ideal gas in the literature. Thus, a second desorption energy is calculated, employing the experimentally determined vibration frequency of $6.4~\cdot~10^{19}$~s$^{-1}$ for EB.\supercite{Chen2019}  The cited literature \cite{Lai2022, Lai2023, Li2022_Lett} primarily uses a heating rate of 2 K/s for measuring TPD spectra. All TPD temperatures, as well as the resulting $E_{des}$ values, are visualized in table \ref{tab.2}.

\begin{table}[h!]
\caption{The TPD temperatures ($T_{des}$) from the literature are used to calculate the resulting desorption energies using eq. \ref{eq.19}. Additionally, the calculated adsorption values from the literature are compared to the values obtained using DFT (including BSSE and ZPE corrections and excluding BSSE).}
\centering
\label{tab.2}
\begin{tabular}{llllll}
\hline
 Adsorbate/site & $T_{des}$ [K] & $E_{des}$ [eV] & $E_{des}$ [eV] & \multicolumn{2}{c}{calculated} \\ 
 & & $\nu$ = 10$^{13}$ s$^{-1}$ & $\nu$ = $6.4 \cdot 10^{19}$ s$^{-1}$ & BSSE & no BSSE\\\hline
 EB/5f-Ti & 245 \supercite{Chen2019, Li2022_Lett, Lai2022} & 0.66 & 0.99, 1.10$^{a, }$ \supercite{Chen2019} & 0.62, 0.71$^b$ & 1.26, 1.38$^b$ \\
 Styrene/pure & 270 \supercite{Lai2022} & 0.73 & 1.09 & 0.77 & 1.66 \\
 Styrene/O$_{br}$ & 435 \supercite{Lai2022} & 1.19 & 1.77 & 1.03 & 1.83 \\
 Styrene/O$_{Ti}$ & 350 \supercite{Lai2022} & 0.95 & 1.42 & 0.70$^b$, 0.76$^b$ & 1.40$^b$, 1.48$^b$ \\ \hline
\end{tabular}
\begin{tablenotes}[]
{\setstretch{0.3}
    \item $^a$ Experimental using the inversion analysis.\supercite{Chen2019}} \\
    \item $^b$ Calculated using the oxidized surface model.
\end{tablenotes}
\end{table}

Significant differences are visible when comparing the calculated values to the experimental values or to those obtained using Redheads equation.  The calculated values, including the BSSE correction, closely resemble the desorption energies obtained from the vibration frequency of ideal gases from literature.\supercite{Redhead1962} The biggest deviation is 0.16~eV for styrene desorption after a reaction involving O$_{br}$ atoms. The exclusion of the BSSE correction leads to an approach of the $E_{des}$ values to the more realistic Redhead and experimental values. The deviation of the values, excluding pure styrene desorption, is now in the range of 0.05–0.16 eV. However, pure styrene desorption is overestimated by 0.57~eV.

In summary, the selected model reproduces the experimental TPD values better without a BSSE correction.\supercite{Mentel2014} While the BSSE correction physically compensates for the artificial energy gain caused by base set overlap, it significantly underestimates the adsorption energies (0.37–0.74~eV) in the present case. This is consistent with the observations of Di Valentin \textit{et al.} \cite{Valentin2006} who found that the PBE functional often underestimates adsorption energies on polar surfaces because it inadequately describes the stabilization of localized electronic states.

In contrast to the above, uncorrected calculations slightly overestimate the experimental values by 0.05–0.16~eV. This is explainable by the delocalization error in charge transfer complexes, as described by Cohen $et$ $al$.\supercite{Cohen2008} and Mori-Sánchez $et$ $al$.\supercite{Mori2008} , which causes artificial energetic stabilization.
 
Thus, the present case study demonstrates advantageous error compensation.\supercite{Kruse2012} The artificial binding gain due to the BSSE and the delocalization contribution, described by Cohen\supercite{Cohen2008} and Mori-Sánchez\supercite{Mori2008} counteracts the fundamental underestimation of binding by PBE.\supercite{Valentin2006} Since the resulting net deviation from the experiment remains consistently small (0.05–0.16~eV), the uncorrected values are used as the effective adsorption energies for further reactivity discussions. The BSSE-corrected values should be considered as the lower physical limit of the employed functional.

Overall, it can be stated that the system accurately describes the dehydrogenation of EB to styrene, allowing for the identification of reasonable, quantitative trends. This is evident in the alignment of the experimental TPD temperatures with the calculated $E_{des}$ values. EB exhibits the lowest desorption energy, followed by pure styrene and styrene produced on the reduced surface \ch{TiO2}(110) (at O$_{br}$ atoms).

In case of styrene on the oxidized surface, a new surface model was created which will be discussed in chapter~\ref{4.4}.

\subsection{Description of Adsorption Properties of Ground and Excited States} \label{4.2}
Characterizing the initial adsorbed state is essential because it establishes the foundation for thermal and photochemical reaction pathways. The adsorption of ethylbenzene (EB) on the stoichiometric rutile \ch{TiO2}(110) surface was investigated using a combination of density functional theory (DFT-PBE-D3) and multireference calculations (CAS(12,12)).

\subsubsection{Electronic Structure of the Singlet Ground State} \label{4.2.1} 
\begin{figure}[h!]
    \centering
    \includegraphics[scale=0.7]{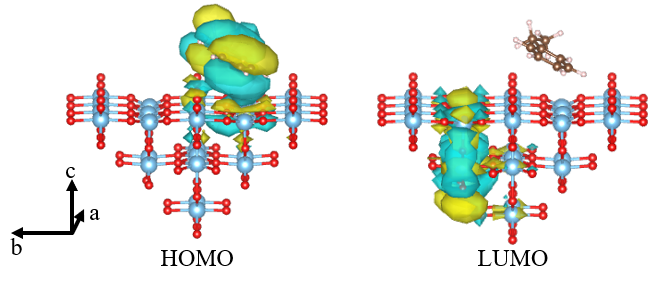}
    \caption{Frontier molecular orbitals of the physisorbed singlet state (RKS-PBE-D3). (Color scheme: red – oxygen, blue – titanium, brown – carbon, white – hydrogen)}
    \label{fig:7}
\end{figure}

An initial geometry optimization was performed in the electronic singlet ground state ($S_0$). The analysis of the frontier orbitals reveals a clear spatial separation between the occupied and unoccupied states. The highest occupied molecular orbital (HOMO) is primarily localized on the ethylbenzene molecule, exhibiting $\pi$ and $\sigma$(CH) character at the $\alpha$- and $\beta$-positions. Only minimal contributions from the surface bridging oxygen atoms (O(2p)) were observed in the HOMO. Conversely, the LUMO consists entirely of Ti(3d) orbitals. The frontier orbitals of the adsorbed singlet state are shown in figure~\ref{fig:7}.

The total electron density shows a delocalization across the entire surface, reflecting the polarization of the lattice. However, an increase in local density at one of the carbon atoms of the phenyl ring indicates the molecules specific orientation. This observation confirms a weak dative Lewis acid-base interaction where the $\pi$-system of the aromatic ring interacts with the electrophilic 5f-Ti centers. This interaction provides the necessary electronic coupling for possible subsequent charge transfer processes.\supercite{Chen2019, Henderson2011}

Furthermore, the adsorption of EB significantly modulates the electronic gap of the system. While the pristine surface model exhibits a band gap of $1.96$~eV, the presence of the adsorbate reduces it greatly to $1.42$~eV. This narrowing is attributed to the introduction of occupied EB $\pi$-states located energetically above the O(2p) valence band edge of the \ch{TiO2} substrate. Consequently, the adsorption effectively sensitizes the catalyst, potentially enabling charge transfer at lower photon energies than in the unmodified semiconductor.

\subsubsection{Charge Transfer and Radical Formation in the Triplet State} \label{4.2.2} 
\begin{figure}[h!]
    \centering
    \includegraphics[scale=0.7]{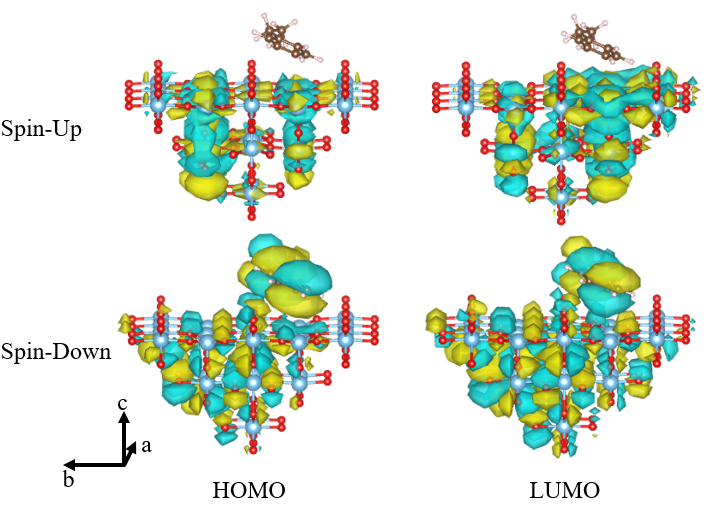}
    \caption{Spin-polarized frontier orbitals showing the energy gap between surface-centered and molecule-centered states. (Color scheme: red – oxygen, blue – titanium, brown – carbon, white – hydrogen)}
    \label{fig:8}
\end{figure}

To simulate the photochemical excitation, the adsorption geometry was investigated in the triplet state ($T_1$) using spin-unrestricted DFT (UKS-PBE-D3). The exceptionally low spin contamination (0.003) indicates a high quality wave function. The triplet state shows a significant energetic and spatial rearrangement. The spin-up (alpha) HOMO and LUMO (approximately 1.41~eV) consist entirely of Ti(3d) character and represent the injected electron in the conduction band. The spin-down (beta) HOMO and LUMO (approximately -0.89~eV) consist of a mixture of molecule-centered $\pi/\sigma$(CH) states and surface O(2p) states, representing the remaining hole. Figure~\ref{fig:8} shows the frontier orbitals of the adsorbed triplet state.

In the triplet configuration, the effective band gap increases from 1.96~eV to approximately $2.31$~eV. This widening reflects the electronic stabilization of the system following the adsorbate to surface charge transfer (ASCT). The localization of the transferred electron into the Ti(3d) states, along with the formation of the organic radical cation, leads to an energetic realignment of the frontier orbitals. This further stabilizes the resulting radical pair and may prevent from immediate recombination.

\begin{figure}[h!]
    \centering
    \includegraphics[scale=0.7]{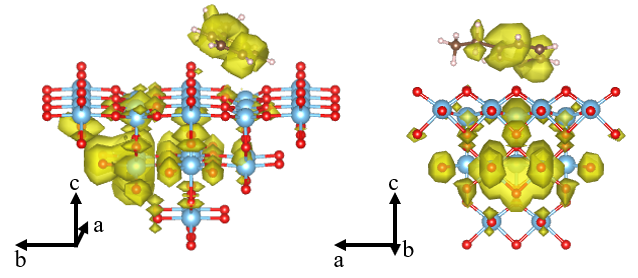}
    \caption{DFT spin density visualization of the formation of the Ti$^{3+}$ site and the organic $\alpha$-radical from two different perspectives. (Color scheme: red – oxygen, blue – titanium, brown – carbon, white – hydrogen)}
    \label{fig:9}
\end{figure}

The resulting spin density (see figure \ref{fig:9}) provides the first mechanistic proof of $\alpha$-CH radical formation. It clearly shows a localized Ti$^{3+}$ center and a radical character at the $\alpha$-carbon of the EB molecule. The $\alpha$-radical is significantly stabilized by the adjacent $\pi$-system, which is typical of benzylic radical species.

\subsubsection{Multireference Validation and Methodological Justification} \label{4.2.3}
While DFT provides a robust framework for geometry optimization, the multireference nature of the excited states requires validation via CAS(12,12). For the singlet adsorption geometry ($R$(OH)~=~2.546~\AA), the electronic ground state ($S_0$) is dominated by a closed-shell configuration with a weight of $93\%$ ("pure singlet"), highlighting the stability of the physisorbed species. In contrast, the first excited states ($S_1$ and the triplet manifold $T_1$) exhibit a nearly pure charge transfer (CT) character. Configuration interaction (CI) strings explicitly identify this transition as the promotion of an electron from the binding $\pi/\sigma$(CH) orbitals of the ethylbenzene molecule to the unoccupied 3d orbitals of a surface titanium atom.

\begin{figure}[h!]
    \centering
    \includegraphics[scale=0.7]{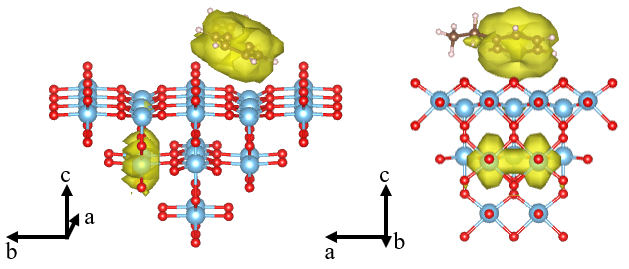}
    \caption{The multireference spin density confirms the radical pair character (EB$^{\bullet+}$ / Ti$^{3+}$) identified by the partial charge analysis. (Color scheme: red – oxygen, blue – titanium, brown – carbon, white – hydrogen)}
    \label{fig:10}
\end{figure}

\begin{figure}[h!]
    \centering
    \includegraphics[scale=0.7]{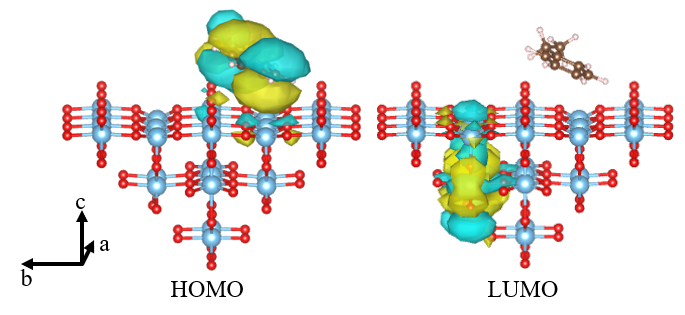}
    \caption{Active space natural orbitals (from CAS(12,12)) that reflect the transfer from the organic $\pi/\sigma$(CH) system to the Ti(3d) conduction band. (Color scheme: red – oxygen, blue – titanium, brown – carbon, white – hydrogen)}
    \label{fig:11}
\end{figure}

The quantitative analysis of the CASSCF (DFT) wave function provides evidence for this assumption. The calculated sum of partial charges for the adsorbate in the first excited state is 0.68 ($0.98$) (Mulliken) and 0.71 ($1.01$) (Löwdin). These results suggest that almost one electron is transferred from the molecule to the substrate.  Consequently, the excited state is formally classified as a spatially separated radical pair, consisting of an organic radical cation ($EB^{\bullet+}$) and a Ti$^{3+}$ center. This can be seen in the spin density and frontier orbitals of the adsorption triplet geometry, as illustrated in figure~\ref{fig:10} and \ref{fig:11}, respectively.

Moreover, the energy splitting between the singlet CT state ($S_1$) and the triplet CT state ($T_1$) is found to be extremely small. This quasi-degeneracy can be explained as a direct physical consequence of the spatial separation and the specific orbital character of the electron-hole pair. According to the theoretical framework discussed by Liao and Carter \cite{Liao2013}, the exchange interaction ($K$), which dictates the magnitude of the singlet-triplet splitting ($2K$), becomes minimal when the unpaired electrons occupy spatially separated orbitals and reside on different atomic centers. In the present system, the hole is localized within the organic $\pi$-system, and the electron is trapped on the Ti(3d) orbital of the surface. This near-degeneracy facilitates highly efficient intersystem crossing (ISC) into the triplet manifold, a process documented in the fundamental photophysical models for \ch{TiO2} surface reactions.\supercite{Linsebigler1995} This ISC allows the system to transition into a more durable reactive triplet state, which serves as the crucial precursor to the subsequent C-H bond cleavage.

The good quantitative agreement between the UKS-DFT spin densities and the CASSCF partial charge analysis justifies the usage of DFT for the intensive task of geometry optimization along the reaction path. Since DFT correctly captures the electronic character, specifically the formation of the radical pair, the resulting geometric coordinates provide a good physical basis for more expensive multireference energy evaluations. This hybrid approach provides a consistent description of bond breaking processes while keeping the computational cost manageable.

In conclusion, the findings suggest, that  adsorption of EB on TiO$_2$(110) is a physisorption process in the electronic ground state that is immediately transformed into a reactive radical pair upon electronic excitation. The \ch{TiO2}(110) surface acts as an efficient electron acceptor, "priming" the molecule for the first bond cleavage. This process is best described as a stepwise proton coupled electron transfer (PCET) initiated by an adsorbate to surface charge transfer (ASCT), in opposite to a direct hydrogen atom transfer (HAT), as frequently suggested in earlier literature.\supercite{Lin2020, Linsebigler1995}

\subsection{Reaction Mechanism} \label{4.3}
After characterizing the adsorption properties, the mechanistic process of dehydrogenation from ethylbenzene to styrene is examined. The investigation is divided into an introductory consideration of the reaction pathways and determined by the usage of density functional theory (DFT), and a subsequent, detailed analysis of the electronic states using multi reference methods (CASSCF).

\subsubsection{Characterization using DFT} \label{4.3.1}
The investigation of the reaction mechanism is initiated by an extensive exploration of the potential energy surface (PES) using DFT at the UKS-PBE-D3 level. In order to allow the mapping of the minimum energy path (MEP) for the dehydrogenation of ethylbenzene (EB) to styrene, the nudged elastic band method with the climbing image enhancement (NEB-CI) was employed, followed by transition state optimizations (OPTTS) using the  Quasi-Newton method implemented in ORCA. All identified stationary points were verified via frequency analysis to ensure the presence of exactly one imaginary frequency for transition states and only real frequencies for local minima.

\begin{figure}[h!]
    \centering
    \includegraphics[scale=0.6]{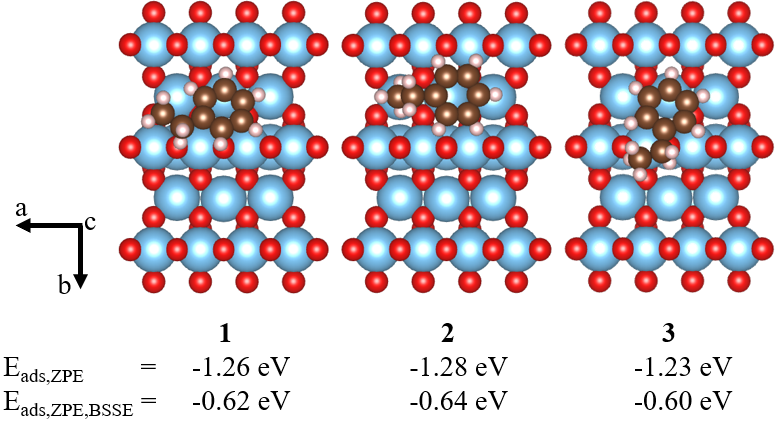} 
    \caption{The adsorption geometries and their adsorption energies that were found are presented. (Color scheme: red – oxygen, blue – titanium, brown – carbon, white – hydrogen)}
    \label{fig:12}
\end{figure}

The reaction begins with an evaluation of the adsorption of EB in its singlet ($S_0$) and triplet ($T_1$) states. While the singlet state represents the physisorbed precursor, the triplet state serves as the starting point for the photocatalytic cycle after the initial adsorbate to surface charge transfer (ASCT). Figure~\ref{fig:12} shows nearly degenerate adsorption energies for multiple adsorption geometries. Due to steric constraints and the specific orientation of the ethyl group relative to the bridging oxygen atom (O$_{br}$), distinct configurations were selected to investigate the regioselectivity of the first hydrogen abstraction.

\begin{figure}[h!]
    \flushleft \hspace*{-0.45cm} 
    \includegraphics[scale=0.55]{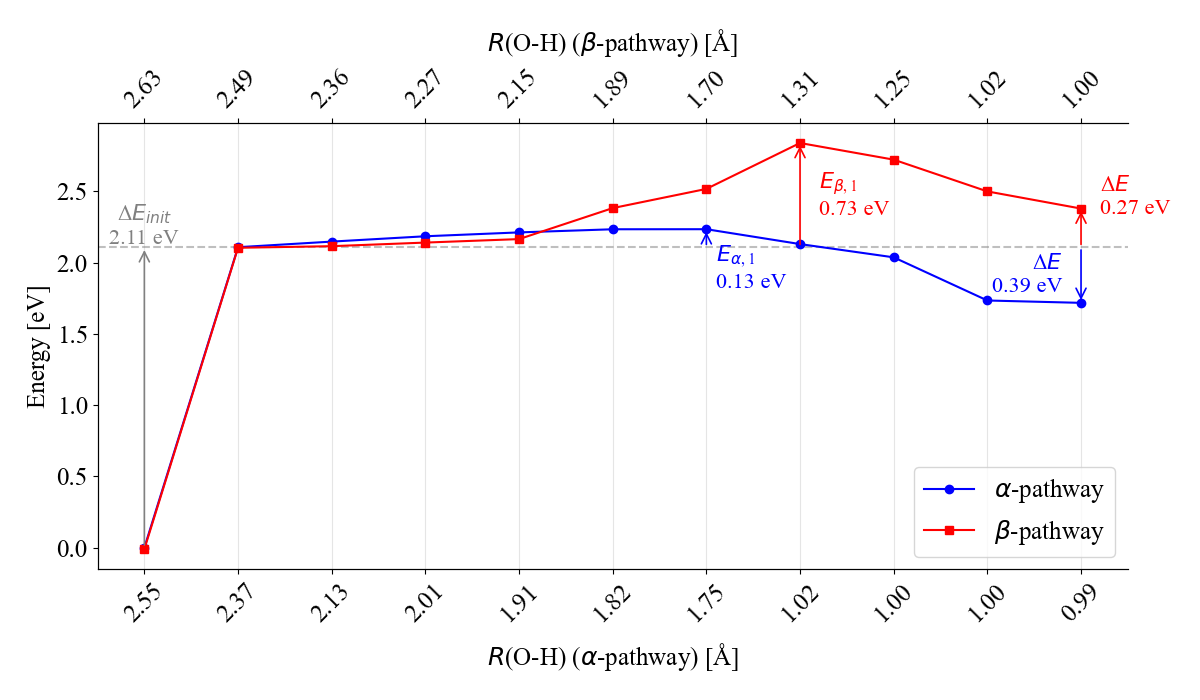} 
    \caption{Energy profiles for the first H-abstraction. A comparison of the favored $\alpha$-abstraction and the unfavorable $\beta$-abstraction. The x-axis shows the O-H bond length as a structural reaction progress indicator.}
    \label{fig:13}
\end{figure}

The first H-cleavage can occur either at the benzylic $\alpha$-carbon or the terminal $\beta$-carbon. Configurations 1 and 2 from figure~\ref{fig:12} were chosen based on the steric accessibility of the ethyl group to the surface $O_{br}$ atoms for the $\alpha$- and $\beta$-pathways, respectively. The energy barriers for these processes differ significantly (see figure \ref{fig:13}). The abstraction of the $\alpha$-hydrogen requires an activation energy ($E_{\alpha,1}$) of only 0.13~eV, whereas the $\beta$-hydrogen abstraction faces a much higher barrier of 0.73~eV. The 0.13~eV barrier for the $\alpha$-pathway is about half the barrier of 0.3-0.4~eV, as reported in the literature for this step.\supercite{Lai2023} Furthermore, the formation of the $\alpha$-phenylethyl radical (intermediate) is exothermic by 0.39~eV relative to the triplet adsorption state, acting as a thermodynamic sink. This sink is supported by the literature, indicating that the intermediate is exothermic by 0.10 to 0.73~eV.\supercite{Lin2020} In contrast to these findings, the $\beta$-abstraction leads to an endothermic intermediate (0.27~eV). Consequently, the $\beta$-pathway is unlikely, so that the following investigations focus exclusively on the $\alpha$-pathway.

The first transition state (TS1) of the $\alpha$-pathway is characterized by an "early" geometric configuration. As found after an extra OPTTS calculation, not shown in figure \ref{fig:13}. The $\alpha$~-~C~-~H bond length remains at 1.10~\AA, identical to the adsorbed state, while the O-H distance is only slightly reduced to 2.34~\AA\ (from 2.37~\AA\ in the adsorption minimum). This minimal geometric distortion is consistent with the Hammond postulate for exothermic elementary steps with low activation barriers.\supercite{Donahue2001} Electronic analysis of TS1 reveals a collective spin population of approximately 0.24 across the aromatic ring carbons (averaging 0.04 per atom), indicating an early delocalization of radical character into the $\pi$-system. Simultaneously, the migrating $\alpha$-hydrogen exhibits a partial positive charge of 0.1, suggesting the onset of a proton like transfer to the surface oxygen, while its negligible spin population (0.01) confirms that the radical character is predominantly hosted by the organic framework.

\begin{figure}[h!]
    \centering
    \includegraphics[scale=0.6]{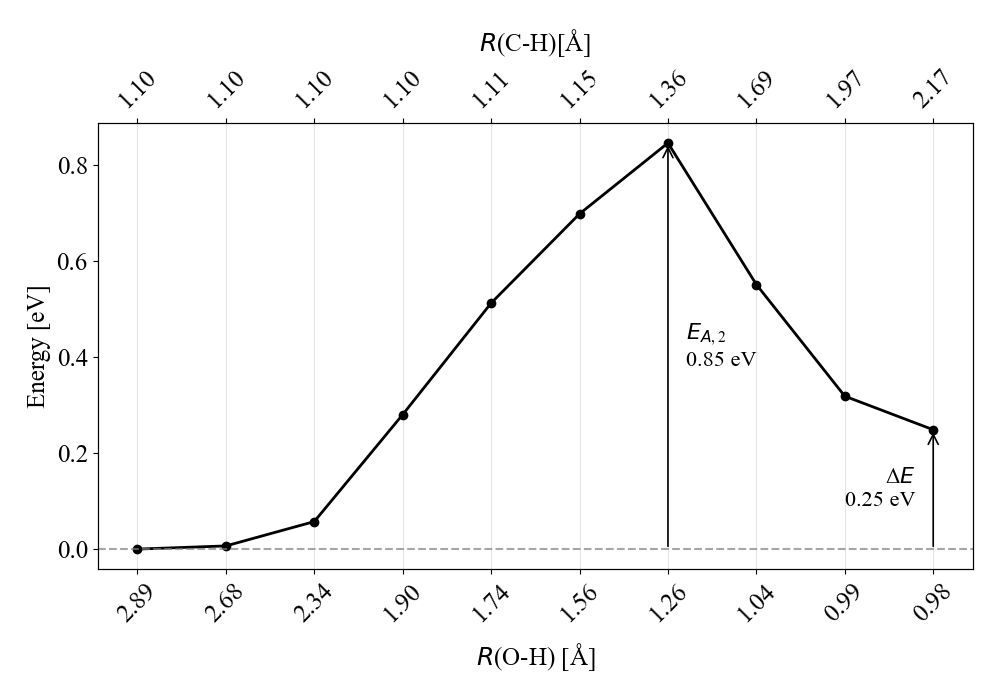} 
    \caption{The reaction profile for Step II (see figure \ref{fig:4}), showing the transition from the $\alpha$-phenylethyl intermediate to the final styrene product, with an activation barrier of 0.85~eV.}
    \label{fig:14}
\end{figure}

Following the formation of the stable $\alpha$-phenylethyl intermediate ($R$(C-H)~=~1.95~\AA, $R$(O-H)~=~0.99~\AA), the second dehydrogenation step involves the abstraction of a $\beta$-hydrogen. This step represents the rate determining step of the process with a barrier of $E_{A,2}$~=~0.85~eV (see figure \ref{fig:14}). This barrier is smaller than the 1.20~eV calculated in the literature, but it remains the rate determining step.\supercite{Lin2020}

In contrast to TS1, the second transition state (TS2) exhibits a much more "late" geometry. The $\beta$-C-H bond stretches from 1.10~\AA\ to 1.36~\AA\ and the O-H distance decreases from 2.89~\AA\ to 1.26~\AA. This increased geometric deformation accounts for the higher activation energy. The Mulliken analysis of TS2 shows a high spin population of 0.28 on the $\alpha$-carbon, reflecting its radical nature. Meanwhile, the increasing charge on the migrating hydrogen atom (0.20) suggests a significant proton-like behavior during the transfer to the basic O$_{br}$ site. The final product state consists of styrene and two surface hydroxyl groups, which is being reached with a C-H distance of 2.17~\AA\ and a final O-H bond of 0.98~\AA.

The numerical reliability of the spin-unrestricted formalism was monitored via the expectation value of the total spin operator $\braket{S^2}$ for all triplet geometries. Across the entire reaction coordinate, the spin contamination remained remarkably low, with values ranging from 0.003 to 0.014. These values suggest that the triplet states are accurately described and are not significantly mixed with higher multiplicity states. While these DFT results provide a clear geometric and thermodynamic overview, the radical nature of the transition states and the involvement of multiple Ti$^{3+}$ centers suggest a complex electronic landscape. The multireference character of these systems requires a more refined treatment, such as CASSCF, to understand the effect of electronic excitation on these barriers.

\begin{figure}[h!]
    \centering
    \includegraphics[scale=0.55]{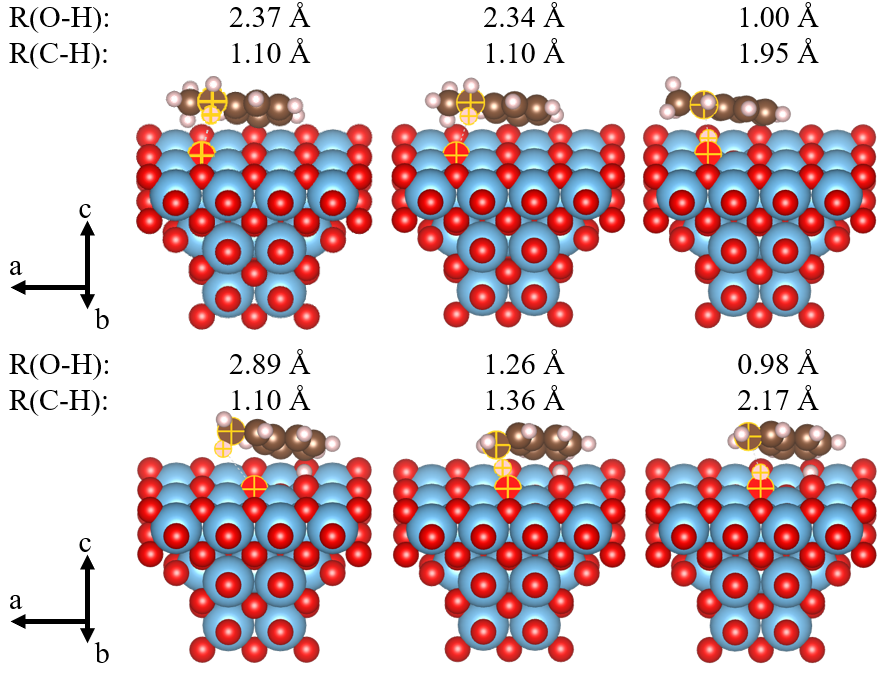} 
    \caption{Geometric evolution of the dehydrogenation mechanism. Top: Adsorption ($S_0$) $\to$ TS1 $\to$ Intermediate. Bottom: Relocated Intermediate $\to$ TS2 $\to$ Product. Distances are given in \AA. The increase in $R$(C-H) at the intermediate stage is due to the relocation of the first abstracted hydrogen to an adjacent $O_{br}$ site. This vacates the reactive center for the second step.}
    \label{fig:15}
\end{figure}

To further clarify the structural evolution along the reaction coordinate, the geometric sequence of the bond breaking and bond forming events is visualized in figure~\ref{fig:15} and table~\ref{tab3}. This sequence illustrates the transformation from the physisorbed ethylbenzene to the final styrene product and highlights the critical C-H and O-H distances.

\begin{table}[h!]
\caption{Selected geometric parameters [\AA] and relative energies [eV] along the PBE-D3 reaction coordinate.}
\centering
\label{tab3}
\begin{tabular}{lccccc}
\hline
Parameter & Adsorption ($T_1$) & TS1 & Intermediate & TS2 & Product \\ \hline
$R$(C-H) & 1.10 & 1.10 & 1.95 $|$ 1.10 & 1.36 & 2.50 \\
$R$(O-H) & 2.37 & 2.34 & 0.99 $|$ 2.89 & 1.26 & 0.98 \\
Rel. Energy & 0.00 & 0.13 & -0.39 & 0.46 & -0.14 \\ \hline
\end{tabular}
\end{table}

A noteworthy character of the structural data is the significant increase in the $R$(C-H) distance at the intermediate stage, which increases from $1.95$~\AA\ to $3.49$~\AA. This is not a result of molecular desorption, but rather a deliberate structural adjustment. To initiate the second dehydrogenation step at the $\beta$-carbon, the first abstracted hydrogen atom must be relocated to a neighboring bridging oxygen atom ($O_{br}$). This relocation ensures that the $O_{br}$ site directly adjacent to the $\beta$-hydrogen remains vacant and sterically accessible for the second abstraction (Step II). Calculations comparing the intermediate with the hydrogen atom in both positions revealed an energy difference of only $0.03$~eV. Given this negligible energetic penalty, the relocation is a physically plausible diffusion step on the surface that prepares the system for the rate determining transition state (TS2).

\subsubsection{Multi Reference Analysis of the Reaction Mechanism} \label{4.3.2}
The DFT calculations presented in the previous chapter provide a robust description of the geometric reaction path and the basic energetic trends. However, the complex nature of the dehydrogenation process, involving multiple bond breaking events and radical intermediates, requires a more sophisticated electronic treatment. The inherent limitations of single determinant methods, such as DFT, in describing quasi-degenerate states and homolytic bond cleavage require the use of multireference techniques.

Consequently, high level $ab$ $initio$ calculations were performed using the state-averaged complete active space self consistent field (SA-CASSCF) method. Based on chemical intuition and the need for a consistent description of the entire reaction, from adsorption to product formation, an active space of 12 electrons in 12 orbitals, denoted as CAS(12,12), was chosen (see chapter~\ref{3.3.5}). To capture the relevant electronic landscape for both thermal and photochemical pathways, the state-averaging procedure encompasses the three lowest singlet ($S_0, S_1, S_2$) and three lowest triplet ($T_1, T_2, T_3$) roots. For more information on the accuracy of the calculations, check out the calculation input example in the \hyperref[Appendix]{Appendix}. This approach provides a balanced description of the energy minimum pathway (MEP) and enables the identification of charge transfer states. It also indicates regions of electronic proximity, such as avoided crossings.

\begin{figure}[h!]
    \flushleft \hspace*{-0.2cm} 
    \includegraphics[scale=0.45]{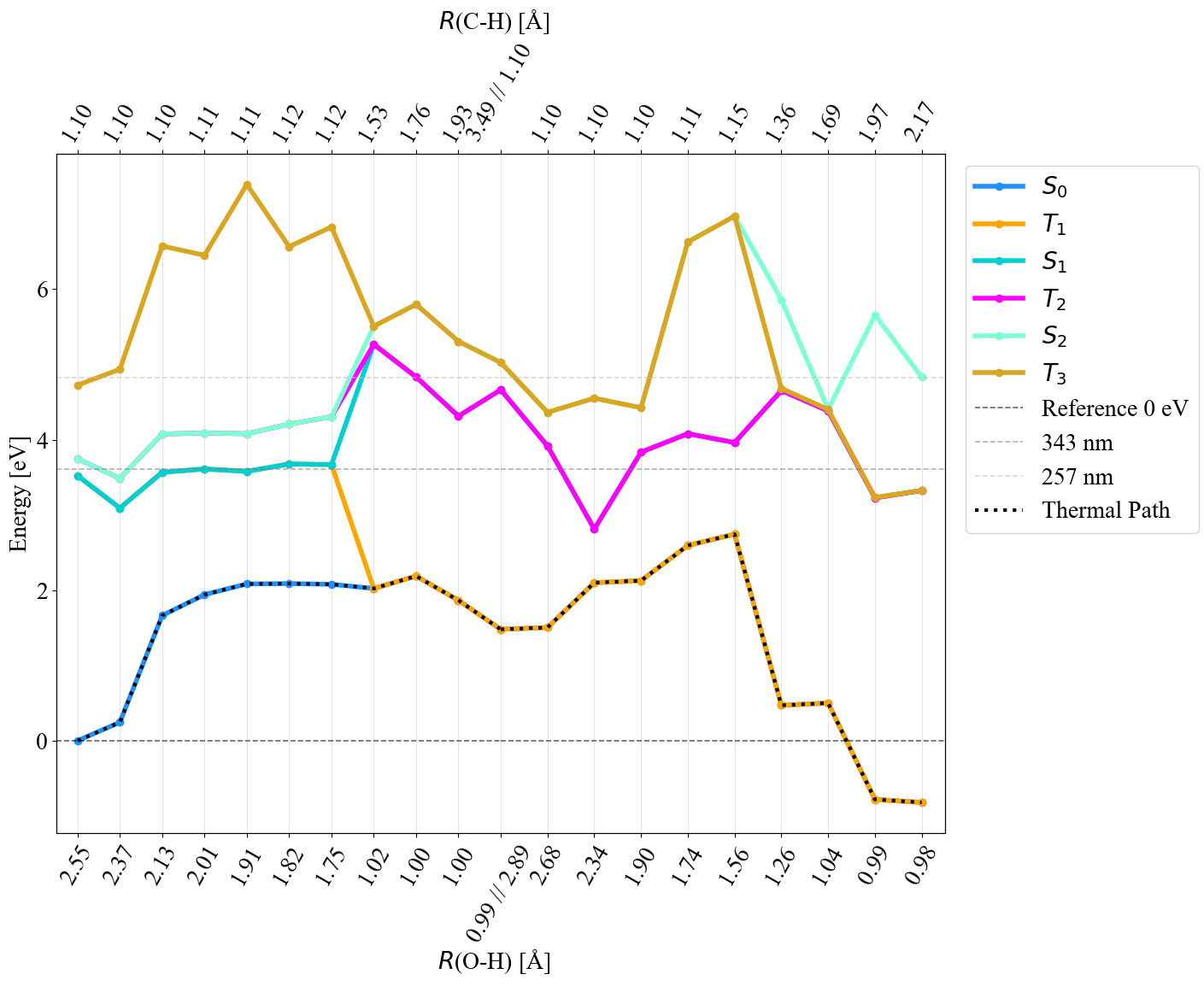} 
    \caption{Energy profile of the six lowest electronic roots ($S_0, S_1, S_2, T_1, T_2, T_3$) calculated at the SA-CASSCF(12,12) level. The dotted line indicates the thermal reaction path, starting from the adsorption state.}
    \label{fig:16}
\end{figure}

\subsubsubsection{Thermal Reaction Pathway} \label{4.3.2.1}
The thermal reaction pathway is conventionally understood as the progression of the system along the electronic ground state energy profile. In the context of the CASSCF analysis, this corresponds to the trajectory followed by the lowest singlet root ($S_0$, see figure~\ref{fig:16}). The following discussion treats the electronic evolution of the system along the minimum energy path previously identified by DFT.

\begin{figure}[h!]
    \centering
    \includegraphics[scale=0.80]{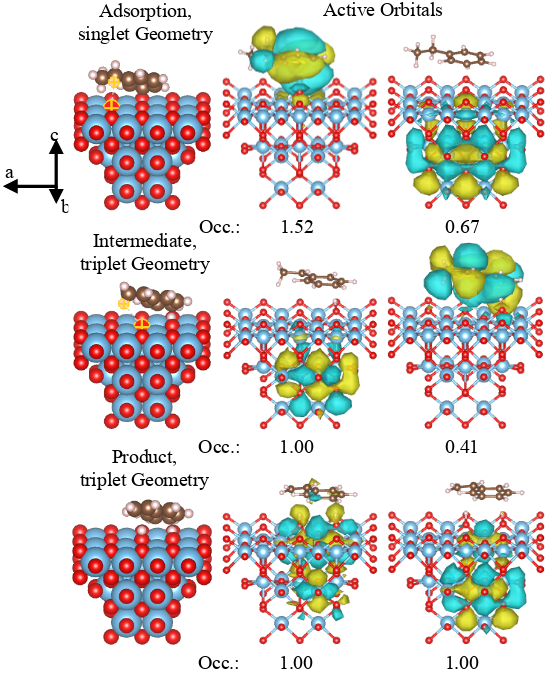} 
    \caption{Visualization of the electronic reorganization that occurs during the dehydrogenation of ethylbenzene on \ch{TiO2}(110) via SA-CASSCF(12,12). The marked atoms and dotted lines in the geometries (left) indicate the hydrogen transfer steps. The center and right columns show the two most active orbitals and their respective state-averaged natural orbital occupancy values (Occ.), which are farthest from 0.00 and 2.00.}
    \label{fig:17}
\end{figure}

At the initial adsorption geometry ($R$(O-H)~=~2.55~\AA), the $S_0$ state is characterized by a predominantly closed-shell configuration. The analysis of the configuration interaction (CI) coefficients reveals a 93\% dominance of the "pure singlet" configuration, (represented by the occupation \texttt{222222000000}). This reflects a stable, physisorbed ethylbenzene molecule, where all electrons remain paired within the organic $\pi$-system and the surface O(2p) valence band. As the reaction progresses and the $\alpha$-hydrogen approaches the bridging oxygen atom ($O_{br}$), the multi reference character of the wave function begins to increase. Figure \ref{fig:17} illustrates the initial electronic coupling. The natural orbital with an occupancy of 1.52 is primarily localized on the EB/$\pi$-system, while the corresponding orbital (Occ. 0.67) shows significant Ti(3d) character. This confirms the adsorbate to surface charge transfer (ASCT) path.

The most significant electronic transition occurs during the first C-H bond cleavage (Step I). As the $R$(OH) distance decreases toward the intermediate stage ($R$(O-H)~=~1.02~\AA), the $S_0$ state undergoes a fundamental change in character. The CI string occupation shifts toward the \texttt{222221100000} configuration, in which two orbitals become singly occupied. This marks the formation of a biradical species consisting of a localized Ti$^{3+}$ center and the organic $\alpha$-phenylethyl radical. In this region, the wave function can no longer accurately be described by a single configuration because the coefficients for the radical pair character begins to dominate. This electronic "unpairing" is the quantum chemical manifestation of bond cleavage, which is energetically stabilized by the surface.

As illustrated in figure~\ref{fig:16}, the thermal path (dotted line) follows the $S_0$ state until the onset of the first C-H bond activation. At a bond distance of $R$(O-H)~=~1.75~\AA, the $S_0$ and $T_1$ states become energetically degenerated. This degeneracy is a direct consequence of the spatial separation of the unpaired electrons in the radical intermediate, leading to a vanishing exchange integral. Consequently, the thermal reaction effectively proceeds on an equally accessible biradical manifold.

At the intermediate stage, the radical nature is evident (see fiagure~\ref{fig:17}). The SOMO on the surface (Occ. 1.00) represents a trapped electron in a Ti$^{3+}$ state. Meanwhile, the orbital on the organic fragment (Occ. 0.41) confirms the formation of the benzylic radical. The occupancy of 0.41 observed for the organic radical orbital suggests that the unpaired electron is not strictly localized at the $\alpha$-carbon. Rather, a multireference description accounts for a significant delocalization, where electron density is donated to the adjacent $\pi$*-system of the phenyl ring and partially to the unoccupied surface Ti(3d) orbitals. This delocalization provides the electronic stabilization required for the benzylic radical intermediate.

After the intermediate forms, the system transitions to the second hydrogen abstraction (Step II). The ground state $S_0$ continues to represent a radical pair manifold. However, near the second transition state (TS2), the electronic landscape becomes increasingly dense. While the thermal pathway in $S_0$ has to overcome the rate determining barrier identified by DFT, CASSCF analysis shows that the stability of the biradical intermediate is essential to prevent immediate recombination/back reaction. At the product stage, the $S_0$ state reflects the formation of two hydroxyl groups and the resulting styrene molecule. Interestingly, the ground state here maintains a high degree of biradical character (approximately 72\% Ti$^{3+}$ pair character in the singlet manifold), suggesting that the two transferred electrons remain localized in the Ti(3d) orbitals. This reduces the two surface titanium atoms to Ti$^{3+}$. Figure~\ref{fig:17} illustrates this with two degenerate SOMOs (Occ. 1.00 each).

This analysis shows that, although the thermal pathway seems straightforward in a ground state picture, the reacting system actually involves a complex transition from a closed-shell system to a highly correlated biradical manifold. Thus, the requirement of CASSCF is justified by the need to capture this transfer in the electronic character, which could only be approximated by single determinant DFT through artificial spin polarization.

\begin{figure}[h!]
    \flushleft \hspace*{-0.2cm} 
    \includegraphics[scale=0.45]{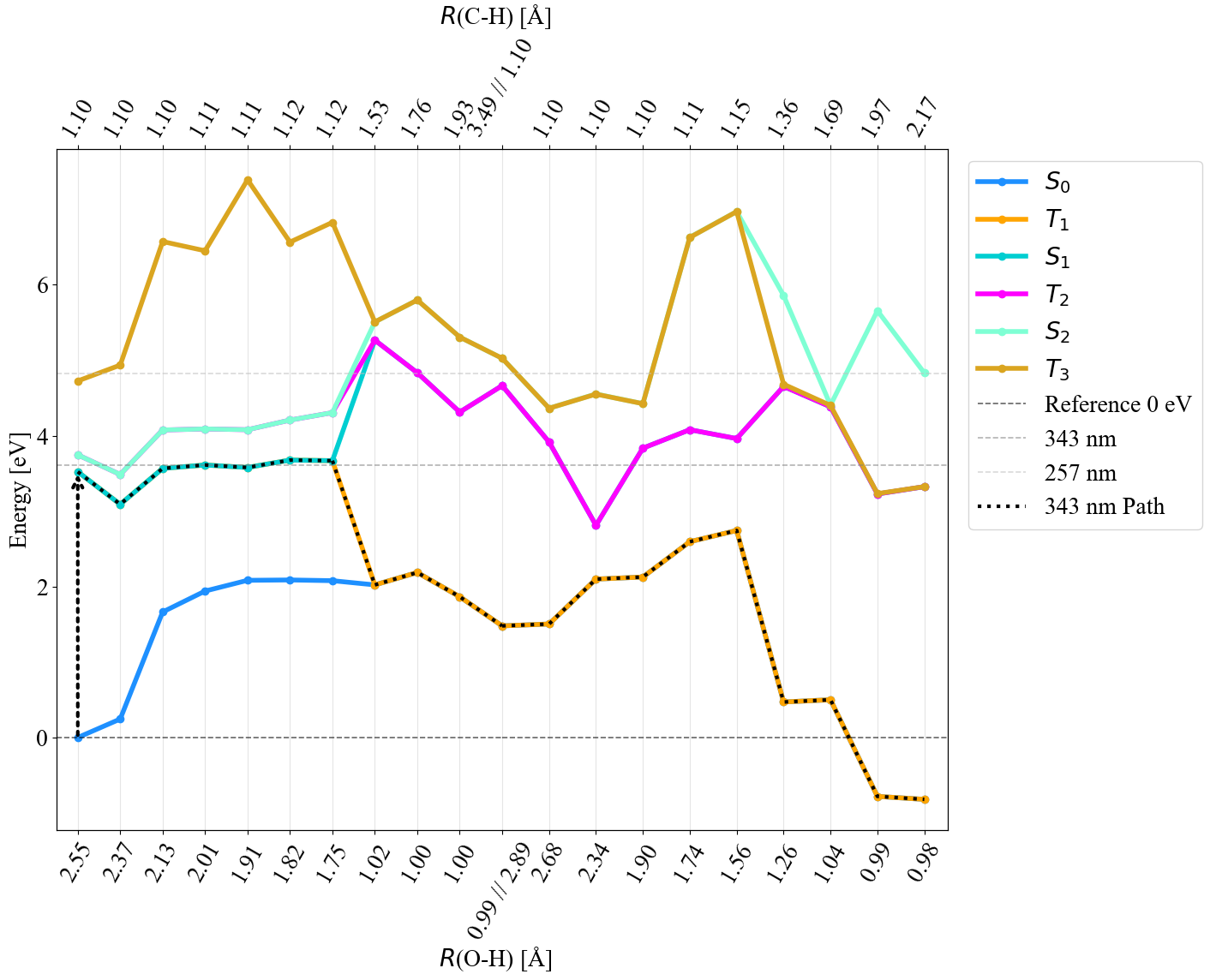} 
    \caption{Energy profile and the proposed reaction path are shown for low energy photoexcitation (343~nm). The dotted line indicates the systems progression, indicating the initial jump to $S_1$ and the subsequent relaxation into the biradical manifold ($S_0$/$T_1$) near the intermediate stage.}
    \label{fig:18}
\end{figure}

\subsubsubsection{Mechanism of Low Energy Photoexcitation} \label{4.3.2.2}
To explain the wavelength dependency observed in experimental studies, the reaction was investigated under low energy photoexcitation conditions. Irradiation at 343~nm corresponds to a photon energy of approximately 3.61~eV. As illustrated in figure~\ref{fig:18}, this energy is sufficient to elevate the system from the physisorbed singlet ground state ($S_0$) at $R$(O-H)~=~2.55~\AA\ into the first excited singlet state ($S_1$).

A critical feature of the initial electronic structure at $R$(O-H)~=~2.55~\AA\ is the energetic quasi-degeneracy of the $S_1$ and $T_1$ roots. Both states represent a charge transfer precursor, in which one electron is promoted from the ethylbenzene molecule into the Ti(3d) states of the surface. As the reaction coordinate progresses toward the first H-abstraction, reaching an O-H distance of $R$(O-H)~=~1.02~\AA, a significant divergence in the energy profile occurs.

In this region, the $T_1$ root drops sharply in energy and becomes degenerate with the $S_0$ state, thereby forming the stable biradical ground state manifold. Conversely, the $S_1$ root increases in energy and eventually merges with the $T_2$ surface. As the $S_1$ state rises, its energy exceeds the 3.61~eV provided by the 343~nm photon excitation. This energetic crossing means that, if it follows the path restricted by the available photon energy, the system will not be able to remain longer on on the $S_1$ surface.

Instead, the close proximity and eventual degeneracy of $T_1$ with $S_0$ facilitate highly efficient non-radiative relaxation channels. The system following the 343~nm path is effectively forced to relax into the lower-lying $T_1$/$S_0$ biradical manifold. This relaxation occurs at the intermediate stage ($R$(O-H)~=~0.99~\AA, $R$(C-H)~=~3.49~\AA), meaning the system loses its electronic excitation before the second reaction step can be initiated.

To complete the second dehydrogenation step (Step II) toward the final styrene product, the system must now overcome the significant thermal barrier located between \linebreak $R$(O-H)~=~1.90~\AA\ and $R$(O-H)~=~1.04~\AA, with its peak at $R$(O-H)~=~1.56~\AA. The loss of electronic excitation energy due to the $S_1$/$T_1$ splitting and subsequent "trapping" in the ground state explains the low quantum yield observed at 343~nm. Although the photon provides enough energy for the initial activation, the topology of the excited state surfaces forces the system back into a "thermal-like" state. There, it encounters the rate determining barrier of the second C-H bond cleavage.\supercite{Lai2023, Lai2022, Li2022_Lett, Lin2020} Because of the kinetic momentum gained when relaxing into the ground state, the reaction may still pass the barrier, resulting in the formation of small amounts of product.

\begin{figure}[h!]
    \flushleft \hspace*{-0.2cm} 
    \includegraphics[scale=0.45]{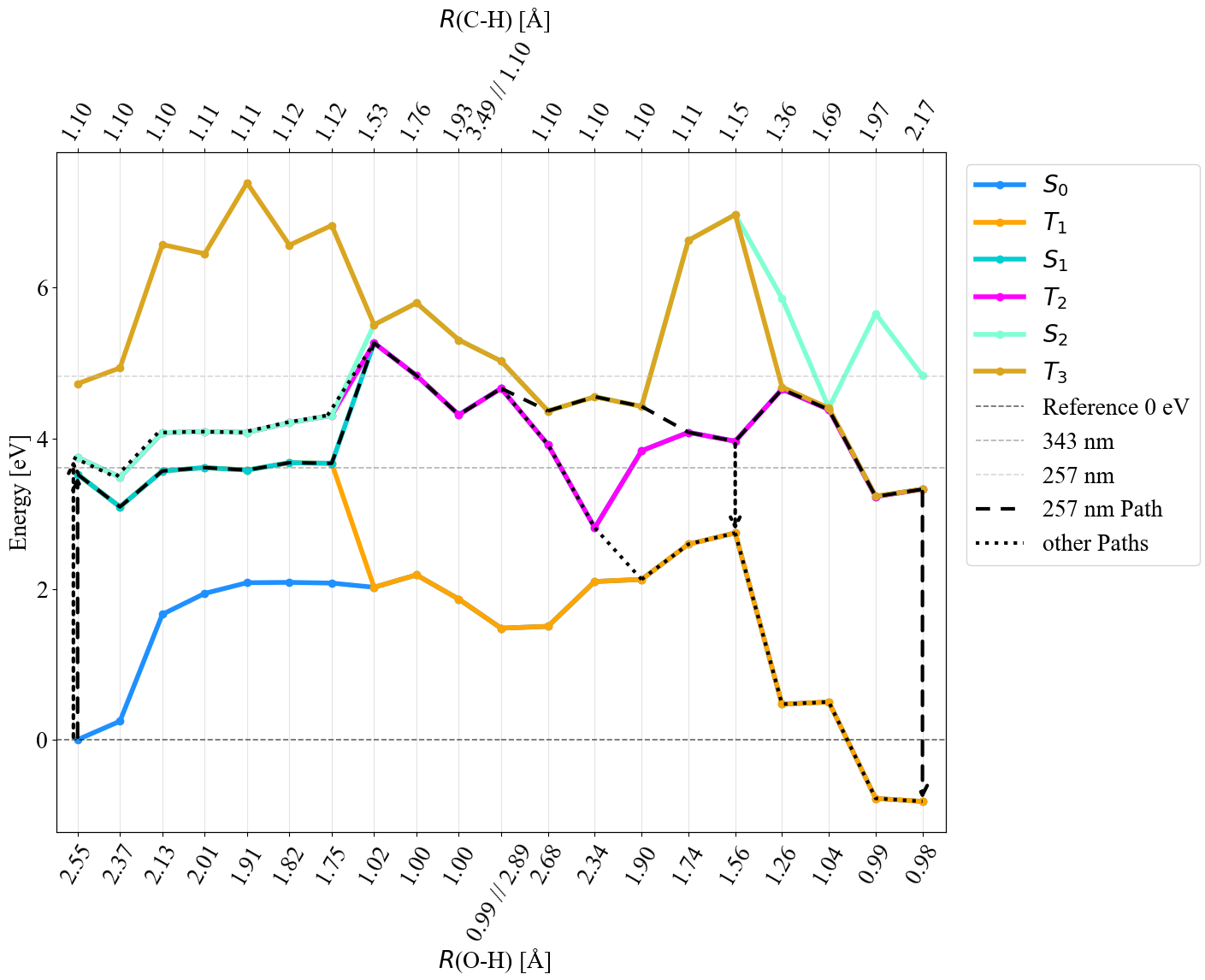} 
    \caption{Energy profile and the proposed reaction paths are shown for high energy photoexcitation (257~nm). The dashed black line represents the primary reaction channel, while dotted lines indicate alternative pathways. The higher energy input allows the system to remain on excited potential energy surfaces ($S_1, S_2, T_2, T_3$), effectively bypassing the rate determining barrier of the ground state.}
    \label{fig:19}
\end{figure}

\subsubsubsection{Mechanism of High Energy Photoexcitation} \label{4.3.2.3}
To investigate the experimental observation that 257~nm irradiation leads to a sevenfold increase in styrene yield compared to 343~nm excitation, the reaction was analyzed under high energy photoexcitation conditions.\supercite{Lai2023} A photon wavelength of 257~nm corresponds to an energy of approximately 4.82~eV. As illustrated in figure \ref{fig:19}, the 257~nm excitation enables the system to navigate through several high energy electronic manifolds, effectively circumventing the thermal traps that limit lower energy pathways.

The primary reaction channel (dashed black line) begins with a vertical excitation from the singlet ground state ($S_0$) to the $S_1$ state at $R$(O-H)~=~2.55~\AA. During this initial stage, the system occupies a charge transfer state, which is characterized by an ASCT. In this state, an electron is promoted from the organic $\pi$-system into the surface Ti(3d) conduction band. This sequence is driven by a single photon process. The energy provided by a single 257~nm photon is sufficient to drive the entire reaction coordinate, making secondary excitation unnecessary. The path follows the $S_1$ root until the O-H distance reaches 1.75~\AA, at which point the electronic character of the root transitions into the $S_1$/$T_2$ biradical manifold. This manifold represents the formation of the $\alpha$-phenylethyl radical cation and a reduced Ti$^{3+}$ center.

Upon reaching the intermediate stage ($R$(O-H) approximately 0.99~//~2.89~\AA), the system has enough vibrational and electronic energy to transition to the $S_2$/$T_3$ manifold. This transition is facilitated by the energetic approach between $S_1$ and $S_2$ observed in the potential energy profile between $R$(O-H)~=~2.68 and 2.34~\AA. The $S_2$ state represents a higher degree of electronic excitation, potentially involving the population of antibonding $\sigma$*-orbitals that directly destabilize the remaining $\beta$-C-H bond. The system remains on the highly activated $S_2$/$T_3$ surface until $R$(O-H)~=~1.90~\AA\ and then returns to the $S_1$/$T_2$ manifold.

Crucially, in this high energy primary path, the system is proposed to remain on the excited $S_1$/$T_2$ manifold until the final product geometry is reached, instead of undergoing non-radiative relaxation at the second transition state (TS2). This persistence on the excited state surface aligns with experimental observations that the reaction is not as strongly exothermic as the deep $S_0$ ground state sink would suggest. By bypassing the early non-radiative funnels, the system avoids the significant energy release associated with the ground state description. The final deactivation to the $S_0$ product manifold likely occurs via radiative relaxation only after the styrene-surface complex has fully formed.

The 257~nm excitation further provides a high degree of kinetic flexibility through several alternative pathways. One such channel begins with an excitation into the $S_2$/$T_2$ manifold and follows the $T_2$ root from $R$(O-H)~=~1.75~\AA\ onwards. Eventually, this pathway merges with the biradical manifold of the primary path. Another, more complex alternative involves the behavior of the system after the intermediate stage. As the path follows the $S_1$/$T_2$ manifold toward $R$(O-H)~=~2.34~\AA, the electronic character of the surface undergoes a significant "root flipping" event. This root flipping may indicate an avoided crossing. Due to the change in the underlying wave function, the path may switch to the $S_0$/$T_1$ ground state at $R$(O-H)~=~1.90~\AA. Even in this scenario, the barrier encountered in the ground state is only about half the height of the barrier faced by the 343~nm pathway. Therefore, it is plausible that the system can overcome this reduced barrier efficiently by retaining a high degree of "kinetic momentum" from a single, high energy photon.

The systems persistence on these excited surfaces is mechanistically significant. This ensures that the electron density is maintained in orbitals that facilitate bond cleavage rather than recombination. Overall, these diverse pathways explain the superior yield of 257~nm irradiation. The short wavelength laser provides access to a redundant network of high energy electronic states through single photon excitation. Even if partial relaxation occurs, the system remains on a trajectory that favors successful dehydrogenation to styrene without ever being limited by the rate determining ground state barrier.\supercite{Lai2023, Lai2022, Li2022_Lett, Lin2020}

\subsubsubsection{Comparison between CASSCF and DFT Results} \label{4.3.2.4}
To evaluate the reliability of the calculated potential energy surfaces, a direct comparison between the DFT results and the multi reference SA-CASSCF(12,12) data was performed. A critical aspect of this comparison is the choice of the energy reference. In this case, all CASSCF roots and the DFT data are normalized to the physisorbed singlet ground state ($S_0$) at $R$(O-H)~=~2.55~\AA. Note that the DFT study in the literature\supercite{Lin2020} uses the triplet adsorption state ($T_1$) as a reference for the subsequent reference energies. The same was done for the absolute values of the intermediate and product energies ($\Delta E_{int}$ and $\Delta E_{prod}$).

\begin{figure}[h!]
    \flushleft \hspace*{-0.2cm} 
    \includegraphics[scale=0.43]{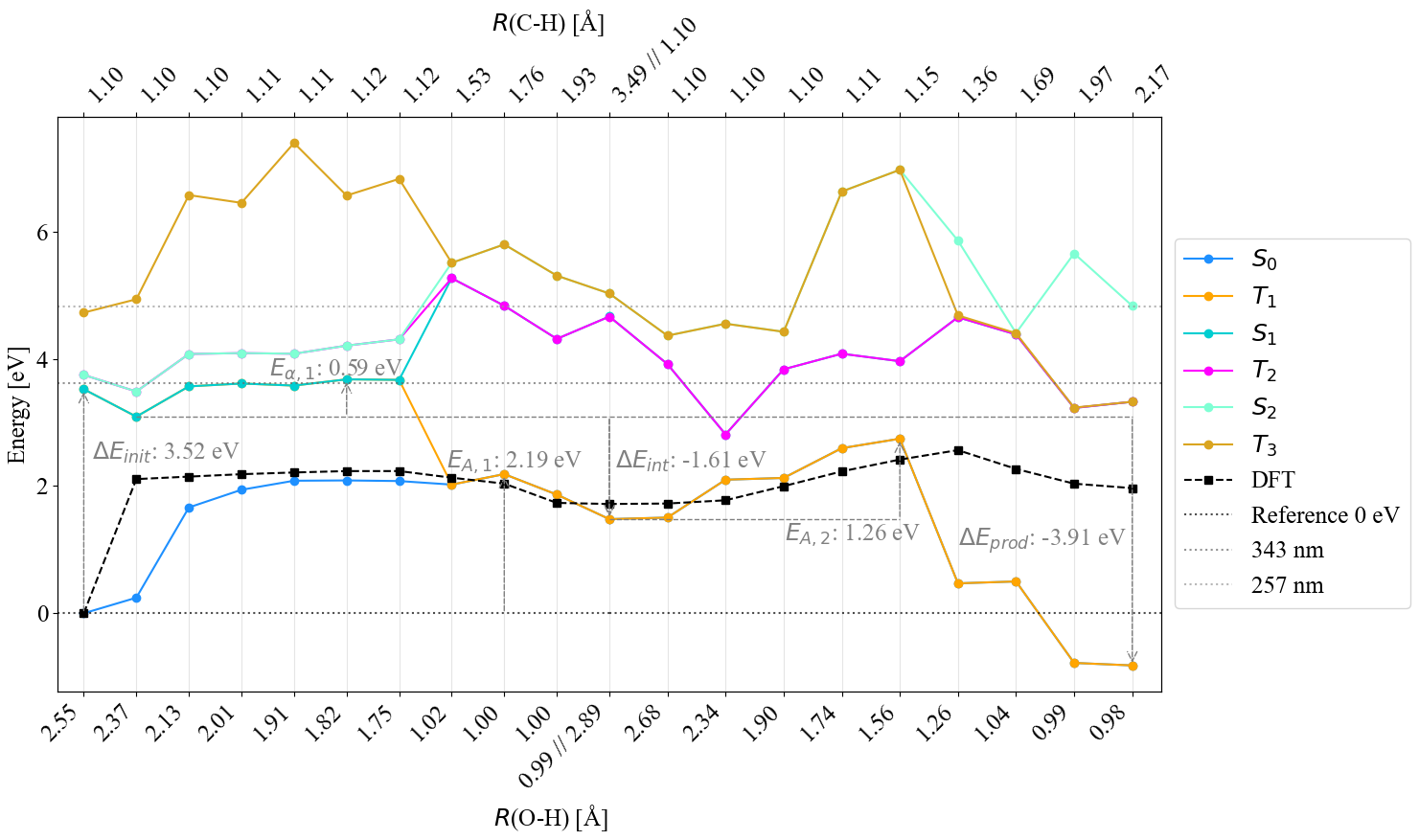} 
    \caption{Comparison of energy profiles calculated using SA-CASSCF(12,12) and DFT (PBE-D3) is shown. All energies are relative to the singlet physisorbed state ($S_0$) at 2.55~\AA. The plot shows the six lowest electronic roots, the DFT profile (black dashed), and the laser excitation levels (343~nm and 257~nm - complete horizontal dashed lines). Annotated arrows indicate activation barriers ($E_A$) and relative stabilization energies ($\Delta E$).}
    \label{fig:20}
\end{figure}

The comparison (see figure~\ref{fig:20}) reveals that, while DFT correctly identifies the geometric minima, the electronic character differs from CASSCF. These differences are evident in terms of absolute energy depths and barrier heights. Key values are being summarized in table~\ref{tab.4}.

\begin{table}[h!]
\centering
\caption{Comparison of activation energies ($E_A$/$E_\alpha$) and relative energies ($\Delta E$) are shown in eV. The $S_0$ state of adsorption singlet geometry was selected as the reference value. For the relative energies, the $T_1$ state of the triplet adsorption geometry was chosen, as done in literature \cite{Lin2020}.}
\label{tab.4}
\begin{tabular}{lccc}
\hline
Parameter & DFT (PBE-D3) & CASSCF(12,12) & Literature\supercite{Lin2020, Lai2023} \\ \hline
$E_{A,1}$ (Step I - Thermal) & 2.24 & 2.19 & ----- \\
$E_{\alpha,1}$ (Step I - Photochemical) & 0.19 & 0.59 & 0.3 to 0.4 \\
$E_{A,2}$ (Step II) & 0.85 & 1.26 & 1.2 \\
$\Delta E_{int}$ (Intermediate) & -0.39 &-1.61 & -0.1 to -0.7 \\
$\Delta E_{prod}$ (Product) & -0.14 & -3.91 & 0.47 \\ \hline
\end{tabular}
\end{table}

The strongest discrepancy was found in the stabilization of the product and intermediate. CASSCF predicts a much deeper energy sink for the styrene-surface complex (-3.91~eV) than DFT (-0.14~eV). These results suggest that the multi reference treatment captures a significantly stronger stabilization of the biradical species on the rutile surface. Additionally, the calculated CASSCF barriers $E_{\alpha,1}$~=~0.59~eV, where $\alpha$ indicates photochemistry and $E_{A,2}$~=~1.26~eV are higher than the DFT results, but are in better alignment with the literature barriers.

The necessity of the CASSCF approach is emphasized by the spin contamination, $\braket{S^2}$, observed in the DFT calculations. The physisorbed state shows a very low contamination of 0.003. This value increases nearly fivefold during the reaction, reaching 0.008 at the first C-H cleavage ($R$(O-H)~=~1.02~\AA) and 0.014 at the final product. Although these absolute values remain within the acceptable range for standard DFT applications, the systematic increase is an indicator of the merging of the singlet and triplet manifolds. In the biradical region, the electronic wavefunction is a mixture of different configurations that cannot be accurately represented by a single slater determinant. CASSCF resolves this by explicitly treating the quasi-degeneracy of the $S_0$ and $T_1$ roots, which effectively merges them into a common biradical manifold where the singlet-triplet gap becomes negligible.

The comparison shows that, while useful the mapping of the MEP, a single reference method like DFT provides only a limited description of the energetic landscape. The significant differences in product stabilization and the rising spin contamination confirm that the dehydrogenation of ethylbenzene is a process with strong multireference character. Therefore, explicitly treating static correlation in CASSCF is therefore essential to understand how the system navigates through the biradical manifold and why high energy photoexcitation at 257~nm is so much more effective than lower energy pathways.

It is noteworthy that, despite the shifts in absolute energy, DFT performs remarkably well in describing the relative progression of the potential energy surface in the region between TS1 and TS2. However, in the initial stage of the reaction (adsorption to TS1), the calculated DFT triplet state curiously tracks the energy level of the CASSCF singlet ground state ($S_0$) rather than the corresponding triplet root. The most significant discrepancy occurs during the final transition from TS2 to the product geometry. This can be seen as a failure of the single reference approach. Here, the curves diverge drastically as DFT fails to capture the energetic stabilization of the styrene-surface complex predicted by the multi reference treatment. This failure is also visible from the deviation of the desorption values for the product (see table \ref{tab.2}).

\subsubsubsection{Failure of Dynamic Correlation: NEVPT2 Analysis} \label{4.3.2.5}
While the SA-CASSCF method is capable of accurately capturing the static correlation necessary for describing bond dissociation and near-degeneracy effects.\supercite{Roos1987} The inclusion of dynamic correlation via $N$-electron valence state perturbation theory (NEVPT2) was found to be numerically unstable for the present surface model. Theoretically, NEVPT2\supercite{Angeli2001_1} is designed to be free of the "intruder state" problems often encountered in alternatives like CASPT2\supercite{Andersson1992, Angeli2001_1}. Despite this, the results obtained for the \ch{Ti27O54[O34]^{-68}} cluster show a breakdown of the perturbative expansion.

As summarized in table~\ref{tab.5}, the second-order energy corrections ($dE$) are physically unreasonable, reaching magnitudes of approximately -25~Hartree. Furthermore, the correction is sensitive to the imaginary level shift parameter, a technique originally proposed to handle singularities in the partially contracted version.\supercite{Roos1995} A variation of the shift from 0.1 to 10.0 leads to a drastic change in $dE$ from -25.1 to -6.7~Hartree. Such behavior indicates that the denominator in the perturbation sum is dominated by states with near-zero energy gaps, preventing the convergence of the correlation energy despite the intruder state free nature of the NEVPT2 formalism.\supercite{Angeli2001_1}

\begin{table}[h]
\centering
\caption{NEVPT2 (SC - strongly contracted, PC - partially contracted) energy corrections ($dE$) and total energy ($E_{tot}$) in Hartree for different level shifts. These illustrate the numerical divergence and instability of the dynamic correlation treatment.}
\label{tab.5}
\begin{tabular}{lccc}
\hline
Method & Im. Levelshift & $dE$ & $E_{tot}$ \\ \hline
SC-NEVPT2 & ------ & -25.131 & -30663.251 \\
PC-NEVPT2 & 0.1 & -25.138 & -30663.259 \\
PC-NEVPT2 & 0.5 & -25.094 & -30663.215 \\
PC-NEVPT2 & 1.0 & -24.672 & -30662.793 \\
PC-NEVPT2 & 5.0 & -13.954 & -30652.074 \\
PC-NEVPT2 & 10.0 & -6.693 & -30644.814 \\ \hline
\end{tabular}
\end{table}

The failure can be attributed to the specific nature of the "charged cluster" model used to simulate the \ch{TiO2}(110) surface.\supercite{Sun2016} The high negative total charge of -68 creates an intense electrostatic potential that facilitates extreme polarization and relaxation effects at the QM-cluster boundary. This likely pulls down a vast number of virtual orbitals, creating an artificial quasi-continuum of low energy virtual states.\supercite{Xu1999, Sauer1989, Seijo1999}

As noted by Xu \textit{et al.} \cite{Xu1999}, the use of charged cluster models can lead to spurious state interactions. This phenomenon is further elucidated by Seijo \textit{et al.} \cite{Seijo1999}, who described the collapse of electronic states onto external sites in the absence of proper quantum embedding terms. This creates an artificial quasi-continuum, which violates the fundamental requirement of Rayleigh-Schrödinger perturbation theory. The energy denominators in the second-order treatment approach zero, leading to the observed divergence in the NEVPT2 calculation. The NEVPT2 failure possibilities mentioned thus far are only hypotheses that would require further research for confirmation.

Despite the failure of NEVPT2, the SA-CASSCF profiles remain a robust basis for quantitative trend predictions. This is the case, because dynamic correlation often contributes as a nearly uniform vertical energy shift across the reaction coordinate when the underlying electronic character is correctly captured.\supercite{SerranoAndres2005, Roos1987} Since the SA-CASSCF method explicitly treats the static correlation required for a possible homolytic C-H bond cleavage and radical formation, the relative topology of the potential energy surfaces is preserved.

\subsection{Influence of Surface Oxidation on the Reaction Mechanism} \label{4.4}
Experimental studies by Li \textit{et al.} \cite{Li2022_Lett} have demonstrated that the oxidation state of the rutile \ch{TiO2}(110) surface is a decisive factor for the catalytic performance of EB dehydrogenation. A key finding is that even under low energy photoexcitation conditions (355~nm~$\approx$ 3.49~eV), the styrene yield increases by approximately a factor of four on an oxidized surface compared to a reduced stoichiometric surface. This drastic increase in activity is primarily attributed to pre-adsorbed oxygen atoms (O$_{Ti}$), which act as efficient hydrogen scavengers. This inhibits the back reaction of the phenylethyl intermediate. However, this gain in reactivity involves a significant trade-off in terms of selectivity. While the reduced surface exhibits nearly 100\% selectivity toward styrene, the oxidized surface promotes the formation of byproducts, specifically 2,3-diphenylbutane and acetophenone. These circumstances decrease the styrene selectivity to approximately 74\% at saturation.\supercite{Li2022_Lett} The following sections provide a detailed quantum chemical analysis of the reaction on an oxidized cluster model to understand the underlying electronic reasons for the increase in yield and the changed branching ratios.

\subsubsection{Construction and Validation of the Oxidized Cluster Model} \label{4.4.1}
\begin{figure}[h!]
    \centering
    \includegraphics[scale=0.6]{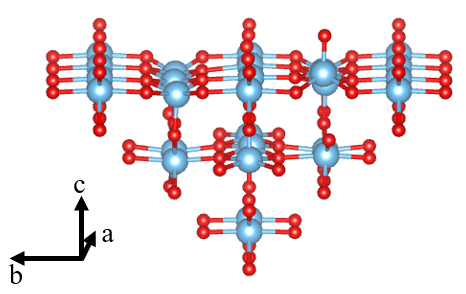}
    \caption{The oxidized surface model, which uses the reduced surface and introduces one O$^{2-}$ atom. (Color scheme: red – oxygen, blue – titanium)}
    \label{fig:21}
\end{figure}

To theoretically represent the oxidized surface conditions observed in the experiments, the stoichiometric cluster model was modified. As illustrated in figure~\ref{fig:21}, an additional oxygen species was introduced by placing one O$^{2-}$ atom at a fivefold coordinated titanium (5f-Ti) site. This species is referred to as O$_{Ti}$. This extra species significantly increases the complexity of the potential energy surface. In order to reach a stable local minimum during geometry optimization, the number of relaxed inner cluster atoms had to be increased from 19 (in the reduced model) to 28.

\newpage

Because of the fundamental changes in the cluster composition and the resulting total charge, the absolute energy values of the oxidized system cannot be directly compared to those of the reduced surface. However, the reliability of the model can be validated by comparing the calculated trends with experimental TPD data. As shown in table~\ref{tab.2}, the model successfully reproduces the following experimental observation. Ethylbenzene (EB) has a smaller desorption energy than the styrene produced on the oxidized \ch{TiO2}(110) surface (at the O$_{Ti}$ sites).

Beyond the energetic trends, the oxidation state impacts on the electronic structure of the adsorbate-surface complex. A key characteristic of the oxidized model is the reduction of the electronic gap. According to the DFT calculations, the bandgap narrows to 0.97~eV. While the physisorbed state on the reduced surface requires high energy for the initial charge transfer, the O$_{Ti}$ species introduces three new O(2p) states within the bandgap region. Changes in the surface oxygen stoichiometry significantly modify the electronic landscape. This aligns with the findings of fundamental studies of rutile \ch{TiO2}(110), which show that the presence and depth of oxygen related species strongly modulate the surface orbital energies and the resulting work function.\supercite{Pabisiak2007} Consequently, this electronic sensitization drastically lowers the excitation threshold, allowing the system to be "primed" for the reaction even under lower energy irradiation. This sensitization is a critical prerequisite for explaining the high styrene yields observed at 355~nm, which will be analyzed in detail in the following sections.

\subsubsection{Characterization using DFT} \label{4.4.2}
\begin{figure}[h!]
    \flushleft \hspace*{-0.2cm} 
    \includegraphics[scale=0.43]{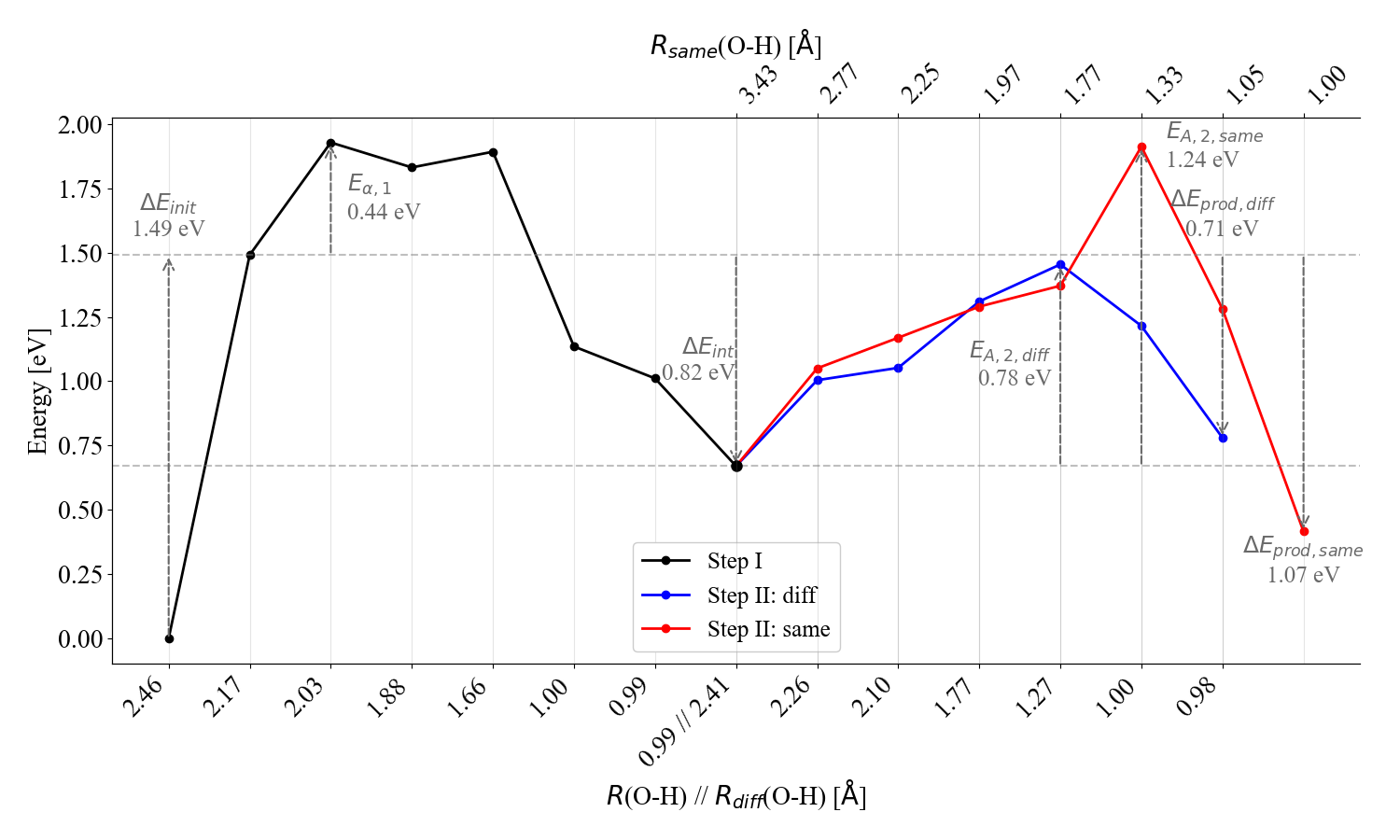}
\caption{DFT calculated energy profile for the dehydrogenation of EB on the oxidized \ch{TiO2}(110) surface. After the intermediate stage, the path branches into a "diff" pathway, (second H transfer to O$_{br}$, leading to two OH groups) and a "same" pathway (second H transfer, leading to water formation at O$_{Ti}$).}
\label{fig:22}
\end{figure}

To understand the impact of surface oxidation on the reaction kinetics and thermodynamics, the energy profile was explored using spin-unrestricted DFT (UKS-PBE-D3). Figure~\ref{fig:22} illustrates the complete reaction coordinate, starting from the physisorbed state and branching into two distinct pathways for the second hydrogen abstraction. The "same" path uses the created hydroxy group as the cleavage agent. In the "diff" path, an O$_{br}$ is utilized instead.

\begin{figure}[h!]
\centering
    \includegraphics[scale=0.7]{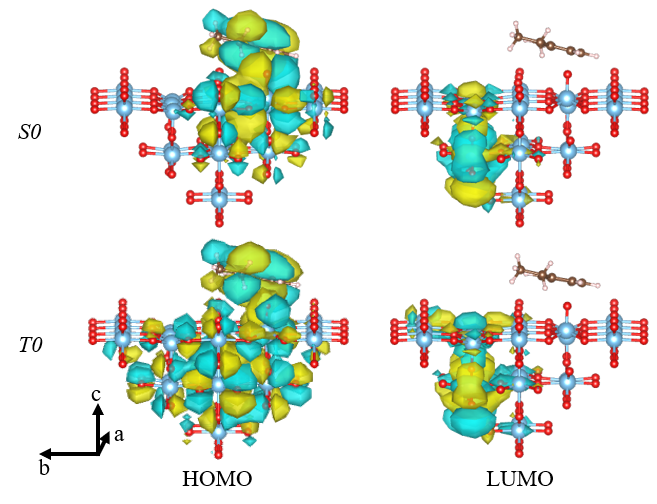}
\caption{Electronic structure analysis of the adsorbed state on the oxidized surface. Left: HOMO and LUMO for $S_0$ and $T_1$, showing the involvement of EB and O$_{Ti}$ in the occupied states. Right: DFT spin density visualization for $T_1$, highlighting the distribution of radical character across the $\alpha$-carbon, the $\pi$-system, O$_{Ti}$, and the reduced Ti(3d) centers.}
\label{fig:23}
\end{figure}

Analyzing the frontier molecular orbitals of the singlet ground state ($S_0$) and the first triplet state ($T_1$) shows that the pre-adsorbed oxygen atom (O$_{Ti}$) significantly participates in the electronic structure. As shown in figure~\ref{fig:23}, the HOMO of the $S_0$ state is characterized by electron density localized on both the ethylbenzene molecule and the O$_{Ti}$ species, while the LUMO consists of pure Ti(3d) character. The triplet state ($T_1$) exhibits an essentially identical pattern, indicating that the "hole" created upon excitation is shared between the organic $\pi$-system and the additional oxygen atom.

The Mulliken population analysis provides further quantitative insight into this arrangement. In the $S_0$ state, the O$_{Ti}$ atom carries a significant negative charge of -0.72 with zero spin. When transitioning to the $T_1$ state, a significant portion of the spin density (0.21) localizes on this oxygen atom, while the remaining spin is distributed among the organic framework (specifically the benzylic $\alpha$-carbon and the phenyl ring) and the surface titanium atoms. These findings confirm that the pre-adsorbed oxygen acts as an active, radical-like center at the start of the photochemical cycle.

The first hydrogen transfer (Step I) proceeds via an "early" transition state (TS1), with a barrier of 0.44~eV. Compared to the reduced surface (0.13~eV), oxidation slightly increases this initial hurdle. The system then reaches a more stable intermediate ($\Delta E_{int}$ = -0.82~eV), which represents a significant thermodynamic sink (0.43 eV deeper than on the reduced surface).

For the second rate determining dehydrogenation (Step II), the oxidized surface provides two competing routes with different kinetic and thermodynamic profiles (see figure~\ref{fig:22}). In the "diff" pathway, an O$_{br}$ site abstracts the second hydrogen. This path passes through TS2$_{diff}$ with a barrier of only 0.78~eV. The resulting product state (styrene and two surface hydroxyl groups) is located at -0.71~eV relative to $T_1$ adsorption. This represents a slight endothermic shift of 0.11~eV compared to the intermediate. The "same" pathway leads directly to the formation of styrene and water, but faces a much higher barrier of 1.50~eV (TS2$_{same}$). Despite this kinetic hurdle, the "same" pathway results in the most stable product configuration at -1.08~eV, which is 0.26~eV lower in energy than the intermediate.

The high barrier of the "same" pathway (1.50~eV) suggests that the reaction would likely stall at the intermediate stage if this were the only route to styrene formation. However, the "diff" pathway provides a much lower kinetic barrier (0.78~eV). Considering that the measured barrier for hydrogen migration on the \ch{TiO2}(110) surface is approximately 0.74~eV,\supercite{Li2008} a two step mechanism becomes highly plausible, if the system possesses insufficient energy to overcome TS2$_{same}$. The system can follow the "diff" route to form the styrene-surface complex with two OH groups. Then, a hydrogen atom can migrate across the surface to the O$_{Ti}$ site forming \ch{H2O} to reach the thermodynamically most stable configuration. The synergy between a lower barrier dehydrogenation on the one hand and surface diffusion on the other hand explains why the reaction proceeds efficiently despite the high "same"-site barrier. This would also effectively explain, why water is found as a coupling byproduct of the reaction.\supercite{Li2022_Lett}

\subsubsection{Multi Reference Analysis of the Reaction Mechanism} \label{4.4.3}
To accurately describe the electronic complexity introduced by the additional oxygen species (O$_{Ti}$), the active space was expanded to a CAS(14,14). This expansion includes the twelve orbitals of the stoichiometric system, as well as the O(2p) orbital of the O$_{Ti}$ and a corresponding Ti(3d) orbital. This expansion became essential because the oxidation state significantly alters the distribution of gap states, which are identified as the primary origin of photocatalytic activity on rutile surfaces.\supercite{Yim2010} A fundamental result of this multi reference treatment is the significant electronic sensitization of the surface. At the initial adsorption geometry, the vertical excitation energy to the $S_1$ state (ASCT character) is reduced to only 1.29~eV, which is significantly lower than the values found for the reduced surface.\supercite{Pabisiak2007}

This "priming" of the system is further characterized by a "double-crossing" or root-swapping event that occurs between $R$(O-H)~=~2.46~\AA\ and 2.03~\AA. Due to the intense electrostatic stabilization provided by the O$_{Ti}$ site, the charge transfer configuration actually becomes the $S_0$ ground state at $R$(OH)~=~2.17~\AA. The closed shell configuration is pushed into the higher-lying $S_2$ manifold. This role reversal, a consequence of the non-crossing rule in polyatomic systems,\supercite{Yarkony1996, Zhu2016} ensures that a reactive biradical character is present from the early stages of the reactant approach.

\begin{figure}[h!]
    \flushleft \hspace*{-0.2cm} 
    \includegraphics[scale=0.415]{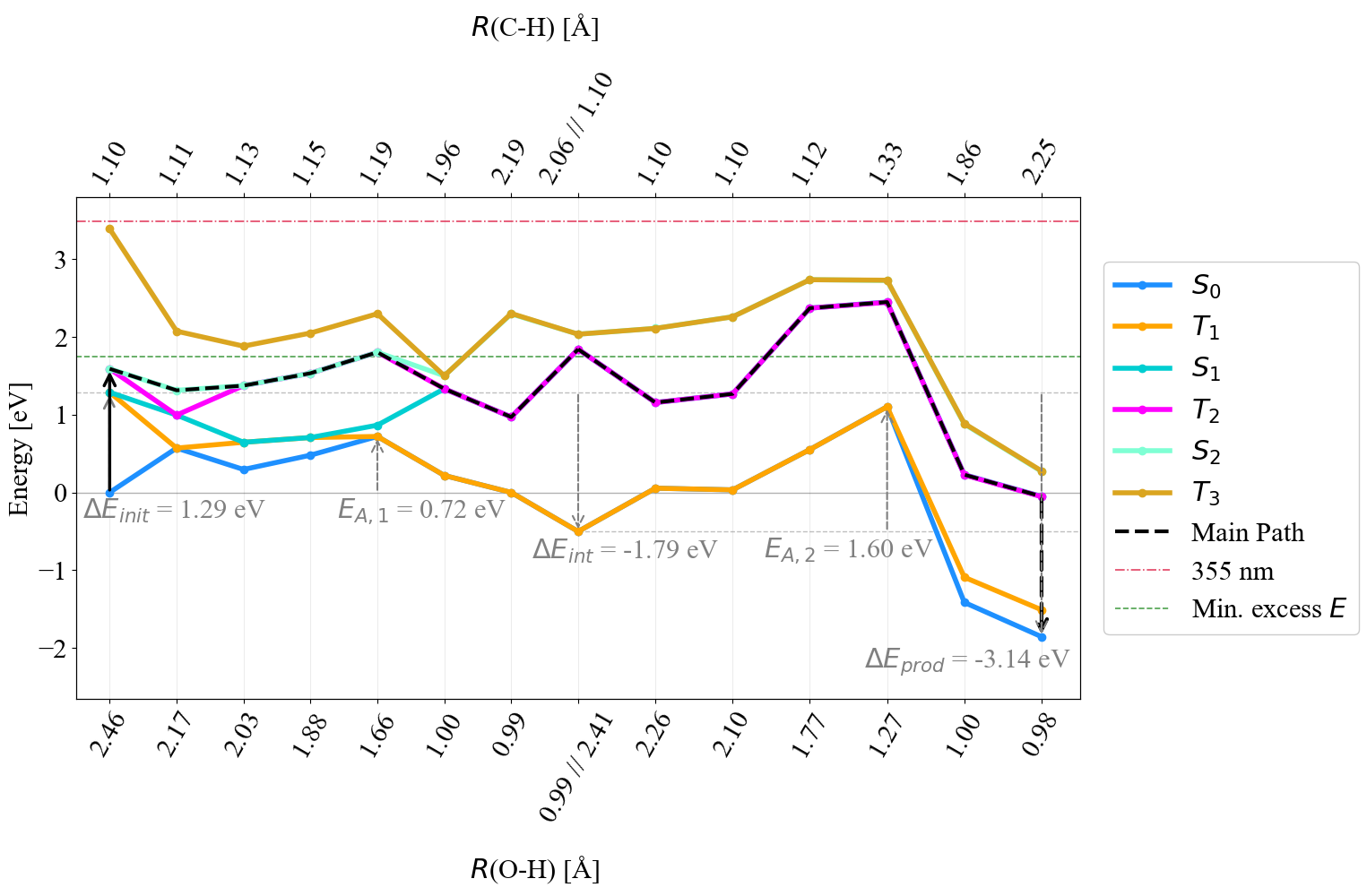}
\caption{Energy profile of the "diff" pathway on the oxidized \ch{TiO2}(110) surface calculated at the SA-CASSCF(14,14) level. All energies are normalized to the $S_0$ state at $R$(OH)~=~2.46~\AA. The black dashed line indicates the proposed "high-road" trajectory.}
\label{fig:24}
\end{figure}

\begin{figure}[h!]
    \flushleft \hspace*{-0.2cm} 
    \includegraphics[scale=0.415]{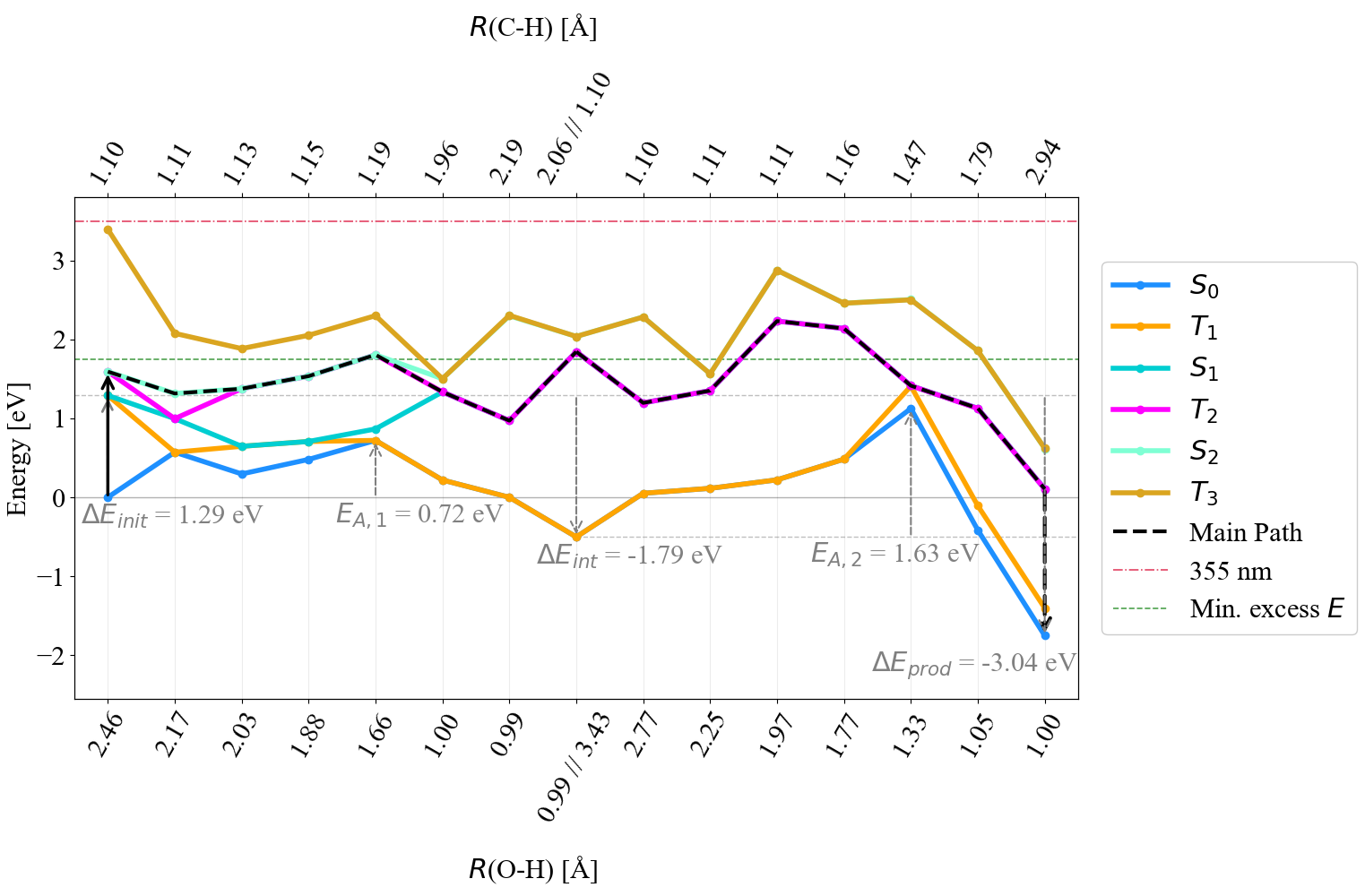}
\caption{Energy profile of the "same" pathway (water formation) on the oxidized \ch{TiO2}(110) surface calculated at the SA-CASSCF(14,14) level. All energies are normalized to the $S_0$ state at $R$(OH)~=~2.46~\AA. The black dashed line indicates the proposed "high-road" trajectory.}
\label{fig:25}
\end{figure}

Under 355~nm (3.49~eV) irradiation, the system follows a "high-road" trajectory (dashed black line in figures~\ref{fig:24} and \ref{fig:25}) that maximizes the expected kinetic efficiency by avoiding local potential traps. This mechanism is driven by a single photon process in which the energy of the 355~nm photon is sufficient to keep the system in the excited state manifold throughout the entire reaction. To verify the physical robustness of this path, a "minimum excess energy" threshold was established (indicated by the green dashed line). This conservative estimate assumes that the initial excitation requires an energy equivalent to the experimental bulk bandgap of 3.03~eV rather than the calculated 1.29~eV. Even under this assumption, the system maintains a significant energetic buffer. As illustrated in the energy profiles, this allows the system to completely bypass the local energy maxima found on the $S_0$ surface. At $R$(OH)~=~1.66~\AA, the $S_1$ and $S_2$ roots reach a state of near-degeneracy, facilitating an efficient internal conversion that keeps the system on an electronically activated biradical trajectory.

Experimental observations show that the reaction is not as strongly exothermic as a ground state relaxation suggests. Thus, the proposed main path follows the excited $S_1$/$T_2$ manifold until the very end of the reaction coordinate. Non-adiabatic funnels are observed at $R$(OH)~=~1.33~\AA\ (for the "same" pathway) and 1.27~\AA\ (for the "diff" pathway). These provide theoretical exit channels for radiationless relaxation.\supercite{Bernardi1996, Ryabinkin2017} The possibility of persistence on the excited manifold is more probable. By remaining on the $S_1$/$T_2$ surface until the styrene-surface complex is fully formed, the system maintains the "kinetic momentum" required to bypass all ground state hurdles.\supercite{Malhado2014} The final deactivation to the $S_0$ manifold likely occurs via radiative relaxation from the excited product state. This mechanism suggests a consistent and plausible explanation for the energy balance of the reaction, because it avoids the artificial thermodynamic sink predicted by the ground state description.

An important observation at these product stages is the energetic splitting between the $S_0$ and $T_1$ roots of approximately 0.35~eV, a pattern not observed on the reduced surface. This splitting is a direct result of the formation of a Ti$^{2+}$-like configuration, which accounts for 31\% of the wavefunction weight. The spatial confinement of the reduced titanium centers increases the exchange interaction and lifts the degeneracy of the biradical state. The drastically lowered excitation threshold, the single photon efficiency, and the ability of the system to remain on these highly activated manifolds until the final product stage explain the fourfold increase in styrene yield observed under 355~nm irradiation on the oxidized surface.

\subsubsection{Comparison between CASSCF and DFT Results} \label{4.4.4}
The direct comparison of the energy profiles calculated using spin-unrestricted DFT (UKS-PBE-D3) and multi reference SA-CASSCF(14,14) reveals significant discrepancies. These discrepancies are especially visible in the description of the electronic states during the "priming" stage and the final product stabilization. Although both methods were normalized to the physisorbed $S_0$ geometry at $R$(O-H)~=~2.46~\AA, their energetic predictions for the reaction coordinate differ remarkably. Furthermore, the spin contamination remains below 0.016 during the entire reaction.

The calculated activation and relative energies for the oxidized surface are summarized in table~\ref{tab.6}. The data reveals that the single reference approach generally underestimates both the stabilization of the radical species, as well as the height of the second reaction barrier.

\begin{table}[h]
\centering
\caption{Comparison of activation energies ($E_A$) and relative energies ($\Delta E$) in eV for the oxidized surface. All values are relative to the $S_0$ adsorption geometry ($S_0$ or $T_1$ state).}
\label{tab.6}
\begin{tabular}{lcc}
\hline
Parameter & DFT (PBE-D3) & CASSCF(14,14) \\ \hline
$E_{A,1}$ (Step I - Thermal) & 1,93 & 0.72 \\ 
$E_{\alpha,1}$ (Step I - Photochemical) & 0.44 & ------ \\      
$E_{A,2}$ (Step II, diff) & 0.78 & 1.60 \\      
$E_{A,2}$ (Step II, same) & 1.24 & 1.63 \\
$\Delta E_{int}$ (Intermediate) & -0.82 & -1.79 \\    
$\Delta E_{prod}$ (Product) & -0.71 / -1.07 & -3.04 / -3.14 \\ \hline
\end{tabular}
\end{table}

As seen in Section 4.1, CASSCF tends to overestimate electronic excitation energies and band gaps, compared to experimental benchmarks. However, on the oxidized surface, even the CASSCF treatment predicts a drastically reduced vertical excitation energy of 1.29~eV (compared to $>$3.5~eV on the reduced surface). Despite the systematic overestimation inherent to CASSCF, this significant reduction confirms the role of the O$_{Ti}$ species as a powerful photosensitizer, lowering the threshold for the initial charge transfer into the Ti(3d) states.

\begin{figure}[b!]
    \flushleft \hspace*{-0.2cm} 
    \includegraphics[scale=0.39]{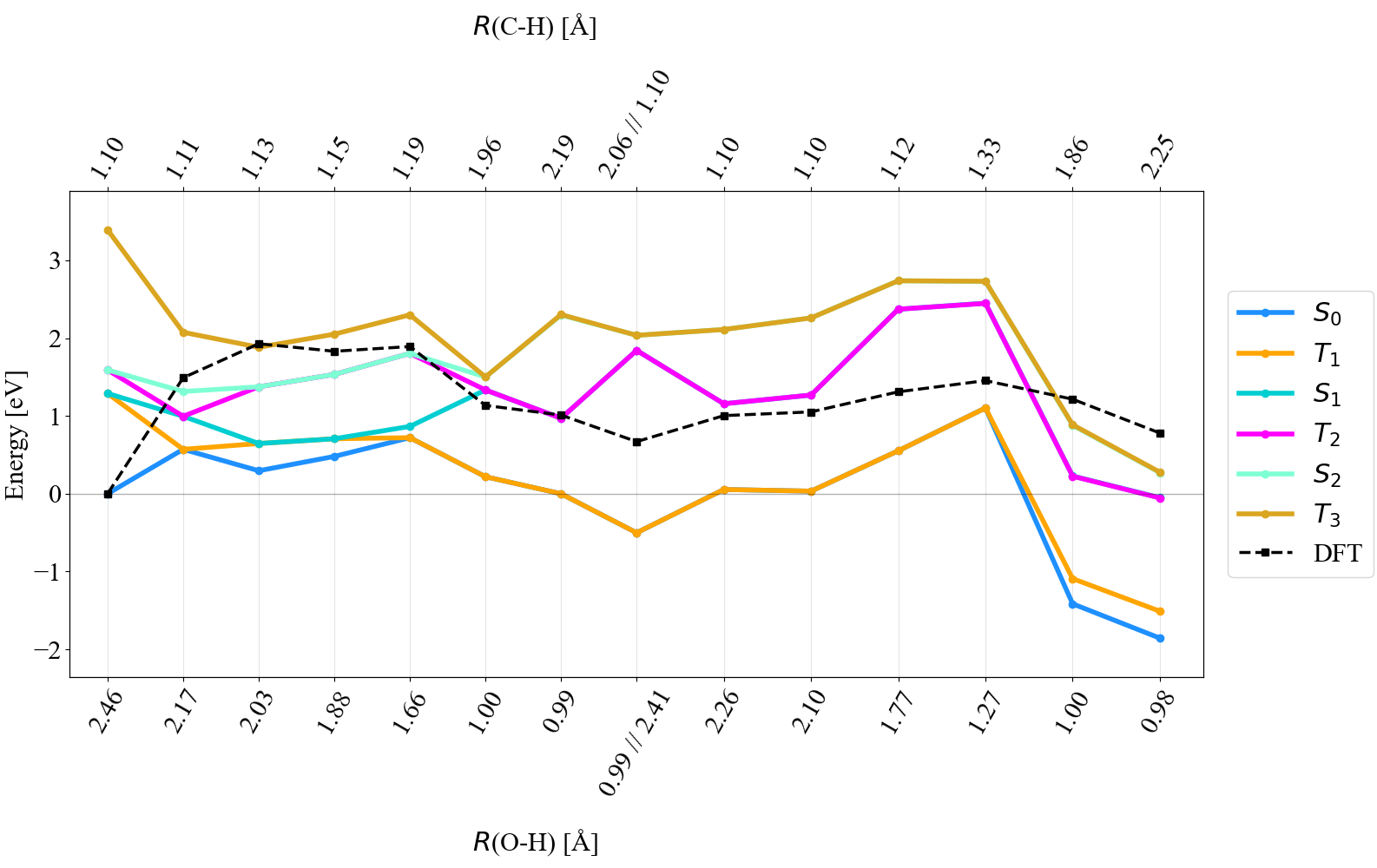}
\caption{Comparison of energy profiles for the "diff" pathway on the oxidized \ch{TiO2}(110) surface, calculated via SA-CASSCF(14,14) and UKS-DFT (PBE-D3). All energies are relative to the physisorbed $S_0$ state at 2.46~\AA. The plot shows the six lowest electronic roots and the DFT profile (black dashed).}
\label{fig:26}
\end{figure}

\begin{figure}[h!]
    \flushleft \hspace*{-0.2cm} 
    \includegraphics[scale=0.39]{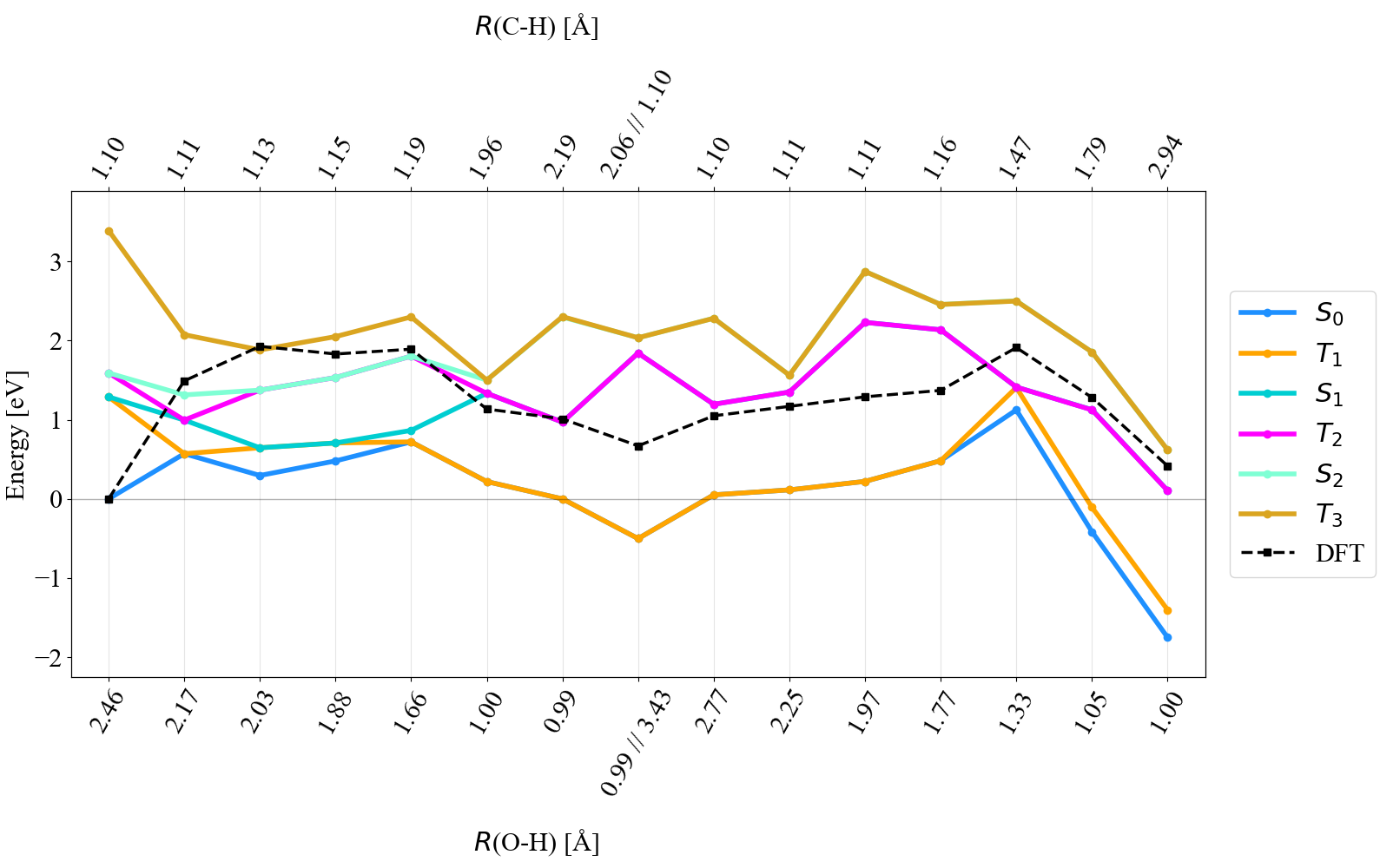}
\caption{A comparison of energy profiles for the "same" pathway (water formation) on the oxidized surface, calculated via SA-CASSCF(14,14) and UKS-DFT (PBE-D3).All energies are relative to the physisorbed $S_0$ state at 2.46~\AA. The plot shows the six lowest electronic roots and the DFT profile (black dashed).}
\label{fig:27}
\end{figure}

Taking a closer look at the electronic character reveals that the UKS-DFT approach fails to correctly identify the lowest triplet root ($T_1$)(see figures \ref{fig:26} and \ref{fig:27}). Initially, the DFT triplet path (black dashed line) can be found halfway between the CASSCF $T_2$ and $T_3$ roots. Beyond $R$(O-H)~=~1.00~\AA, the DFT energy purely tracks the $T_2$ state until the end of the reaction. However, the DFT triplet does not align with the CASSCF $T_1$ root, which is the state that it is formally supposed to represent. This discrepancy highlights the limitations of the single reference wave function in DFT. Therefore, this single reference method is unable to resolve the complex root-swapping and near-degeneracy events that characterize the multi reference landscape of the oxidized surface.

Further differences are observed in the second dehydrogenation step. In the CASSCF description, the barriers for the "same" and "diff" pathways are nearly identical ($\approx$~1.6~eV). In contrast, DFT predicts a significant kinetic advantage for the "diff" pathway (0.78~eV vs. 1.24~eV).

Similarly, the methods diverge regarding the final product stability. CASSCF predicts a very deep energy sink, with both the "same" and "diff" configurations being almost degenerate and highly stable ($\approx$~-3.1~eV). However, the DFT model shows that only the "same" product is more stable than the intermediate by 0.25~eV, while the "diff" product is less stable by 0.11~eV.  These results indicate that the PBE functional fails to capture the significant stabilization of the styrene-surface complex.  This failure is likely linked to the highly correlated Ti$^{2+}$ character (31\% weight) identified in the CASSCF wave function. Correctly describing the thermodynamic driving force requires an explicit treatment of the exchange interaction between the reduced titanium centers.

\newpage

In conclusion, the comparison reveals that pre-oxidation creates an electronic landscape fundamentally beyond the reach of single reference DFT methods. DFT provides an inaccurate picture of the branching ratios because it misses the "priming" effect, tracks higher lying triplet states incorrectly, and fails to account for the true depth of the product sink. Thus, CAS(14,14) analysis is essential for understanding why 355~nm irradiation is so effective. The real energy profile offers a more balanced and accessible pathway for styrene formation on oxygen preactivated rutile surfaces.

\subsubsection{Role of Surface Oxidation in the Reaction Mechanism} \label{4.4.5}
The dramatic increase in catalytic activity resulting from surface pre-oxidation is not just a matter of quantity. Rather, it leads to a fundamental shift in the reaction mechanism across both dehydrogenation steps. A comparative analysis of Mulliken spin densities and atomic charges provides physical evidence of this transition.

On the stoichiometric (reduced) surface, the bridging oxygen atoms (O$_{br}$) remain in a closed shell state (O$^{2-}$) with spin densities close to 0.00. Consequently, the O$_{br}$ site acts as a Lewis base. The Mulliken analysis of the first migrating hydrogen shows a positive charge (up to 0.31), confirming a stepwise proton coupled electron transfer (PCET). The electron is injected into the Ti(3d) orbitals while only the proton moves to the O$_{br}$ site.

On the oxidized surface, the O$_{Ti}$ species acts as a pre-activated radical center with a spin density of 0.14. This allows for a direct hydrogen atom transfer (HAT). During the early stages of transfer, the first hydrogen atom remains nearly neutral. It carries its electron directly to the O$_{Ti}$ radical, which fills the electronic "hole." This concerted homolytic cleavage is fundamentally more efficient than the spatially separated charge transfer on the reduced surface.

Although the second H-abstraction (step \Romannum{2}) is the rate determining step for both surfaces, the electronic environment differs significantly. On the reduced surface, the second step follows a different and more demanding PCET pathway. According to the Mulliken data, the now cleaving hydrogen (charge 0.22, spin 0.00) must decouple an electron and a proton again. Without radical stabilization, this step faces the high activation barrier (0.85~eV) identified in chapter~\ref{4.3.1}. On the oxidized surface, step II benefits from the "thermodynamic sink" created in step I. Even as the reaction branches into the "diff" or "same" pathways, the organic fragment maintains a high radical character (spin density of ~0.38). This stabilization of the phenylethyl intermediate lowers the energy required for the second C-H cleavage. In the "diff" pathway, although the transfer to a neighboring O$_{br}$ site may revert to a PCET-like electronic signature (H charge of 0.28). Overall, the process is kinetically facilitated by a lack of recombinative reactions, because the first hydrogen is already "trapped" at the O$_{Ti}$ site. Furthermore, the CASSCF analysis reveals that the "same" pathway is nearly energetically degenerate to the "diff" route, exhibiting the same activation barrier of 1.63~eV (compared to 1.60~eV for the "diff" path). The product of the "same" pathway is slightly less stable by 0.1~eV ($\Delta E_{prod}$~=~-3.04~eV vs.~-3.14~eV). Both configurations lead to a massive thermodynamic stabilization of the styrene-surface complex. This near-degeneracy of the two pathways provides the system with multiple accessible relaxation channels. This ensures the reaction can efficiently proceed toward a deep energy sink, regardless of the specific proton acceptor site.

The high efficiency of the first steps HAT and the radical stabilization of the intermediate in the second step both explain the fourfold increase in styrene yield. The stoichiometric surface is hindered by two consecutive, demanding PCET steps. However, the oxidized surface uses the radical nature of O$_{Ti}$ to bypass these electronic barriers. This allows the oxidized surface to effectively act as a "hydrogen scavenger," driving the reaction toward completion under significantly milder conditions.

\section{Conclusion and Outlook} \label{5}
This thesis presents a comprehensive quantum chemical investigation of the thermal and photochemical dehydrogenation of ethylbenzene on the rutile \ch{TiO2}(110) surface. A combination of density functional theory (DFT) and a multi-reference method (CASSCF) was used. This approach revealed the complex interplay between surface oxidation, photon energy, and electronic states.

A central finding of this work is the pronounced multireference character of the reaction, especially during bond cleavage and in the product stages. DFT provides a robust framework for geometry optimizations. However, DFT fails to capture the electronic complexity of the biradical species and incorrectly tracks the stabilization of the product manifold.  Therefore, CASSCF was necessary to describe the electronic landscape of the reaction coordinate.

Due to the immense computational complexity of the active spaces, CAS(12,12) for the stoichiometric and CAS(14,14) for the oxidized surface, a complete mapping of the \linebreak 3$N$~-~5 dimensional potential energy surface (PES) was not possible with currently available methodologies. Instead, the investigation focused on one dimensional energy profiles along the found minimum energy paths (MEP), which were evaluated at stationary points and along predefined DFT trajectories. While this approach does not provide an absolute guarantee of the global MEP topology, it provides a consistent and physically sound explanation for the experimental observations.

This study differentiates between the stoichiometric and oxidized surfaces regarding the reaction mechanism. On the stoichiometric surface, the process is dominated by proton coupled electron transfer (PCET), in which the electron and proton are both transferred in a coupled yet spatially distinct manner. In contrast, the oxidized surface enables direct hydrogen atom transfer (HAT) in the first step, which is facilitated by pre-activated O$_{Ti}$ radical species. This concerted homolytic cleavage is fundamentally more efficient than the spatially separated charge transfer that occurs on the reduced surface.

The significant wavelength dependence (sevenfold yield at 257~nm vs. 343~nm) was successfully explained by the access to higher electronic manifolds ($S_2$,$T_3$). This investigation identified the reaction as a highly efficient single photon process. One single 257~nm photon provides sufficient energy for the system to navigate through excited state manifolds, effectively bypassing the rate determining barriers of the ground state. This provides strong theoretical evidence for the "hot hole" theory discussed in literature.\supercite{Xu2021, Lai2023}

A key refinement of the mechanistic model is the deactivation into the product state. Although non-adiabatic funnels were identified as potential exit channels, the experimental evidence of low reaction exothermicity suggests that the system likely persists on the excited $S_1$/$T_2$ manifold until the styrene-surface complex is fully formed. Therefore, the final relaxation to the $S_0$ ground state is proposed to occur via radiative relaxation rather than early non-radiative tunneling. This prevents the system from falling into the deep, thermodynamic sink predicted by the ground state description, thus aligning the theoretical model with experimental energetics.

The fourfold increase in yield on the oxidized surface is attributed to electronic sensitization. The O$_{Ti}$ species lowers the excitation threshold and acts as a "hydrogen scavenger," which prevents recombinative back reactions and drives the reaction toward completion under milder conditions.

The insights gained in this work open several paths for future research to further optimize the catalytic process. While literature assumes that the dehydrogenation mechanism is identical on anatase and rutile,\supercite{Lin2020} an explicit quantum chemical comparison is still missing. Investigating the anatase phase or other facets is crucial to verifying whether the adsorption characteristics and the identified PCET/HAT pathways are indeed a general feature of \ch{TiO2} catalysts or are specific to the rutile(110) facet.

Another promising approach is to introduce transition metal dopants or clusters on the \ch{TiO2} surface. These could facilitate the initial C-H bond cleavage and more importantly, promote the recombinative desorption of hydrogen as \ch{H2}. This would prevent the formation of water as a byproduct, which can lead to surface degradation and catalyst deactivation.

Future studies could also investigate the impact of heterojunctions (e.g., \ch{TiO2})/\ch{VO_x} or /\ch{CeO2}.\supercite{Chen2022} Strategic charge separation could collect photogenerated holes and protons in different areas of the catalyst. This could potentially influence selectivity and improve overall efficiency.

\newpage
\thispagestyle{plain}
\clearpage
\phantomsection
\addcontentsline{toc}{section}{References}
\printbibliography

@article{Chen2019,
author = {Chen, L. and Zhang, S. and Persaud, R. R. and Smith, R. S. and Kay, B. D. and Dixon, D. and Dohnálek, Z.},
%title = {{Understanding the Binding of Aromatic Hydrocarbons on Rutile TiO$_{2}$(110)}},
journaltitle = {J. Phys. Chem. C},
volume = {123},
pages = {16766-16777},
year = {2019},
doi = {10.1021/acs.jpcc.9b03355}}

@book{Jensen,
author = {Jensen, F.},
title = {{Introduction to Computational Chemistry}},
publisher = {WILEY},
year = {2017},
location = {Chichester, UK},
edition={3rd}}

@book{Gerhards,
author = {Gerhards, L.},
title = {{Quantum Chemical Investigation of the Photocatalytic Sulfoxidation of Hydrocarbons – A Potential New Reaction for Industry}},
location = {Oldenburg},
year = {2022},
publisher = {Dr.~Dissertation}}

@article{Gerhards2021,
author = {Gerhards, L. and Klüner, T.},
%title = {{Quantum Chemical Investigation of Photocatalytical Sulfoxidation of Hydrocarbons on TiO$_{2}$}},
journaltitle = {J. Phys. Chem. C},
volume = {125},
pages = {13313-13323},
year = {2021},
doi = {https://doi.org/10.1021/acs.jpcc.1c03377}}

@article{Gerhards2022,
author = {Gerhards, L. and Klüner, T.},
%title = {{Theoretical investigation of CH-bond activation by photocatalytic excited SO$_{2}$ and the effects of C-, N-, S-, and Se-doped TiO$_{2}$}},
journaltitle = {Phys. Chem. Chem. Phys.},
volume = {24},
pages = {2051-2069},
year = {2022},
doi = {10.1039/d1cp04335h}}

@article{Lai2022,
author = {Lai, Y. and Pu, Z. and Liu, P. and Li, F. and Zeng, J. and Yang, X. and Guo, Q.},
%title = {{Low-Temperature C-H Bond Activation: Ethylbenzene-to-Styrene Conversion on Rutile TiO$_{2}$(110)}},
journaltitle = {J. Phys. Chem. C},
volume = {126},
pages = {6231-6240},
year = {2022},
doi = {https://doi.org/10.1021/acs.jpcc.2c00244}}

@article{Lin2020,
author = {Lin, H. and Wang, Z. and Wang, H. and Gao, J. and Ding, H. and Xu, Y. and Li, Q. and Guo, Q. and Ma, Z. and Yang, X. and Pan, M.},
%title = {{In Situ Observation of Stepwise C-H Bond Scission: Deciphering the Catalytic Selectivity of Ethylbenzene-to-Styrene Conversionon TiO$_{2}$}},
journaltitle = {J. Phys. Chem. Lett.},
volume = {11},
pages = {9850-9855},
year = {2020},
doi = {https://dx.doi.org/10.1021/acs.jpclett.0c02729}}

@article{Li2022_Lett,
author = {Li, F. and Chen, X. and Lai, Y. and Wang, T. and Yang, X. and Guo, Q.},
%title = {{Low-Temperature C-H Bond Activation via Photocatalysis: Highly Efficient Ethylbenzene Dehydrogenation into Styrene on Rutile TiO$_{2}$(110)}},
journaltitle = {J. Phys. Chem. Lett.},
volume = {13},
pages = {9186-9194},
year = {2022},
doi = {https://doi.org/10.1021/acs.jpclett.2c02269}}

@article{Lai2023,
author = {Lai, Y. and Zeng, Y. and Li, F. and Chen, X. and Wang, T. and Yang, X. and Guo, Q.},
%title = {{Low-Temperature Ethylbenzene Conversion on Rutile TiO$_{2}$(100) via Photocatalysis: The Strong Photon Energy Dependence}},
journaltitle = {J. Phys. Chem. Lett.},
volume = {14},
pages = {6286-6294},
year = {2023},
doi = {https://doi.org/10.1021/acs.jpclett.3c01491}}

@book{Fan,
author = {Fan, H.-X. and Rajendran, A. and Li, W.-Y.},
title = {{Industrial Arene Chemistry: Markets, Technologies, Sustainable Processes and Cases Studies of Aromatic Commodities}},
publisher = {Chapter 42, WILEY-VCH},
year = {2023},
location = {Weinheim, Germany},
edition={1st}}

@article{Soares2025,
author = {Soares, J. de O. and Barbosa, F. F. and Braga, T. P.},
%title = {{Oxidative Dehydrogenation of Ethylbenzene over Iron-Based Catalysts: Challenges, Prospects, and Future Trends}},
journaltitle = {ChemCatChem},
volume = {17},
pages = {e00224},
year = {2025},
doi = {https://doi.org/10.1002/cctc.202500224}}

@article{Li2022_JACS,
author = {Li, F. and Wang, B. and Chen, X. and Lai, Y. and Wang, T. and Fan, H. and Yang, X. and Guo, Q.},
%title = {{Photocatalytic Oxidative Dehydrogenation of Propane for Selective Propene Production with TiO$_2$}},
journaltitle = {JACS Au},
volume = {2},
pages = {2607-2616},
year = {2022},
doi = {https://doi.org/10.1021/jacsau.2c00512}}

@article{Sanz2015,
author = {Sanz, S. G. and McMillan, L. and McGregor, J. and Zeitler, J. A. and Al-Yassir, N. and Al-Khattaf, A. and Gladden, L. F.},
%title = {{A new perspective on catalytic dehydrogenation of ethylbenzene: the influence of side-reactions on catalytic performance}},
journaltitle = {Catal. Sci. Technol.},
volume = {5},
pages = {3782-3797},
year = {2015},
doi = {10.1039/c5cy00457h}}

@article{Wang2022,
author = {Wang, X. and Wan, L. and Wang, Z. and Liu, X. and Gao, Y. and Wang, L. and Liu, J. and Guo, Q. and Hu, W. and Yang, J.},
%title = {{Identifying Photocatalytic Active Sites of C$_2$H$_6$ C-H Bond Activation on TiO$_2$ via Combining First-Principles Ground-State and Excited State Electronic Structure Calculations}},
journaltitle = {J. Phys. Chem.  Lett.},
volume = {13},
pages = {6532-6540},
year = {2022},
doi = {https://doi.org/10.1021/acs.jpclett.2c01100}}

@article{Xu2021,
author = {Xu, C. and Xu, F. and Chen, X. and Li, Z. and Luan, Z. and Wang, Y. and Guo, Q. and Yang, X.},
%title = {{Wavelength-Dependent Water Oxidation on Rutile TiO$_2$(110)}},
journaltitle = {J. Phys. Chem.  Lett.},
volume = {12},
pages = {1066-1072},
year = {2021},
doi = {https://dx.doi.org/10.1021/acs.jpclett.0c03726}}

@article{Zhou2022,
author = {Zhou, M. and Wang, H.},
%title = {{Optimally Selecting Photo- and Electrocatalysis to Facilitate CH$_4$ Activation on TiO$_2$(110) Surface: Localized Photoexcitation versus Global Electric-Field Polarization}},
journaltitle = {JACS Au},
volume = {2},
pages = {188-196},
year = {2022},
doi = {https://doi.org/10.1021/jacsau.1c00466}}

@article{Amtout1995,
author = {Amtout, A. and Leonelli, R.},
%title = {{Optical properties of rutile near its fundamental band gap}},
journaltitle = {Phys. Rev. B},
volume = {51},
pages = {6842-6851},
year = {1995},
doi = {https://doi.org/10.1103/PhysRevB.51.6842}}

@article{Schneider2014,
author = {Schneider, J. and Matsuoka, M. and Takeuchi, M. and Zhang, J. and Horiuchi, Y. and Anpo, M. and Bahnemann, D.},
%title = {{Understanding TiO$_2$ Photocatalysis: Mechanisms and Materials}},
journaltitle = {Chem. Rev.},
volume = {114},
pages = {9919-9986},
year = {2014},
doi = {https://doi.org/10.1021/cr5001892}}

@article{Tang1995,
author = {Tang, H. and Lévy, F. and Berger, H. and Schmid, P. E.},
%title = {{Urbach tail of anatase TiO$_2$}},
journaltitle = {Phys. Rev. B},
volume = {52},
pages = {7771-7774},
year = {1995},
doi = {https://doi.org/10.1103/PhysRevB.52.7771}}

@article{Zhang2006,
author = {Zhang, J. and Li, M. and Feng, Z. and Chen, J. and Li, C.},
%title = {{UV Raman Spectroscopic Study on TiO$_2$. I. Phase Transformation at the Surface and in the Bulk}},
journaltitle = {J. Phys. Chem. B},
volume = {110},
pages = {927-935},
year = {2006},
doi = {10.1021/jp0552473}}

@article{Neese2012,
author = {Neese, F.},
%title = {The ORCA program system},
journal = {Wiley Interdiscip. Rev. Comput. Mol. Sci.},
volume = {2},
pages = {73-78},
DOI = {10.1002/wcms.81},
year = {2012},}

@article{Neese2025,
author = {Neese, F.},
%title = {Software Update: The ORCA Program System—Version 6.0},
journal = {Wiley Interdiscip. Rev. Comput. Mol. Sci.},
volume = {15},
pages = {e70019},
doi = {10.1002/wcms.70019},
year = {2025}}

@article{Kresse1_1996,
author = {Kresse, G. and Furthmüller, J.},
%title = {{Efficiency of ab-initio total energy calculations formetals and semiconductors using a plane-wave basis set}},
journaltitle = {Comput.  Mat. Sci.},
volume = {6},
pages = {15–50},
year = {1996},
doi = {https://doi.org/10.1016/0927-0256(96)00008-0}}

@article{Kresse2_1996,
author = {Kresse, G. and Furthmüller, J.},
%title = {{Efficient iterative schemes for ab initio total-energy calculations using a plane-wave basis set}},
journaltitle = {Phys. Rev. B - Condens. Matter Mater. Phys.},
volume = {54},
pages = {11169–11186},
year = {1996},
doi = {https://doi.org/10.1103/PhysRevB.54.11169}}

@article{Low2017,
author = {Low, J. and Yu, J. and Jaroniec, M. and Wageh, S. and Al-Ghamdi, A. A.},
%title = {{Heterojunction Photocatalysts}},
journaltitle = {Adv. Mater.},
volume = {29},
pages = {1601694},
year = {2017},
doi = {10.1002/adma.201601694}}

@book{Koch,
author = {Koch, W. and Holthausen, M. C.},
title = {{A Chemist’s Guide to Density Functional Theory}},
publisher = {Wiley-VCH},
year = {2001},
location = {Weinheim},
edition={2nd}}

@article{Born1927,
author = {Born, M. and Oppenheimer, R.},
%title = {{Zur Quantentheorie der Molekeln}},
journaltitle = {Ann. Phys.},
volume = {20},
pages = {457-484},
year = {1927},
doi = {https://doi.org/10.1002/andp.19273892002}}

@book{Parr,
author = {Parr, R. G. and Yang, W.},
title = {{Density-Functional Theory of Atoms and Molecules}},
publisher = {Oxford University Press},
year = {1989},
location = {Oxford}}

@misc{Awoonor2018,
author = {Awoonor-Williams, E.},
howpublished = {\\https://awoonor.github.io/DFT/},
title = {{Brief Overview on Density Functional Theory}},
note = {Accessed: 2025-09-12},
year = {2018}}

@article{Perdew2001,
author = {Perdew, J. P. and Schmidt, K.},
%title = {{Jacob’s ladder of density functional approximations for the exchange-correlation energy}},
journaltitle = {AIP Conf. Proc.},
volume = {577},
pages = {1-20},
year = {2001},
doi = {https://doi.org/10.1063/1.1390175}}

@article{PBE1996,
author = {Perdew, J. P. and Burke, K. and Ernzerhof, M.},
%title = {{Generalized gradient approximation made simple}},
journaltitle = {Phys. Rev. Lett.},
volume = {77},
pages = {3865-3868},
year = {1996},
doi = {https://doi.org/10.1103/PhysRevLett.77.3865}}

@book{Cramer,
author = {Cramer, C. J.},
title = {{Essentials of computational chemistry: theories and models}},
publisher = {{John Wiley \& Sons}},
year = {2013},
location = {Chichester},
edition={2nd}}

@article{Peintinger2013,
author = {Peintinger, M. F. and Oliveira, D. V. and Bredow, T.},
%title = {{Consistent Gaussian basis sets of triple‐zeta valence with polarization quality for solid‐state calculations}},
journaltitle = {J. Comput. Chem.},
volume = {34},
pages = {451-459},
year = {2013},
doi = {10.1002/jcc.23153}}

@article{Hohenberg1964,
author = {Hohenberg, P. and Kohn, W.},
%title = {{Inhomogeneous electron gas}},
journaltitle = {Phys. Rev.},
volume = {136},
pages = {B864},
year = {1964},
doi = {https://doi.org/10.1103/PhysRev.136.B864}}

@article{Perdew1996,
author = {Perdew, J. P. and Ernzerhof, M. and Burke, K.},
%title = {{Rationale for mixing exact exchange with density functional approximations}},
journaltitle = {J. Chem. Phys.},
volume = {105},
pages = {9982-9985},
year = {1996},
doi = {https://doi.org/10.1063/1.472933}}

@article{Silverstein2010,
author = {Silverstein, D. W. and Jensen, L.},
%title = {{Assessment of the accuracy of long-range corrected functionals for describing the electronic and optical properties of silver clusters}},
journaltitle = {J. Chem. Phys.},
volume = {132},
pages = {194302},
year = {2010},
doi = {10.1063/1.3429883}}

@article{Roos1980,
author = {Roos, B. O. and Taylor, P. R. and Sigbahn, P. E.},
%title = {{A complete active space SCF method (CASSCF) using a density matrix formulated super-CI approach}},
journaltitle = {Chem. Phys.},
volume = {48},
pages = {157-173},
year = {1980},
doi = {https://doi.org/10.1016/0301-0104(80)80045-0}}

@article{Grimme2010,
author = {Grimme, S. and Antony, J. and Ehrlich, S. and Krieg, H.},
%title = {{A consistent and accurate ab initio parametrization of density functional dispersion correction (DFT-D) for the 94 elements H-Pu}},
journaltitle = {J. Chem. Phys.},
volume = {132},
pages = {154104},
year = {2010},
doi = {10.1063/1.3382344}}

@book{Jónsson,
author = {Jónsson, H. and Mills, G. and Jacobsen, K. W.},
title = {{Classical And Quantum Dynamics In Condensed Phase Simulations: Nudged elastic band method for finding minimum energy paths of transitions}},
publisher = {World Scientific},
year = {1998},
pages = {385-404}}

@article{Angeli2001_1,
author = {Angeli, C. and Cimiraglia, R. and Evangelisti, S. and Leininger, T. and Malrieu, J.-P.},
%title = {{Introduction of n-electron valence states for multireference perturbation theory}},
journaltitle = {J. Chem. Phys.},
volume = {114},
pages = {10252-10264},
year = {2001},
doi = {https://doi.org/10.1063/1.1361246}}

@article{Angeli2001_2,
author = {Angeli, C. and Cimiraglia, R. and Malrieu, J.-P.},
%title = {{N-electon valence state pertubation theory: a fast implementation of the strongly contracted variant}},
journaltitle = {Chem. Phys. Lett.},
volume = {350},
pages = {297-305},
year = {2001},
doi = {https://doi.org/10.1016/S0009-2614(01)01303-3}}

@article{Angeli2002,
author = {Angeli, C. and Cimiraglia, R. and Malrieu, J.-P.},
%title = {{ n-electron valence state perturbation theory: A spinless formulation and an efficient implementation of the strongly contracted and of the partially contracted variants}},
journaltitle = {J. Chem. Phys.},
volume = {117},
pages = {9138–9153},
year = {2002},
doi = {https://doi.org/10.1063/1.1515317}}

@article{Bloch1929,
author = {Bloch, F.},
%title = {{Über die Quantenmechanik der Elektronen in Kristallgittern}},
journaltitle = {Z. Phys.},
volume = {52},
pages = {555-600},
year = {1929},
doi = {https://doi.org/10.1007/BF01339455}}

@article{Casarin2005,
author = {Casarin, M. and Ferrigato, F. and Maccato, C. and Vittadini, A.},
%title = {{SO$_2$ on TiO$_2$ (110) and Ti$_2$O$_3$ (1012) nonpolar surfaces: A DFT study}},
journaltitle = {J. Phys. Chem. B},
volume = {109},
pages = {12596-12602},
year = {2005},
doi = {10.1021/jp050314e}}

@article{Casarin1998,
author = {Casarin, M. and Maccato, C. and Vittadini, A.},
%title = {{Molecular chemisorption on TiO$_2$ (110): A local point of view}},
journaltitle = {J. Phys. Chem. B},
volume = {102},
pages = {10745-10752},
year = {1998},
doi = {https://doi.org/10.1021/jp981377i}}

@book{Staemmler,
author = {Staemmler, V.},
title = {{The cluster approach for the adsorption of small molecules on oxide surfaces}},
publisher = {Springer-Verlag},
year = {2005},
location = {Berlin Heidelberg},
pages={219-256}}

@article{He2010,
author = {He, H. and Zapol, P. and Curtiss, L. A.},
%title = {{A Theoretical Study of CO$_2$ Anions on Anatase (101) Surface}},
journaltitle = {J. Phys. Chem. C},
volume = {114},
pages = {21474-21481},
year = {2010},
doi = {https://doi.org/10.1021/jp106579b}}

@article{Teusch2020,
author = {Teusch, T. and Klüner, T.},
%title = {{Photodesorption mechanism of water on WO$_3$(001) – a combined embedded cluster, computational intelligence and wave packet approach}},
journaltitle = {Phys. Chem. Chem. Phys.},
volume = {22},
pages = {19267-19274},
year = {2020},
doi = {10.1039/d0cp02809f}}

@article{Petersen2020_1,
author = {Petersen, T. and Klüner, T.},
%title = {{Photodesorption of H$_2$O from Anatase-TiO$_2$ (101): A Combined Quantum Chemical and Quantum Dynamical Study}},
journaltitle = {J. Phys. Chem. C},
volume = {124},
pages = {11444-11455},
year = {2020},
doi = {https://doi.org/10.1021/acs.jpcc.0c01926}}

@article{Petersen2020_2,
author = {Petersen, T. and Klüner, T.},
%title = {{Water adsorption on ideal anatase-TiO$_2$ (101)–An embedded cluster model for accurate adsorption energetics and excited state properties}},
journaltitle = {Z. Phys. Chem.},
volume = {234},
pages = {813-834},
year = {2020},
doi = {https://doi.org/10.1515/zpch-2019-1425}}

@article{Mitschker2015,
author = {Mitschker, J. and Kluner, T.},
%title = {{Photodesorption of water from rutile(110): ab initio calculation of five-dimensional potential energy surfaces of ground and excited electronic states and wave packet studies}},
journaltitle = {Phys. Chem. Chem. Phys.},
volume = {17},
pages = {268-275},
year = {2015},
doi = {10.1039/c4cp04593a}}

@article{Boys1970,
author = {Boys, S. F. and Bernardi, F.},
%title = {{The calculation of small molecular interactions by the differences of separate total energies. Some procedures with reduced errors}},
journaltitle = {Mol. Phys.},
volume = {19},
pages = {553-566},
year = {1970},
doi = {https://doi.org/10.1080/00268977000101561}}

@article{Dononelli2021,
author = {Dononelli, W. and Klüner, T.},
%title = {{Analyzing the local basis set superposition error for CO adsorbed on rutile (110)}},
journaltitle = {Int. J. Quantum Chem.},
volume = {121},
pages = {e26428},
year = {2021},
doi = {https://doi.org/10.1002/qua.26428}}

@article{Kubas2016,
author = {Kubas, A. and Berger, D. and Oberhofer, H. and Maganas, D. and Reuter, K. and Neese, F.},
%title = {{Surface Adsorption Energetics Studied with “Gold Standard” Wave-Function-Based Ab Initio Methods: Small-Molecule Binding to TiO$_2$(110)}},
journaltitle = {J. Phys. Chem. Lett.},
volume = {7},
pages = {4207-4212},
year = {2016},
doi = {10.1021/acs.jpclett.6b01845}}

@article{Berger2014,
author = {Berger, D. and Logsdail, A. J. and Oberhofer, H. and Farrow, M. R. and Catlow, C. R. and Sherwood, P. and Sokol, A. A. and Blum, V. and Reuter, K.},
%title = {{Embedded-cluster calculations in a numeric atomic orbital density-functional theory framework}},
journaltitle = {J. Chem. Phys.},
volume = {141},
pages = {024105},
year = {2014},
doi = {10.1063/1.4885816}}

@article{Bredow1998,
author = {Bredow, T. and Aprà, E. and Catti, M. and Pacchioni, G.},
%title = {{Cluster and periodic ab-initio calculations on K/TiO$_2$ (110)}},
journaltitle = {Surf. Sci.},
volume = {418},
pages = {150-165},
year = {1998},
doi = {10.1016/S0039-6028(98)00712-2}}

@article{Giordano2001,
author = {Giordano, L. and Pacchioni, G. and Bredow, T. and Sanz, J. F.},
%title = {{Cu, Ag, and Au atoms adsorbed on TiO$_2$(110): cluster and periodic calculations}},
journaltitle = {Surf. Sci.},
volume = {471},
pages = {21-31},
year = {2001},
doi = {https://doi.org/10.1016/S0039-6028(00)00879-7}}

@article{Pacchioni1994,
author = {Pacchioni, G. and Ricart, J. M. and Illas, F.},
%title = {{Ab initio cluster model calculations on the chemisorption of CO$_2$ and SO$_2$ probe molecules on MgO and CaO (100) surfaces. A theoretical measure of oxide basicity}},
journaltitle = {J. Am. Chem. Soc.},
volume = {116},
pages = {10152-10158},
year = {1994},
doi = {https://doi.org/10.1021/ja00101a038}}

@article{Kick2019,
author = {Kick, M. and Oberhofer, H.},
%title = {{Towards a transferable design of solid-state embedding models on the example of a rutile TiO$_2$ (110) surface}},
journaltitle = {J. Chem. Phys.},
volume = {151},
pages = {184114},
year = {2019},
doi = {https://doi.org/10.1063/1.5125204}}

@article{Kohn1965,
author = {Kohn, W. and Sham, L. J.},
%title = {{Self-Consistent Equations Including Exchange and Correlation Effects}},
journaltitle = {Phys. Rev.},
volume = {140},
pages = {A1133–A1138},
year = {1965},
doi = {DOI: https://doi.org/10.1103/PhysRev.140.A1133}}

@article{Mulliken1955,
author = {Mulliken, R. S.},
%title = {{Electronic population analysis on LCAO–MO molecular wave functions. I}},
journaltitle = {J. Chem. Phys.},
volume = {23},
pages = {1833-1840},
year = {1955},
doi = {https://doi.org/10.1063/1.1740588}}

@article{Guo2021,
author = {Guo, Y. and Sivalingam, K. and Neese, F.},
%title = {{Approximations of density matrices in N-electron valence state second-order perturbation theory (NEVPT2). I. Revisiting the NEVPT2 construction}},
journaltitle = {J. Chem. Phys.},
volume = {154},
pages = {214111},
year = {2021},
doi = {https://doi.org/10.1063/5.0051211}}

@article{Ásgeirsson2021,
author = {Ásgeirsson, V. and Birgisson, B. O. and Bjornsson, R. and Becker, U. and Nesse, F. and Riplinger, C. and Jónsson, H.},
%title = {{Nudged Elastic Band Method for Molecular Reactions Using Energy Weighted Springs Combined wit hEigenvector Following}},
journaltitle = {J. Chem. Theory Comput.},
volume = {17},
pages = {4929-4945},
year = {2021},
doi = {https://doi.org/10.1021/acs.jctc.1c00462}}

@article{Henderson2011,
author = {Henderson, M. A.},
%title = {{A surface science perspective on TiO$_2$ photocatalysis}},
journaltitle = {Surf. Sci. Rep.},
volume = {66},
pages = {185-297},
year = {2011},
doi = {https://doi.org/10.1016/j.surfrep.2011.01.001}}

@article{Eichkorn1995,
author = {Eichkorn, K. and Treutler, O. and Öhm, H. and Häser, M. and Ahlrichs, R.},
%title = {{Auxiliary basis sets to approximate Coulomb potentials}},
journaltitle = {Chem. Phys. Lett.},
volume = {240},
pages = {283-290},
year = {1995},
doi = {https://doi.org/10.1016/0009-2614(95)00621-A}}

@article{Neese2003,
author = {Neese, F.},
%title = {{An Improvement of the Resolution of the Identity Approximation for the Formation of the Coulomb Matrix}},
journaltitle = {J. Comput. Chem.},
volume = {24},
pages = {1740-1747},
year = {2003},
doi = {https://doi.org/10.1002/jcc.10318}}

@article{Weigend2002,
author = {Weigend, F. and Köhn, A. and Hättig, C.},
%title = {{Efficient use of the correlation consistent basis sets in resolution of the identity MP2 calculations}},
journaltitle = {J. Chem. Phys.},
volume = {116},
pages = {3175-3183},
year = {2002},
doi = {https://doi.org/10.1063/1.1445115}}

@article{Weigend2005,
author = {Weigend, F. and Ahlrichs, R.},
%title = {{Balanced basis sets of split valence, triple zeta valence and quadruple zeta valence quality for H to Rn: Design and assessment of accuracy \textdagger}},
journaltitle = {Phys. Chem. Chem. Phys.},
volume = {7},
pages = {3297-3305},
year = {2005},
doi = {http://dx.doi.org/10.1039/b508541a}}

@article{Hellweg2007,
author = {Hellweg, A. and Hättig, C. and Höfer, S. and Klopper, W.},
%title = {{Optimized accurate auxiliary basis sets for RI-MP2 and RI-CC2 calculations for the atoms Rb to Rn}},
journaltitle = {Theor. Chem. Acc.},
volume = {117},
pages = {587-597},
year = {2007},
doi = {10.1007/s00214-007-0250-5}}

@article{Redhead1962,
author = {Redhead, P. A.},
%title = {{Thermal Desorption of Gases}},
journaltitle = {Vacuum},
volume = {12},
pages = {203-211},
year = {1962},
doi = {https://doi.org/10.1016/0042-207X(62)90978-8}}

@article{Cohen2008,
author = {Cohen, A. J. and Mori-Sánchez, P. and Yang, W.},
%title = {{Insights into Current Limitations of Density Functional Theory}},
journaltitle = {Science},
volume = {321},
pages = {792-794},
year = {2008},
doi = {10.1126/science.1158722}}

@article{Valentin2006,
author = {Di Valentin, C. and Pacchioni, G. and Selloni, A.},
%title = {{Electronic Structure of Defect States in Hydroxylated and Reduced Rutile \che{TiO2}(110) Surfaces}},
journaltitle = {Phys. Rev. Lett.},
volume = {97},
pages = {166803},
year = {2006},
doi = {https://doi.org/10.1103/PhysRevLett.97.166803}}

@article{Sun2016,
author = {Sun, Q. and Chan, G. K.-L.},
%title = {{Quantum Embedding Theories}},
journaltitle = {Acc. Chem. Res.},
volume = {49},
pages = {2705-2712},
year = {2016},
doi = {10.1021/acs.accounts.6b00356}}

@article{Mentel2014,
author = {Mentel, Ł. M. and Baerends, E. J.},
%title = {{Can the Counterpoise Correction for Basis Set Superposition Effect Be Justified?}},
journaltitle = {J. Chem. Theory Comput.},
volume = {10},
pages = {252-267},
year = {2014},
doi = {https://doi.org/10.1021/ct400990u}}

@article{Mori2008,
author = {Mori-Sánchez, P. and Cohen, A. J. and Yang, W.},
%title = {{Localization and Delocalization Errors in Density Functional Theory and Implications for Band-Gap Prediction}},
journaltitle = {Phys. Rev. Lett.},
volume = {100},
pages = {146401},
year = {2008},
doi = {10.1103/PhysRevLett.100.146401}}

@article{Kruse2012,
author = {Kruse, H. and Grimme, S.},
%title = {{A geometrical correction for the inter- and intra-molecular basis set superposition error in Hartree-Fock and density functional theory calculations for large systems}},
journaltitle = {J. Chem. Phys.},
volume = {136},
pages = {154101},
year = {2012},
doi = {https://doi.org/10.1063/1.3700154}}

@article{Linsebigler1995,
author = {Linsebigler, A. L. and Lu, G. and Yates Jr., J. T. },
%title = {{Photocatalysis on \ch{Ti02} Surfaces: Principles, Mechanisms, and Selected Results}},
journaltitle = {Chem. Rev.},
volume = {95},
pages = {735-758},
year = {1995},
doi = {https://doi.org/10.1021/cr00035a013}}

@article{Liao2013,
author = {Liao, P. and Carter, E. A.},
%title = {{New concepts and modeling strategies to design and evaluate photo-electro-catalysts based on transition metal oxides†}},
journaltitle = {Chem. Soc. Rev.},
volume = {42},
pages = {2401-2422},
year = {2013},
doi = {10.1039/c2cs35267b}}

@article{Donahue2001,
author = {Donahue, N. M.},
%title = {{Revisiting the Hammond Postulate: The Role of Reactant and Product Ionic States in Regulating Barrier Heights, Locations, and Transition State Frequencies †}},
journaltitle = {J. Phys. Chem. A},
volume = {105},
pages = {1489-1497},
year = {2001},
doi = {https://doi.org/10.1021/jp001004t}}

@article{Andersson1992 ,
author = {Andersson, K. and Malmqvist, P. and Roos, B. O.},
%title = {{Second‐order perturbation theory with a complete active space self‐consistent field reference function}},
journaltitle = {J. Chem. Phys.},
volume = {96},
pages = {1218-1226},
year = {1992},
doi = {https://doi.org/10.1063/1.462209}}

@article{Roos1995,
author = {Roos, B. O. and Anderson, K.},
%title = {{Multiconfigurational perturbation theory with level shift - the Cr$_2$ potential revisited}},
journaltitle = {Chem. Phys. Lett.},
volume = {245},
pages = {215-223},
year = {1995},
doi = {https://doi.org/10.1016/0009-2614(95)01010-7}}

@article{SerranoAndres2005,
author = {Serrano-Andrés, L. and Merchán, M.},
%title = {{Quantum chemistry of the excited state: 2005 overview}},
journaltitle = {THEOCHEM},
volume = {729},
pages = {99-108},
year = {2005},
doi = {https://doi.org/10.1016/j.theochem.2005.03.020}}

@inbook{Roos1987,
author = {Roos, B. O.},
title = {{The Complete Active Space Self-Consistent Field Method and its Applications in Electronic Structure Calculations}},
publisher = {John Wiley \& Sons, Ltd.},
booktitle = {Advances in Chemical Physics},
pages = {399-445},
year = {1987},
doi = {https://doi.org/10.1002/9780470142943.ch7}}

@article{Li2008,
author = {Li, S.-C. and Zhang, Z. and Sheppard, D. and Kay, B. D. and White, J. M. and Du, Y. and Lyubinetsky, I. and Henkelman, G. and Dohnálek, Z.},
%title = {{Intrinsic Diffusion of Hydrogen on Rutile \ch{TiO2}(110)}},
journaltitle = {J. Am. Chem. Soc.},
volume = {130},
pages = {980-988},
year = {2008},
doi = {10.1021/ja8012825}}

@article{Yim2010,
author = {Yim, C. M. and Pang, C. L. and Thornton, G.},
%title = {{Oxygen Vacancy Origin of the Surface Band-Gap State of \ch{TiO2}(110)}},
journaltitle = {Phys. Rev. Lett.},
volume = {104},
pages = {036806},
year = {2010},
doi = {https://doi.org/10.1103/PhysRevLett.104.036806}}

@article{Ryabinkin2017,
author = {Ryabinkin, I. G. and Joubert-Doriol, L. and Izmaylov, A. F.},
%title = {{Geometric Phase Effects in Nonadiabatic Dynamics near Conical Intersections}},
journaltitle = {Acc. Chem. Res.},
volume = {50},
pages = {1785-1793},
year = {2017},
doi = {10.1021/acs.accounts.7b00220}}

@article{Malhado2014,
author = {Malhado, J. P. and Bearpark, M. J. and Hynes, J. T.},
%title = {{Non-adiabatic dynamics close to conical intersections and the surface hopping perspective}},
journaltitle = {Front. Chem.},
volume = {2},
pages = {1-21},
year = {2014},
doi = {https://doi.org/10.3389/fchem.2014.00097}}

@article{Bernardi1996,
author = {Bernardi, F. and Olivucci, M. and Robb, M. A.},
%title = {{Potential Energy Surface Crossings in Organic Photochemistry}},
journaltitle = {Chem. Soc. Rev.},
volume = {25},
pages = {321-328},
year = {1996},
doi = {https://doi.org/10.1039/CS9962500321}}

@article{Zhu2016,
author = {Zhu, X. and Yarkony, D. R.},
%title = {{Non-adiabaticity: the importance of conical intersections}},
journaltitle = {Mol. Phys.},
volume = {114},
pages = {1983-2013},
year = {2016},
doi = {https://doi.org/10.1080/00268976.2016.1170218}}

@article{Pabisiak2007,
author = {Pabisiak, T. and Kiejna, A.},
%title = {{Energetics of oxygen vacancies at rutile \ch{TiO2}(110) surface}},
journaltitle = {Solid State Commun.},
volume = {144},
pages = {324-328},
year = {2007},
doi = {10.1016/j.ssc.2007.08.043}}

@article{Yarkony1996,
author = {Yarkony, D. R.},
%title = {{Diabolical conical intersections}},
journaltitle = {Rev. Mod. Phys.},
volume = {68},
pages = {985-1013},
year = {1996},
doi = {https://doi.org/10.1103/RevModPhys.68.985}}

@article{Xu1999,
author = {Xu, X. and Nakatsuji, H. and Lu, X. and Ehara, M. and Cai, Y. and Wang, N. Q. and Zhang, Q. E.},
%title = {{Cluster modeling of metal oxides: case study of MgO and the CO/MgO adsorption system}},
journaltitle = {Theor. Chem. Acc.},
volume = {102},
pages = {170-179},
year = {1999},
doi = {10.1007/s00214980m140}}

@inbook{Seijo1999,
author = {Seijo, L. and Barandiarán, Z.},
title = {{The \textit{Ab Initio} Model Potential Method: A Common Strategy for Effective Core Potential and Embedded Cluster Calculations}},
publisher = {World Scientific},
booktitle = {Computational Chemistry: Reviews of Current Trends},
pages = {55-152},
year = {1999},
doi = {https://doi.org/10.1142/9789812815156_0002}}

@article{Sauer1989,
author = {Sauer, J.},
%title = {{Molecular Models in ab Initio Studies of Solids and Surfaces: From Ionic Crystals and Semiconductors to Catalysts}},
journaltitle = {Chem. Rev.},
volume = {89},
pages = {199-255},
year = {1989},
doi = {https://doi.org/10.1021/cr00091a006}}

@article{Chen2022,
author = {Chen, Y. and Dang, D. and Yan, B. and Cheng, Y.},
%title = {{Mixed Metal Oxides of M1 MoVNbTeOx and TiO2 as Composite Catalyst for Oxidative Dehydrogenation of Ethane}},
journaltitle = {Catalysts},
volume = {12},
pages = {71},
year = {2022},
doi = {https://doi.org/10.3390/catal12010071}}

\newpage
\pagestyle{plain}
\clearpage
\phantomsection
\addcontentsline{toc}{section}{List of Abbreviations}
\paragraph*{\Large{List of Abbreviations}}\mbox{}\\

5f-Ti  - Five-fold coordinated Titanium\\
6f-Ti  - Six-fold coordinated Titanium\\
ABS    - Acrylonitrile Butadiene Styrene \\
AO     - Atomic Orbitals \\
ASCT   - Adsorbate-to-Surface Charge Transfer \\
BR     - Boundary Region \\
BSIE   - Basis Set Incompleteness Error \\
BSSE   - Basis Set Superposition Error \\
CASPT2 - Complete Active Space Peturbation Theory \\
CASSCF - Complete Active Space–Self-Consistent Field \\
CB     - Conduction Band  \\
CI     - Configuration Interaction \\
CSF    - Configuration State Functions \\
CT     - Charge Transfer \\
D3     - Grimme's D3 Method (Dispersion Correction) \\
def2-TZVP - def2 Basis Set of Triple Zeta Valence Polarized Quality \\
def2-TZVP/C - def2-TZVP /Correlation-Consistent Auxiliary Basis Set \\
DFT    - Density Functional Theory \\
$\Delta E_{prod}$/$\Delta E_{int}$ - Energy Difference of Adsorbate and Product/Intermediate ($T1$ States) \\
$E_{ads}$/$E_{des}$ - Adsorption/Desorption Energy \\
$E_{A}$ - Activation Barrier (Thermal and possibly Photochemical Simultaneously) \\
$E_{\alpha}$/$E_{\beta}$ - Activation Barrier (Photochemical, $\alpha$/$\beta$ Pathway) \\
EB     - Ethylbenzene \\
ECP    - Effective Core Potential  \\
GGA    - Generalized Gradient Approximation \\
HAT    - Hydrogen Atom Transfer \\
HOMO   - Highest Occupied Molecular Orbital  \\
ISC    - Intersystem Crossing \\
LCAO   - Linear Combination of Atomic Orbitals \\
LDA    - Local Density Approximation \\
LUMO   - Lowest Unoccupied Molecular Orbital \\
MEP    - Minimum Energy Path \\
MO     - Molecular Orbital \\
NEB    - Nudged Elastic Band \\
NEB-CI - NEB with Climbing Image enhancement \\
NEVPT2 - N-Electron Valence State Perturbation Theory 2nd Order  \\
O$_{br}$ - Bridging Oxygen \\
O$_{Ti}$ - Pre-adsorbed / Atomic Oxygen bound to 5f-Ti center \\
Occ.   - Occupation (referring to Occupied Orbitals) \\
OPTTS  - Transition State Optimization using a Quasi-Newton Method (ORCA) \\
ORCA   - \textit{Ab initio} Quantum Chemistry Program Package \\
P25    - Commercial \ch{TiO2} Photocatalyst (typically 80\% Anatase / 20\% Rutile) \\
PBC    - Periodic Boundary Conditions  \\
PBE    - Perdew-Burke-Ernzerhof \\
PBE0   - Perdew-Burke-Ernzerhof with 25\% Hartree-Fock Exchange \\
PCET   - Proton-Coupled Electron Transfer \\
PC-NEVPT2 - Partially Contracted NEVPT2 \\
PCF    - Point Charge Field \\
PES    - Potential Energy Surface \\
POB-TZVP - Peintinger-Oliveira-Bredow Triple Zeta Valence Polarized Basis Set \\
PS     - Polystyrene \\
RI     - Resolution of Identity \\
RKS    - Restricted Kohn-Sham \\
RMSE   - Root Mean Square Error \\
SA-CASSCF - State-Averaged CASSCF \\
SBR    - Styrene-Butadiene Rubber \\
SC-NEVPT2 - Strongly Contracted NEVPT2 \\
SCF    - Self-Consistent Field \\
SDD    - Stuttgart/Dresden Effective Core Potential \\
SOMO   - Singly Occupied Molecular Orbital \\
STM    - Scanning Tunneling Microscopy \\
TPD    - Temperature-Programmed Desorption spectroscopy \\
TS     - Transition State \\
UKS    - Unrestricted Kohn-Sham \\
VASP   - Vienna \textit{Ab initio} Simulation Package \\
VB     - Valence Band \\
vdW    - van der Waals interactions \\
ZPE    - Zero-Point Energy

\newpage
\thispagestyle{plain}
\clearpage
\phantomsection
\label{Appendix}
\addcontentsline{toc}{section}{Appendix: ORCA Input Example}
\paragraph*{Appendix: ORCA Input Example}
\begin{footnotesize}
General input for PBE-DFT-D3 calculations. General comments are given after a \#. \\
Oxidized surface and CASSCF specifics are commented after \#\# and \#\#\#, respectively.

! PBE UKS KDIIS SOSCF KEEPDENS defgrid3 noTrah TightSCF D3  \\
\# depending on the calculation following keywords are added: OPT, Freq, NEB-CI \\
\#\#\# ! CASSCF defgrid3 noTrah NormalSCF MORead SlowConv \\
! XYZFILE  \\
! NORMALPRINT \\
\%pal  \\
  nprocs 16 \\
end  \\
\%maxcore 15000 \\
\%freq \\
  QuasiRRHO false \\
end

\%basis \begin{multicols}{2} 
 NewGTO O  \\
   S 6  \\
     1 27032.382631 0.00021726302465  \\
     2 4052.3871392 0.00168386621990 \\
     3 922.32722710 0.00873956162650 \\
     4 261.24070989 0.03523996880800 \\
     5 85.354641351 0.11153519115000 \\
     6 31.035035245 0.25588953961000 \\
   S 2 \\
     1 12.260860728 0.39768730901000 \\
     2 4.9987076005 0.24627849430000 \\
   S 1 \\
     1 1.0987136000 1.00000000000000 \\
   S 1 \\
     1 0.3565870100 1.00000000000000 \\
   P 4 \\
     1 63.274954801 0.0060685103418 \\
     2 14.627049379 0.0419125758240 \\
     3 4.4501223456 0.1615384108800 \\
     4 1.5275799647 0.3570695131100 \\
   P 1 \\
     1 0.5489735000 1.0000000000000 \\
   P 1 \\
     1 0.1858671100 1.0000000000000 \\
   D 1 \\
     1 0.2534621300 1.0000000000000 \\
 end
 
 NEWGTO C \\
  S 6 \\
    1 13575.349682 0.00022245814352 \\
    2 2035.2333680 0.00172327382520 \\
    3 463.22562359 0.00892557153140 \\
    4 131.20019598 0.03572798450200 \\
    5 42.853015891 0.11076259931000 \\
    6 15.584185766 0.24295627626000 \\
  S 2 \\
    1 6.2067138508 0.41440263448000 \\
    2 2.5764896527 0.23744968655000 \\
  S 1 \\
    1 0.4941102000 1.00000000000000 \\
  S 1 \\
    1 0.1644071000 1.00000000000000 \\
  P 4 \\
    1 34.697232244 0.00533336578050 \\
    2 7.9582622826 0.03586410909200 \\
    3 2.3780826883 0.14215873329000 \\
    4 0.8143320818 0.34270471845000 \\
  P 1 \\
    1 0.5662417100 1.00000000000000 \\
  P 1 \\
    1 0.2673545000 1.00000000000000 \\
  D 1 \\
    1 0.8791584200 1.00000000000000 \\
end

 NewGTO Ti \\
   S 8 \\
     1 211575.690250 0.00023318151011 \\
     2 31714.9450580 0.00180796908510 \\
     3 7217.54765430 0.00939843113520 \\
     4 2042.93942470 0.03815685361800 \\
     5 665.128962080 0.12374757197000 \\
     6 238.749422640 0.29208551143000 \\
     7 92.5086910010 0.41226800855000 \\
     8 36.4039192090 0.21090534061000 \\
   S 4 \\
     1 232.726246070 -0.0249201407380 \\
     2 71.7912097110 -0.1174649008700 \\
     3 11.1585346150 0.56503342318000 \\
     4 4.65481354160 0.56211101812000 \\
   S 2 \\
     1 6.80346291740 -0.2301142550300 \\
     2 1.12010764030 0.72103186735000 \\
   S 1 \\
     1 1.74504592000 1.00000000000000 \\
   S 1 \\
     1 0.63463883000 1.00000000000000 \\
   S 1 \\
     1 0.21154627700 1.00000000000000 \\
   P 6 \\
     1 1063.14747320 0.00246908393200 \\
     2 251.565070610 0.01977334552300 \\
     3 80.4085548540 0.09098797667200 \\
     4 29.7681932690 0.25559900413000 \\
     5 11.7368305560 0.40489386764000 \\
     6 4.71423752300 0.23693402558000 \\
   P 3 \\
     1 17.7968037040 -0.0278786396150 \\
     2 2.42726986800 0.55672914668000 \\
     3 0.96823445537 1.00554473500000 \\
   P 1 \\
     1 1.73304386000 1.00000000000000 \\
   P 1 \\
     1 0.37056694000 1.00000000000000 \\
   D 4 \\
     1 37.7133847230 0.01151383509200 \\
     2 10.6929311840 0.06724634399600 \\
     3 3.67284469900 0.21484207775000 \\
     4 1.35885903030 0.38890892779000 \\
   D 1 \\
     1 0.86367514000 1.00000000000000 \\
   D 1 \\
     1 0.43183757000 1.00000000000000 \\
   F 1 \\
     1 0.56200000000 1.00000000000000 \\
 end

NewGTO H \\
  S 3 \\
    1 34.0613410000  0.00602519780 \\
    2  5.1235746000  0.04502109400 \\
    3  1.1646626000  0.20189726000 \\
  S 1 \\
    1 0.41574551000  1.00000000000 \\
  S 1 \\
    1 0.17951110000  1.00000000000 \\
  P 1 \\
    1 0.80000000000  1.00000000000 \\
end
 
\end{multicols}

\#\#\# AuxC "Def2-TZVP/C"

End

\#\#\# \%moinp "ads.gbw"

\#\#\# \%casscf\\
\#\#\#   nel 14\\
\#\#\#   norb 14\\
\#\#\#   mult 3,1\\
\#\#\#   nroots 3,3\\
\#\#\#   MaxIter 400\\
\#\#\#   TrafoStep RI\\
\#\#\#   shiftup 10 \# example\\
\#\#\#   shiftdn 10 \# example\\
\#\#\#   Orbstep DIIS\\
\#\#\#   Switchstep DIIS\\
\#\#\#   ActConstraints 1\\
\#\#\#   GTol 1e-2\\
\#\#\# end

\%coords \\
CTyp     xyz    \# type of coords (xyz or internal) \\
Charge  -68     \# total charge of system  \#\# -70 \\
Mult     1      \# multiplicity: 2$S$+1 \\
Units    Angs   \# unit of length: angs or bohrs \\
coords    \\
  O   -0.00002869993132     -0.00000501752815      0.04985018490068 \\
  O    0.00000003548573      6.46646201446929     -0.00000012337872 \\
  O    0.00000003584935     -6.46646201236914     -0.00000012412092 \\
  O    2.95157863965652     -0.00000662950963      0.05593983976143 \\
  O    2.95137305186511      6.46646192623315      0.00000006692748 \\
  O    2.95137305206080     -6.46646192545515      0.00000006758148 \\
  O    5.90274502112850     -0.00000000470918     -0.00000000480353 \\
  O    5.90274503013806      6.46646200517757      0.00000002920414 \\
  O    5.90274503042381     -6.46646200459092      0.00000002966005 \\
  O   -5.90274500412913     -0.00000001250567     -0.00000002977473 \\
  O   -5.90274501802972      6.46646205181564      0.00000001287015 \\
  O   -5.90274501777916     -6.46646205096723      0.00000001271490 \\
  O   -2.95152785939932     -0.00001016135223      0.05593611660748 \\
  O   -2.95137298977527      6.46646199688773     -0.00000001715061 \\
  O   -2.95137298925295     -6.46646199631268     -0.00000001744171 \\
  Ti   0.00000969482847      3.26163910094455     -1.47790960606893 \\
  Ti   0.00000801597440     -3.26164304090473     -1.47790696703827 \\
  Ti   2.95874007571960      3.19344883743790     -1.49407107878696 \\
  Ti   2.95873607898483     -3.19344539178197     -1.49406411480603 \\
  Ti  -2.95877115401205      3.19345287482095     -1.49409717308641 \\
  Ti  -2.95876727298748     -3.19344745455611     -1.49409101654292 \\
  Ti   1.48305284646558     -0.00000173814701     -1.03107199529118 \\
  Ti   1.47568604803774      6.46646202546514     -1.04868091732561 \\
  Ti   1.47568604864065     -6.46646202908155     -1.04868091657446 \\
  Ti   4.42705883494130      0.00000004180929     -1.04868079515141 \\
  Ti   4.42705877753022      6.46646209992871     -1.04868121753900 \\
  Ti   4.42705877611220     -6.46646210058013     -1.04868121925936 \\
  Ti  -4.42705909871805      0.00000002921465     -1.04868077863073 \\
  Ti  -4.42705889706200      6.46646191018565     -1.04868110246516 \\
  Ti  -4.42705889793604     -6.46646191182003     -1.04868110221003 \\
  Ti  -1.48304241058635     -0.00000036313585     -1.03110319008360 \\
  Ti  -1.47568619009492      6.46646200174657     -1.04868072505281 \\
  Ti  -1.47568619223373     -6.46646200607445     -1.04868072310909 \\
  O    1.47568601501814      4.44575710797494     -1.14384895385609 \\
  O    1.47568599892356     -8.48716698839037     -1.14384901682778 \\
  O    1.47340683303361     -2.01424810976738     -1.10246714438457 \\
  O    4.42705900266735      4.44575700964357     -1.14384886297115 \\
  O    4.42705902092908     -8.48716700284780     -1.14384898799025 \\
  O    4.42705904652100     -2.02070500907198     -1.14384920157314 \\
  O   -4.42705895630249      4.44575701907159     -1.14384889862863 \\
  O   -4.42705901060607     -8.48716699950732     -1.14384899692013 \\
  O   -4.42705898987677     -2.02070500500653     -1.14384915062442 \\
  O   -1.47568597759622      4.44575713074093     -1.14384913564841 \\
  O   -1.47568599103004     -8.48716698743733     -1.14384903802719 \\
  O   -1.47342932846807     -2.01418330328451     -1.10266145913369 \\
  O    1.47340941485777      2.01424959129708     -1.10245701604157 \\
  O    1.47568599870760      8.48716698859109     -1.14384901700486 \\
  O    1.47568601465898     -4.44575711023944     -1.14384895493706 \\
  O    4.42705904523879      2.02070500369331     -1.14384919910433 \\
  O    4.42705902121992      8.48716700348174     -1.14384898783613 \\
  O    4.42705900259338     -4.44575701073203     -1.14384886102370 \\
  O   -4.42705898859756      2.02070500004570     -1.14384914877519 \\
  O   -4.42705901074874      8.48716699960451     -1.14384899685186 \\
  O   -4.42705895586464     -4.44575702019817     -1.14384889723095 \\
  O   -1.47343423940872      2.01418145998414     -1.10264594936948 \\
  O   -1.47568599117941      8.48716698767123     -1.14384903795224 \\
  O   -1.47568597674766     -4.44575713312841     -1.14384913749149 \\
  O    0.00000000921509     -0.00000003455969     -2.53598797691956 \\
  O    0.00000001619517      6.46646197253917     -2.53598804072308 \\
  O    0.00000001641093     -6.46646197183536     -2.53598804097881 \\
  O    2.95137304077938      0.00000001255845     -2.53598803156503 \\
  O    2.95137303156725      6.46646202815219     -2.53598799039176 \\
  O    2.95137303163085     -6.46646202762150     -2.53598799025660 \\
  O    5.90274502495889     -0.00000003136874     -2.53598801500686 \\
  O    5.90274501674794      6.46646200262026     -2.53598798423866 \\
  O    5.90274501675803     -6.46646200306094     -2.53598798407056 \\
  O   -5.90274500726011     -0.00000001888304     -2.53598801049268 \\
  O   -5.90274500917213      6.46646200580420     -2.53598799466864 \\
  O   -5.90274500900708     -6.46646200579677     -2.53598799467603 \\
  O   -2.95137300481187      0.00000002939317     -2.53598804202209 \\
  O   -2.95137301170746      6.46646206942397     -2.53598801000032 \\
  O   -2.95137301132879     -6.46646206810451     -2.53598801027489 \\
  O    0.00000000705230      3.23323112484994     -3.27448899338440 \\
  O    0.00000000692697     -3.23323113318748     -3.27448899316328 \\
  O    2.95137303932347      3.23323061567947     -3.27448898713846 \\
  O    2.95137303923722     -3.23323061350478     -3.27448898744769 \\
  O   -2.95137300989793      3.23323066616216     -3.27448897343784 \\
  O   -2.95137300962215     -3.23323066460472     -3.27448897365959 \\
  Ti  -0.00000001162195      0.00000000719210     -4.35009306272108 \\
  Ti   2.95137300931777     -0.00000000303581     -4.35009302899235 \\
  Ti  -2.95137300063178     -0.00000000698152     -4.35009303417005 \\
  Ti   1.47568606828165      3.23323109305462     -4.67344493872511 \\
  Ti   1.47568606734888     -3.23323109384659     -4.67344493913363 \\
  Ti  -1.47568603993947      3.23323093587749     -4.67344493079870 \\
  Ti  -1.47568603862959     -3.23323093726958     -4.67344493166032 \\
  O    1.47568600071726      1.27335999445197     -4.53203000152849 \\
  O    1.47568599203540     -5.19310099026989     -4.53203002928642 \\
  O    4.42705898371532      1.27336000923284     -4.53202998727539 \\
  O   -4.42705898248748      1.27336000387305     -4.53202999242882 \\
  O   -1.47568600009486      1.27336000862481     -4.53202999996074 \\
  O   -1.47568599087681     -5.19310100491913     -4.53203002906535 \\
  O    1.47568600020210     -1.27335999473593     -4.53203000168384 \\
  O    1.47568599218255      5.19310199070383     -4.53203002910117 \\
  O    4.42705898528121     -1.27336000858462     -4.53202999155182 \\
  O   -4.42705898356567     -1.27336000456272     -4.53202999447398 \\
  O   -1.47568599975200     -1.27336000872102     -4.53203000027780 \\
  O   -1.47568599098003      5.19310200538912     -4.53203002893806 \\
  O   -0.00000000054268      3.23323112591555     -5.80005802968366 \\
  O   -0.00000000059716     -3.23323112613562     -5.80005802974106 \\
  O    2.95137299257233      3.23323099824656     -5.80005801410365 \\
  O    2.95137299244160     -3.23323099769881     -5.80005801397542 \\
  O   -2.95137299795451      3.23323102020050     -5.80005801531541 \\
  O   -2.95137299792611     -3.23323101934686     -5.80005801523682 \\
  O    0.00000000021030      0.00000000102177     -6.59439599341023 \\
  O    2.95137299920840     -0.00000000024690     -6.59439600105114 \\
  O   -2.95137299925378     -0.00000000161609     -6.59439599894911 \\
  Ti   1.47568600078677     -0.00000000070055     -7.75408200292984 \\
  Ti  -1.47568600058458      0.00000000009657     -7.75408200227316 \\
  O    1.47568600056250     -1.97526899128993     -7.79713600109999 \\
  O   -1.47568599789120     -1.97526899486407     -7.79713600158007 \\
  O    1.47568600040552      1.97526899132828     -7.79713600110573 \\
  O   -1.47568599808738      1.97526899488193     -7.79713600151935 \\
  O    0.00000000001120      0.00000000182383     -9.15393900193838 \\
  O    2.95137300045044     -0.00000000084414     -9.15393900249790 \\
  O   -2.95137300000963     -0.00000000056382     -9.15393900275358 \\
Ti$>$ 2.723404255    0.00000000    9.69969300   -1.49033600 NEWECP "SDD" END \\ 
Ti$>$ 2.723404255    0.00000000   -9.69969300   -1.49033600 NEWECP "SDD" END \\
Ti$>$ 2.723404255    2.95137300    9.69969300   -1.49033600 NEWECP "SDD" END \\
Ti$>$ 2.723404255    2.95137300   -9.69969300   -1.49033600 NEWECP "SDD" END \\
Ti$>$ 2.723404255    5.90274500    3.23323100   -1.49033600 NEWECP "SDD" END \\
Ti$>$ 2.723404255    5.90274500    9.69969300   -1.49033600 NEWECP "SDD" END \\
Ti$>$ 2.723404255    5.90274500   -9.69969300   -1.49033600 NEWECP "SDD" END \\
Ti$>$ 2.723404255    5.90274500   -3.23323100   -1.49033600 NEWECP "SDD" END \\
Ti$>$ 2.723404255   -5.90274500    3.23323100   -1.49033600 NEWECP "SDD" END \\
Ti$>$ 2.723404255   -5.90274500    9.69969300   -1.49033600 NEWECP "SDD" END \\
Ti$>$ 2.723404255   -5.90274500   -9.69969300   -1.49033600 NEWECP "SDD" END \\
Ti$>$ 2.723404255   -5.90274500   -3.23323100   -1.49033600 NEWECP "SDD" END \\
Ti$>$ 2.723404255   -2.95137300    9.69969300   -1.49033600 NEWECP "SDD" END \\
Ti$>$ 2.723404255   -2.95137300   -9.69969300   -1.49033600 NEWECP "SDD" END \\
Ti$>$ 2.723404255    7.37843200    0.00000000   -1.04868100 NEWECP "SDD" END \\
Ti$>$ 2.723404255    7.37843200    6.46646200   -1.04868100 NEWECP "SDD" END \\
Ti$>$ 2.723404255    7.37843200   -6.46646200   -1.04868100 NEWECP "SDD" END \\
Ti$>$ 2.723404255   -7.37843200    0.00000000   -1.04868100 NEWECP "SDD" END \\
Ti$>$ 2.723404255   -7.37843200    6.46646200   -1.04868100 NEWECP "SDD" END \\
Ti$>$ 2.723404255   -7.37843200   -6.46646200   -1.04868100 NEWECP "SDD" END \\
Ti$>$ 2.723404255    0.00000000    6.46646200   -4.35009300 NEWECP "SDD" END \\
Ti$>$ 2.723404255    0.00000000   -6.46646200   -4.35009300 NEWECP "SDD" END \\
Ti$>$ 2.723404255    2.95137300    6.46646200   -4.35009300 NEWECP "SDD" END \\
Ti$>$ 2.723404255    2.95137300   -6.46646200   -4.35009300 NEWECP "SDD" END \\
Ti$>$ 2.723404255    5.90274500    0.00000000   -4.35009300 NEWECP "SDD" END \\
Ti$>$ 2.723404255    5.90274500    6.46646200   -4.35009300 NEWECP "SDD" END \\
Ti$>$ 2.723404255    5.90274500   -6.46646200   -4.35009300 NEWECP "SDD" END \\
Ti$>$ 2.723404255   -5.90274500    0.00000000   -4.35009300 NEWECP "SDD" END \\
Ti$>$ 2.723404255   -5.90274500    6.46646200   -4.35009300 NEWECP "SDD" END \\
Ti$>$ 2.723404255   -5.90274500   -6.46646200   -4.35009300 NEWECP "SDD" END \\
Ti$>$ 2.723404255   -2.95137300    6.46646200   -4.35009300 NEWECP "SDD" END \\
Ti$>$ 2.723404255   -2.95137300   -6.46646200   -4.35009300 NEWECP "SDD" END \\
Ti$>$ 2.723404255    4.42705900    3.23323100   -4.67344500 NEWECP "SDD" END \\
Ti$>$ 2.723404255    4.42705900   -3.23323100   -4.67344500 NEWECP "SDD" END \\
Ti$>$ 2.723404255   -4.42705900    3.23323100   -4.67344500 NEWECP "SDD" END \\
Ti$>$ 2.723404255   -4.42705900   -3.23323100   -4.67344500 NEWECP "SDD" END \\
Ti$>$ 2.723404255    0.00000000    3.23323100   -7.92502000 NEWECP "SDD" END \\
Ti$>$ 2.723404255    0.00000000   -3.23323100   -7.92502000 NEWECP "SDD" END \\
Ti$>$ 2.723404255    2.95137300    3.23323100   -7.92502000 NEWECP "SDD" END \\
Ti$>$ 2.723404255    2.95137300   -3.23323100   -7.92502000 NEWECP "SDD" END \\
Ti$>$ 2.723404255   -2.95137300    3.23323100   -7.92502000 NEWECP "SDD" END \\
Ti$>$ 2.723404255   -2.95137300   -3.23323100   -7.92502000 NEWECP "SDD" END \\
Ti$>$ 2.723404255    4.42705900    0.00000000   -7.75408200 NEWECP "SDD" END \\
Ti$>$ 2.723404255   -4.42705900    0.00000000   -7.75408200 NEWECP "SDD" END \\
Ti$>$ 2.723404255    0.00000000    0.00000000  -11.01674500 NEWECP "SDD" END \\
Ti$>$ 2.723404255    2.95137300    0.00000000  -11.01674500 NEWECP "SDD" END \\
Ti$>$ 2.723404255   -2.95137300    0.00000000  -11.01674500 NEWECP "SDD" END \\
Ti$>$ +2.000         8.85411800    3.23323100   -1.49033600 NEWECP "SDD" END \\
Ti$>$ +2.000         8.85411800    9.69969300   -1.49033600 NEWECP "SDD" END \\
Ti$>$ +2.000         8.85411800   -9.69969300   -1.49033600 NEWECP "SDD" END \\
Ti$>$ +2.000         8.85411800   -3.23323100   -1.49033600 NEWECP "SDD" END \\
Ti$>$ +2.000        -8.85411800    3.23323100   -1.49033600 NEWECP "SDD" END \\
Ti$>$ +2.000        -8.85411800    9.69969300   -1.49033600 NEWECP "SDD" END \\
Ti$>$ +2.000        -8.85411800   -9.69969300   -1.49033600 NEWECP "SDD" END \\
Ti$>$ +2.000        -8.85411800   -3.23323100   -1.49033600 NEWECP "SDD" END \\
Ti$>$ +2.000        -1.47568600   12.93292400   -1.04868100 NEWECP "SDD" END \\
Ti$>$ +2.000        -1.47568600  -12.93292400   -1.04868100 NEWECP "SDD" END \\
Ti$>$ +2.000         1.47568600   12.93292400   -1.04868100 NEWECP "SDD" END \\
Ti$>$ +2.000         1.47568600  -12.93292400   -1.04868100 NEWECP "SDD" END \\
Ti$>$ +2.000         4.42705900   12.93292400   -1.04868100 NEWECP "SDD" END \\
Ti$>$ +2.000         4.42705900  -12.93292400   -1.04868100 NEWECP "SDD" END \\
Ti$>$ +2.000         7.37843200   12.93292400   -1.04868100 NEWECP "SDD" END \\
Ti$>$ +2.000         7.37843200  -12.93292400   -1.04868100 NEWECP "SDD" END \\
Ti$>$ +2.000        -7.37843200   12.93292400   -1.04868100 NEWECP "SDD" END \\
Ti$>$ +2.000        -7.37843200  -12.93292400   -1.04868100 NEWECP "SDD" END \\
Ti$>$ +2.000        -4.42705900   12.93292400   -1.04868100 NEWECP "SDD" END \\
Ti$>$ +2.000        -4.42705900  -12.93292400   -1.04868100 NEWECP "SDD" END \\
Ti$>$ +2.000         8.85411800    0.00000000   -4.35009300 NEWECP "SDD" END \\
Ti$>$ +2.000         8.85411800    6.46646200   -4.35009300 NEWECP "SDD" END \\
Ti$>$ +2.000         8.85411800   -6.46646200   -4.35009300 NEWECP "SDD" END \\
Ti$>$ +2.000        -8.85411800    0.00000000   -4.35009300 NEWECP "SDD" END \\
Ti$>$ +2.000        -8.85411800    6.46646200   -4.35009300 NEWECP "SDD" END \\
Ti$>$ +2.000        -8.85411800   -6.46646200   -4.35009300 NEWECP "SDD" END \\
Ti$>$ +2.000        -1.47568600    9.69969300   -4.67344500 NEWECP "SDD" END \\
Ti$>$ +2.000        -1.47568600   -9.69969300   -4.67344500 NEWECP "SDD" END \\
Ti$>$ +2.000         1.47568600    9.69969300   -4.67344500 NEWECP "SDD" END \\
Ti$>$ +2.000         1.47568600   -9.69969300   -4.67344500 NEWECP "SDD" END \\
Ti$>$ +2.000         4.42705900    9.69969300   -4.67344500 NEWECP "SDD" END \\
Ti$>$ +2.000         4.42705900   -9.69969300   -4.67344500 NEWECP "SDD" END \\
Ti$>$ +2.000         7.37843200    3.23323100   -4.67344500 NEWECP "SDD" END \\
Ti$>$ +2.000         7.37843200    9.69969300   -4.67344500 NEWECP "SDD" END \\
Ti$>$ +2.000         7.37843200   -9.69969300   -4.67344500 NEWECP "SDD" END \\
Ti$>$ +2.000         7.37843200   -3.23323100   -4.67344500 NEWECP "SDD" END \\
Ti$>$ +2.000        -7.37843200    3.23323100   -4.67344500 NEWECP "SDD" END \\
Ti$>$ +2.000        -7.37843200    9.69969300   -4.67344500 NEWECP "SDD" END \\
Ti$>$ +2.000        -7.37843200   -9.69969300   -4.67344500 NEWECP "SDD" END \\
Ti$>$ +2.000        -7.37843200   -3.23323100   -4.67344500 NEWECP "SDD" END \\
Ti$>$ +2.000        -4.42705900    9.69969300   -4.67344500 NEWECP "SDD" END \\
Ti$>$ +2.000        -4.42705900   -9.69969300   -4.67344500 NEWECP "SDD" END \\
Ti$>$ +2.000         5.90274500    3.23323100   -7.92502000 NEWECP "SDD" END \\
Ti$>$ +2.000         5.90274500   -3.23323100   -7.92502000 NEWECP "SDD" END \\
Ti$>$ +2.000        -5.90274500    3.23323100   -7.92502000 NEWECP "SDD" END \\
Ti$>$ +2.000        -5.90274500   -3.23323100   -7.92502000 NEWECP "SDD" END \\
Ti$>$ +2.000        -1.47568600    6.46646200   -7.75408200 NEWECP "SDD" END \\
Ti$>$ +2.000        -1.47568600   -6.46646200   -7.75408200 NEWECP "SDD" END \\
Ti$>$ +2.000         1.47568600    6.46646200   -7.75408200 NEWECP "SDD" END \\
Ti$>$ +2.000         1.47568600   -6.46646200   -7.75408200 NEWECP "SDD" END \\
Ti$>$ +2.000         4.42705900    6.46646200   -7.75408200 NEWECP "SDD" END \\
Ti$>$ +2.000         4.42705900   -6.46646200   -7.75408200 NEWECP "SDD" END \\
Ti$>$ +2.000         7.37843200    0.00000000   -7.75408200 NEWECP "SDD" END \\
Ti$>$ +2.000        -7.37843200    0.00000000   -7.75408200 NEWECP "SDD" END \\
Ti$>$ +2.000        -4.42705900    6.46646200   -7.75408200 NEWECP "SDD" END \\
Ti$>$ +2.000        -4.42705900   -6.46646200   -7.75408200 NEWECP "SDD" END \\
Ti$>$ +2.000         5.90274500    0.00000000  -11.01674500 NEWECP "SDD" END \\
Ti$>$ +2.000        -5.90274500    0.00000000  -11.01674500 NEWECP "SDD" END \\
Ti$>$ +2.000        -1.47568600    3.23323100  -11.01674500 NEWECP "SDD" END \\
Ti$>$ +2.000        -1.47568600   -3.23323100  -11.01674500 NEWECP "SDD" END \\
Ti$>$ +2.000         1.47568600    3.23323100  -11.01674500 NEWECP "SDD" END \\
Ti$>$ +2.000         1.47568600   -3.23323100  -11.01674500 NEWECP "SDD" END \\
Ti$>$ +2.000         4.42705900    3.23323100  -11.01674500 NEWECP "SDD" END \\
Ti$>$ +2.000         4.42705900   -3.23323100  -11.01674500 NEWECP "SDD" END \\
Ti$>$ +2.000        -4.42705900    3.23323100  -11.01674500 NEWECP "SDD" END \\
Ti$>$ +2.000        -4.42705900   -3.23323100  -11.01674500 NEWECP "SDD" END

\# here goes the adsorbate

end \\
end
\begin{multicols}{2} 
\# include PCF: \\
\%pointcharges "pcf.xyz"

\%scf \\
maxiter 500\\
TolE 1e-6 \\
TolRMSP 5e-5 \\
TolMaxP 1e-4 \\
TolErr 5e-5 \\
CNVSHIFT true \\
LeveLshift 1.0 \\
SHIFTErr 0.05 \\
MAXITER 500

\#\#\# for CASSCF only: \\
\#\#\# ROTATE \\
\#\#\#  $\{ x, y, 90 \}$ \# orbital $x$ and $y$ change position \\
\#\#\# END \\
end

\%geom \\
Maxiter 300  \\
constraints   \# Which atoms should be held frozen \\
\#\{ C 0 C \} \\
\{ C 1:2 C \} \\
\#\{ C 3 C \} \\
\{ C 4:11 C \} \\
\#\{ C 12 C \} \\
\{ C 13:14 C \} \\
\#\{ C 15 C \} \\
\#\{ C 16 C \} \\
\#\{ C 17 C \} \\
\#\{ C 18 C \} \\
\#\{ C 19 C \} \\
\#\{ C 20 C \} \\
\#\{ C 21 C \} \\
\{ C 22:29 C \} \\
\#\{ C 30 C \} \\
\{ C 31:34 C \} \\
\#\{ C 35 C \} \\
\{ C 36:43 C \} \\
\#\{ C 44 C \} \\
\#\{ C 45 C \} \\
\{ C 46:53 C \} \\
\#\{ C 54 C \} \\
\{ C 55:56 C \} \\
\#\{ C 57 C \} \\
\{ C 58:77 C \} \\
\#\{ C 78 C \} \\
\{ C 79:96 C \} \\
\#\{ C 97 C \} \\
\{ C 98:102 C \} \\
\#\{ C 103 C \} \\
\{ C 104:114 C \} \\
\{ C 115:227 C \} \#ECPs

\#\# atoms 24, 33, 60, 74, 77, 81-84

end \\
end
\end{multicols}
\end{footnotesize}

\newpage
\thispagestyle{plain}
\clearpage
\phantomsection
\addcontentsline{toc}{section}{Erklärung}
\paragraph*{Erklärung}$~$ \vspace{1cm}\\
Hiermit versichere ich an Eides statt, dass ich diese Arbeit selbstständig verfasst und keine
anderen als die angegebenen Quellen und Hilfsmittel benutzt habe. Außerdem versichere
ich, dass ich die allgemeinen Prinzipien wissenschaftlicher Arbeit und Veröffentlichung,
wie sie in den Leitlinien guter wissenschaftlicher Praxis der Carl von Ossietzky Universität
Oldenburg festgehalten sind, befolgt habe.
\vspace{2cm}
\\
\begin{tabular}{@{}l@{}}\hline
Nico Yannik Merkt
\end{tabular}
\end{document}